\pdfoutput=1

\documentclass[final]{siamltex}

\def\d{\delta} 
 
\def\e{{\epsilon}}

\def\norm#1{\|#1\|}

\usepackage{algorithm}
\usepackage{algorithmic}
\usepackage{amsmath}
\usepackage{amssymb}
\usepackage{caption}
\usepackage{comment}
\usepackage{graphicx}
\usepackage{float}
\usepackage[latin1]{inputenc}
\usepackage{mathrsfs}
\usepackage{multicol}
\usepackage{multirow}
\usepackage{showkeys,cite}
\usepackage{subfigure}
\usepackage{wrapfig}
\usepackage{xcolor}
\usepackage[textsize=footnotesize]{todonotes}
\usepackage{ulem}
\graphicspath{{ }}

\begin{document}

\title{IMEX HDG-DG: A coupled implicit hybridized discontinuous Galerkin and explicit discontinuous Galerkin  approach for shallow water systems
      \thanks{This research was
    partially supported by DOE grants DE-SC0010518 and
    DE-SC0011118, and the NSF Grant NSF-DMS1620352. We are grateful for the supports.}}

    \author{Shinhoo Kang\thanks{Department of Aerospace Engineering and Engineering Mechanics, The University of Texas at Austin, Austin, TX 78712, USA.} \and Francis X. Giraldo\thanks{Department of Applied Mathematics, Naval Postgraduate School, Monterey, CA 93940, USA} \and Tan Bui-Thanh\thanks{Department of Aerospace Engineering and Engineering Mechanics, and the Institute for Computational Engineering and Sciences, The University of Texas at Austin, Austin, TX 78712, USA.} }

\bibliographystyle{siam}

\newcommand{\TODO}[1]{ \fbox{\parbox{3in}{\bf TODO: #1}}}

\newcommand{\grbf}[1] {\mbox{\boldmath${#1}$\unboldmath}}
\newcommand{\gbf}[1] {\mathbf{#1}}

\newcommand{\beq} {\begin{equation}}
\newcommand{\eeq} {\end{equation}}
\newcommand{\bdm} {\begin{displaymath}}
\newcommand{\edm} {\end{displaymath}}
\newcommand{\bit}{\begin{itemize}}
\newcommand{\eit}{\end{itemize}}
\newcommand{\bde}{\begin{description}}
\newcommand{\ede}{\end{description}}
\newcommand{\bce}{\begin{center}}
\newcommand{\ece}{\end{center}}
\newcommand{\ben} {\begin{enumerate}}
\newcommand{\een} {\end{enumerate}}
\newcommand{\bea} {\begin{eqnarray}}
\newcommand{\eea} {\end{eqnarray}}
\newcommand{\barr} {\begin{array}}
\newcommand{\earr} {\end{array}}
\newcommand{\bean} {\begin{eqnarray*}}
\newcommand{\eean} {\end{eqnarray*}}
\newcommand{\edoc} {\end{document}}
\newcommand{\On}{\hspace{0.2cm} \mbox{on} \hspace{0.2cm}}
\newcommand{\In}{\hspace{0.2cm} \mbox{in} \hspace{0.2cm}}
\newcommand{\Minimize}{\hspace{0.2cm} \mbox{minimize} \hspace{0.2cm}}
\newcommand{\Maximize}{\hspace{0.2cm} \mbox{maximize} \hspace{0.2cm}}
\newcommand{\SubjectTo}{\hspace{0.2cm} \mbox{subject} \hspace{0.15cm}
            \mbox{to} \hspace{0.2cm}}

\def\etal{{\it et al.~}}

\newcommand{\point}[1] {\ensuremath{\boldsymbol{#1}}}
\newcommand{\vvec}[1] {\ensuremath{\boldsymbol{#1}}}
\renewcommand{\vec}[1] {\ensuremath{\boldsymbol{#1}}}
\newcommand{\cvec}[1] {\ensuremath{\boldsymbol{{#1}}}}
\newcommand{\ten}[1] {\ensuremath{\boldsymbol{#1}}}
\newcommand{\tenfour}[1] {\ensuremath{\boldsymbol{\mathsf{#1}}}}
\newcommand{\comp}[1]{\begin{pmatrix}{ #1 }\end{pmatrix}}
\newcommand{\vv}{\ensuremath{\vec{v}}}
\newcommand{\vr}{\left(\ensuremath{\vec{r}}\right)}
\newcommand{\att}{\left(t\right)}
\newcommand{\tvv}{\ensuremath{\tilde{\vec{v}}}}
\newcommand{\ww}{\ensuremath{\vec{w}}}
\newcommand{\GG}{\ensuremath{\ten{G}}}
\newcommand{\FF}{\ensuremath{\boldsymbol{\mathfrak{F}}}}
\newcommand{\tEE}{\ensuremath{\tilde{\vec{E}}}}
\newcommand{\nn}{\ensuremath{\vec{n}}}
\renewcommand{\SS}{\ensuremath{\ten{S}}}
\newcommand{\tSS}{\ensuremath{\tilde{\ten{S}}}}
\newcommand{\cp}{\ensuremath{{c_p}}}
\newcommand{\cs}{\ensuremath{{c_s}}}
\newcommand{\tcs}{\ensuremath{{\tilde{c}_s}}}
\newcommand{\Vp}{\ensuremath{{V^e_p}}}
\newcommand{\Vs}{\ensuremath{{V^e_s}}}
\newcommand{\Ss}{\ensuremath{{S^e_s}}}
\newcommand{\nxnx}[1] {\ensuremath{\nn\times\left(\nn\times{#1}\right)}}

\newcommand{\dd}[2] {\ensuremath{\frac{\partial {#1}}{\partial {#2}}}}
\newcommand{\DD}[2] {\ensuremath{\frac{d {#1}}{d {#2}}}}
\newcommand{\Grad} {\ensuremath{\nabla}}
\newcommand{\Gradv}[1]{\ensuremath{\nabla_{#1}}}
\newcommand{\Div} {\ensuremath{\nabla\cdot}}
\newcommand{\Divv}[1] {\ensuremath{\nabla_{#1}\cdot}}
\newcommand{\Curl} {\ensuremath{\nabla\times}}
\newcommand{\Cten}{\ensuremath{\tenfour{C}}}

\newcommand{\eval}[2][\right]{\relax
  \ifx#1\right\relax \left.\fi#2#1\rvert}

\providecommand{\abs}[1]{\left\lvert#1\right\rvert}
\providecommand{\norm}[1]{\left\lVert#1\right\rVert}

\newcommand{\Nel} {\ensuremath{{N_\text{el}}}}
\newcommand{\Ned} {\ensuremath{{N_\text{ed}}}}
\newcommand{\De} {\ensuremath{{\mathsf{D}^e}}}
\newcommand{\Dep} {\ensuremath{{\mathsf{D}^{e'}}}}
\newcommand{\Dhat} {\ensuremath{\hat{\mathsf{D}}}}
\newcommand{\Dehat} {\ensuremath{{\hat{\mathsf{D}}^e}}}
\newcommand{\half} {\ensuremath{\frac{1}{2}}}
\newcommand{\IN} {\ensuremath{{\mathcal{I}_N}}}
\newcommand{\INe} {\ensuremath{{\mathcal{I}_{N_e}}}}

\newcommand{\mc}[1]{\mathcal{#1}}
\newcommand{\mcC}[1]{\mathcal{#1}}
\newcommand{\mb}[1]{\mathbf{#1}}
\newcommand{\mbb}[1]{\mathbb{#1}}
\newcommand{\mi}[1]{\mathit{#1}}
\newcommand{\mf}[1]{\mathfrak{#1}}
\newcommand{\Oi}[1]{\int_{\mc{D}} #1 \, d\vec{x}}
\newcommand{\TOi}[1]{\int_{0}^T \int_{\mc{D}} #1 \, d\vec{x}dt}
\newcommand{\Ti}[1]{\int_{0}^T #1 \, dt}
\newcommand{\TDei}[1]{\int_{0}^T\int_{\De} #1 \, d\vec{x}dt}
\newcommand{\TDeid}[1]{\int_{0}^T\int_{\De,N} #1 \, d\vec{x}dt}
\newcommand{\Dei}[1]{\int_{\De} #1 \, d\vec{x}}
\newcommand{\DeI}[1]{\int_{D} #1 \, d\vec{x}}
\newcommand{\DeiM}[1]{\int_{\Dhat} #1 \, d\vec{r}}
\newcommand{\DeMd}[1]{\int_{\De,N_e} #1 \, d\vec{x}}
\newcommand{\DeiMd}[1]{\int_{\Dhat,N_e} #1 \, d\vec{r}}
\newcommand{\DeMdN}[1]{\int_{D,N} #1 \, d\vec{x}}
\newcommand{\DeiMdN}[1]{\int_{\Dhat,N} #1 \, d\vec{r}}
\newcommand{\DeiMdNC}[2]{\int_{\Dhat,N_{#2}} #1 \, d\vec{r}}
\newcommand{\pDei}[1]{\int_{\partial \De} #1 \, d\vec{x}}
\newcommand{\pDeiM}[1]{\int_{\partial \Dhat} #1 \, d\vec{r}}
\newcommand{\pDeiMd}[1]{\int_{\partial \Dhat,N_e} #1 \, d\vec{r}}
\newcommand{\pDeiMdN}[1]{\int_{\partial \Dhat,N} #1 \, d\vec{r}}
\newcommand{\pDeiMdNC}[2]{\int_{\partial \Dhat,N_{#2}} #1 \, d\vec{r}}
\newcommand{\pMortar}[2]{\int_{\mc{M}_{#2}} #1 \, d\vec{r}}
\newcommand{\pMortard}[2]{\int_{\mc{M}^{#2},N_{#2}} #1 \, d\vec{r}}
\newcommand{\pMortardd}[3]{\int_{\mc{M}^{#2},N_{#3}} #1 \, d\vec{r}}
\newcommand{\pMortardh}[3]{\int_{\mc{M}_{#2},N_{#3}} #1 \, d\vec{r}}
\newcommand{\pMortardhn}[4]{\int_{{#2}_{#3},N_{#4}} #1 \, d\vec{r}}
\newcommand{\TpDei}[1]{\int_{0}^T\int_{\partial \De} #1 \, d\vec{x}}
\newcommand{\TpDeid}[1]{\int_{0}^T\int_{\partial \De,N} #1 \, d\vec{x}}
\newcommand{\Deid}[1]{\int_{\De,N_e} #1 \, d\vec{x}}
\newcommand{\DeidN}[1]{\int_{\De,N} #1 \, d\vec{x}}
\newcommand{\pDeidN}[1]{\int_{\partial \De,N} #1 \, d\vec{x}}
\newcommand{\pDeid}[1]{\int_{\partial \De,N_e} #1 \, d\vec{x}}
\newcommand{\nor}[1]{\left\| #1 \right\|}
\newcommand{\nort}[1]{{\left\vert\kern-0.25ex\left\vert\kern-0.25ex\left\vert #1 
    \right\vert\kern-0.25ex\right\vert\kern-0.25ex\right\vert}}
\newcommand{\snor}[1]{\left| #1 \right|}
\newcommand{\LRp}[1]{\left( #1 \right)}
\newcommand{\LRs}[1]{\left[ #1 \right]}
\newcommand{\LRa}[1]{\left< #1 \right>}
\newcommand{\LRc}[1]{\left\{ #1 \right\}}
\newcommand{\pp}[2]{\frac{\partial #1}{\partial #2}}
\newcommand{\ppcp}[1]{\frac{\partial #1}{\partial \vec{c_p}}}
\newcommand{\ppcpj}[1]{\frac{\partial #1}{\partial c_p}}
\newcommand{\ppcpmf}[1]{\frac{\partial #1}{\partial \vec{\mf{c_p}}}}
\newcommand{\ppm}[1]{\frac{\partial #1}{\partial \vec{m}}}
\newcommand{\ppho}[3]{\frac{\partial^{#3} #1}{{\partial #2}^{#3}}}
\newcommand{\ppsom}[3]{\frac{\partial^{2} #1}{{\partial #2}{\partial #3}}}
\newcommand{\ppmi}[1]{\frac{\partial #1}{\partial m_i}}
\newcommand{\ppmj}[1]{\frac{\partial #1}{\partial m_j}}
\newcommand{\ppcpm}[1]{\frac{\partial #1}{\partial \vec{c_p}^-}}
\newcommand{\ppcpp}[1]{\frac{\partial #1}{\partial \vec{c_p}^+}}
\newcommand{\ppcs}[1]{\frac{\partial #1}{\partial \vec{c_s}}}
\newcommand{\ppcsm}[1]{\frac{\partial #1}{\partial \vec{c_s}^-}}
\newcommand{\ppcsp}[1]{\frac{\partial #1}{\partial \vec{c_s}^+}}
\newcommand{\td}[2]{\frac{d #1}{d #2}}
\newcommand{\Rvert}[1]{\left. #1 \right|}
\newcommand{\bs}[1]{\boldsymbol{#1}}
\newcommand{\ipDed}[3]{\LRp{ #1, #2}_{\De,N}^{#3}}

\newcommand{\bigO}{\mathcal{O}}

\newcommand{\iOm}[1]{\int_{\Omega} #1 \, d\Omega}
\newcommand{\iGb}[1]{\int_{\Gamma_{b}} #1 \, ds}
\newcommand{\iGs}[1]{\int_{\Gamma_{s}} #1 \, ds}
\newcommand{\iGsy}[1]{\int_{\Gamma_{s}} #1 \, ds(\mb{y})}
\newcommand{\RE}{\mbox{{I}}\hspace{-0.5ex}\mbox{{R}}}
\newcommand{\iC}[1]{\int_{0}^{2\pi} #1 \, d\theta}
\newcommand{\iGr}[1]{\int_{\Gamma_{\infty}} #1 \, ds}

\newcommand{\jump}[1] {\ensuremath{\LRs{\!\!\LRs{{#1}}\!\!}}}
\newcommand{\diff}[1] {\ensuremath{\LRs{#1}}}
\newcommand{\average}[1] {\ensuremath{\LRc{\!\!\{#1\}\!\!}}}

\newcounter{tablecell}

\newcommand*{\numbercell}{
  \stepcounter{tablecell}
  \thetablecell. 
}

\newtheorem{propo}{Proposition}
\newtheorem{coro}{Corollary}
\newtheorem{theo}{Theorem}
\newtheorem{lem}{Lemma}

\newtheorem{defi}{definition}

\newtheorem{rema}{remark}

\newcommand{\figlab}[1]{\label{fig:#1}}
\newcommand{\eqnlab}[1]{\label{eq:#1}}
\newcommand{\theolab}[1]{\label{theo:#1}}
\newcommand{\corolab}[1]{\label{coro:#1}}
\newcommand{\propolab}[1]{\label{propo:#1}}
\newcommand{\lemlab}[1]{\label{lem:#1}}
\newcommand{\defilab}[1]{\label{defi:#1}}
\newcommand{\remalab}[1]{\label{rema:#1}}
\newcommand{\tablab}[1]{\label{tab:#1}}

\newcommand{\figref}[1]{\ref{fig:#1}}
\newcommand{\theoref}[1]{\ref{theo:#1}}
\newcommand{\defiref}[1]{\ref{defi:#1}}
\newcommand{\remaref}[1]{\ref{rema:#1}}
\newcommand{\cororef}[1]{\ref{coro:#1}}
\newcommand{\proporef}[1]{\ref{propo:#1}}
\newcommand{\lemref}[1]{\ref{lem:#1}}
\newcommand{\eqnref}[1]{\eqref{eq:#1}}
\newcommand{\alglab}[1]{\label{alg:#1}}
\newcommand{\algref}[1]{\ref{alg:#1}}
\newcommand{\seclab}[1]{\label{sect:#1}}
\newcommand{\secref}[1]{\ref{sect:#1}}
\newcommand{\tabref}[1]{\ref{tab:#1}}

\renewcommand{\v}{v}
\renewcommand{\u}{u}
\newcommand{\un}{\u_h}
\renewcommand{\e}{e}
\newcommand{\w}{w}
\newcommand{\wb}{\mb{\w}}
\newcommand{\J}{J}
\newcommand{\Jb}{\mb{J}}
\newcommand{\q}{q}
\renewcommand{\r}{r}
\newcommand{\qh}{\hat{\q}}
\newcommand{\qs}{\q^*}
\newcommand{\qht}{\tilde{\qh}}
\newcommand{\qbar}{\overline{\q}}
\newcommand{\qb}{{\mb{\q}}}
\newcommand{\sig}{{\sigma}}
\newcommand{\thetah}{{\hat{\theta}}}
\newcommand{\thetas}{{\theta^*}}
\newcommand{\thetahn}{{\thetah_h}}
\newcommand{\sign}[1]{\text{ sgn}\!\LRp{#1}}
\newcommand{\sigh}{\hat{\sig}}
\newcommand{\sigs}{\sig^*}
\newcommand{\sigb}{{\bs{\sig}}}
\newcommand{\sigbn}{\sigb_h}
\newcommand{\sigbs}{\sigb^*}
\newcommand{\taub}{{\bs{\tau}}}
\newcommand{\kappab}{{\bs{\kappa}}}
\newcommand{\gamb}{{\bs{\gamma}}}
\newcommand{\ut}{\tilde{\u}}
\newcommand{\sigbt}{\tilde{\sigb}}
\newcommand{\vt}{\tilde{\v}}
\newcommand{\taubt}{\tilde{\taub}}
\newcommand{\vb}{\mb{\v}}
\newcommand{\ub}{\mb{\u}}
\newcommand{\ubp}{\mb{\u}^+}
\newcommand{\ubm}{\mb{\u}^-}
\newcommand{\um}{\u^-}
\newcommand{\up}{\u^+}
\newcommand{\eub}{\mb{e}_{\mb{\u}}}
\newcommand{\eq}{e_{\q}}
\newcommand{\eqh}{e_{\qh}}

\newcommand{\m}{{m}}

\newcommand{\ud}{{\dot{\u}}}
\newcommand{\vd}{{\dot{\v}}}
\newcommand{\sigbd}{{\dot{\sigb}}}
\newcommand{\taubd}{\dot{\taub}}

\renewcommand{\d}{{d}}
\newcommand{\R}{{\mathbb{R}}}
\newcommand{\kap}{\kappa}
\newcommand{\F}{\mb{F}}
\newcommand{\f}{f}
\newcommand{\g}{g}
\newcommand{\fb}{\mb{\f}}
\newcommand{\gb}{\mb{\g}}
\newcommand{\gD}{g_D}
\newcommand{\pOmega}{\partial \Omega}
\newcommand{\pOh}{{\pOmega_h}}
\newcommand{\K}{K}
\newcommand{\Kp}{K^+}
\newcommand{\Km}{K^-}
\newcommand{\pK}{{\partial \K}}
\newcommand{\Kj}{{\K_j}}
\newcommand{\Gh}{{\mc{E}_h}}
\newcommand{\Gho}{{\mc{E}_h^o}}
\newcommand{\Ghb}{{\mc{E}_h^{\partial}}}
\newcommand{\Poly}{{\mc{P}}}
\newcommand{\poly}[1]{\mc{P}_{#1}}
\newcommand{\polyb}[1]{\bs{\mc{P}}_{#1}}
\newcommand{\D}{{D}}
\newcommand{\p}{{p}}
\newcommand{\Ts}{{\mb{T}}}
\newcommand{\Vb}{{\mb{V}}}
\newcommand{\V}{{V}}
\newcommand{\Lam}{{\Lambda}}
\newcommand{\Lamb}{\bs{\Lambda}}
\newcommand{\mub}{\bs{\mu}}
\newcommand{\lambdab}{\bs{\lambda}}
\newcommand{\Th}{{\Ts_h}}
\newcommand{\Vh}{{\V_h}}
\newcommand{\Vbh}{{\Vb_h}}
\newcommand{\Lamh}{{\Lam_h}}
\newcommand{\Lambh}{{\Lamb_h}}
\newcommand{\Lamho}{{\overline{\Lam}_h}}
\newcommand{\Lambht}{{\Lamb_h^t}}
\newcommand{\Ltwo}{{L^2\LRp{\Omega}}}
\newcommand{\Ltwoe}{{L^2\LRp{\e}}}
\newcommand{\Ltw}{{L^2\LRp{\K}}}
\newcommand{\VhK}{{\V_h\LRp{\K}}}
\newcommand{\ThK}{{\Ts_h\LRp{\K}}}
\newcommand{\VbhK}{{\Vb_h\LRp{\K}}}
\newcommand{\Lamhe}{{\Lam_h\LRp{\e}}}
\newcommand{\Lambhe}{{\Lamb_h\LRp{\e}}}
\renewcommand{\L}{\mc{L}}
\renewcommand{\D}{\mc{D}}
\newcommand{\T}{\mc{T}}
\newcommand{\Tt}{\tilde{\T}}
\newcommand{\A}{\mb{A}}
\newcommand{\B}{\mb{B}}
\renewcommand{\D}{\mb{D}}
\newcommand{\Am}{\A^-}
\newcommand{\Ap}{\A^+}
\newcommand{\Aa}{\snor{\A}}
\newcommand{\Rb}{\mb{R}}
\newcommand{\C}{\mb{C}}
\renewcommand{\S}{\mb{S}}
\newcommand{\Sm}{\S^{-}}
\newcommand{\Sp}{\S^{+}}
\newcommand{\Spm}{\S^{\pm}}
\newcommand{\Q}{\mb{Q}}

\newcommand{\Lte}{{L^2\LRp{\Gh}}}
\newcommand{\n}{\mb{n}}
\newcommand{\nm}{\n^-}
\newcommand{\np}{\n^+}
\newcommand{\uh}{{\hat{\u}}}
\newcommand{\us}{{\u^*}}
\newcommand{\uhn}{{\uh_h}}
\newcommand{\ubh}{{\hat{\ub}}}
\newcommand{\ubs}{{\ub^*}}
\newcommand{\ubht}{{\tilde{\ubh}}}
\newcommand{\uht}{{\tilde{\uh}}}
\newcommand{\uhtn}{{\uht_h}}
\newcommand{\uhd}{{\dot{\uh}}}
\newcommand{\sigbh}{{\hat{\sigb}}}
\newcommand{\Fh}{{\hat{\F}}}
\newcommand{\Fs}{\F^*}
\renewcommand{\c}{{c}}

\newcommand{\zb}{\mb{z}}
\renewcommand{\sb}{\mb{s}}
\newcommand{\rb}{\mb{r}}
\newcommand{\betab}{\bs{\beta}}
\newcommand{\veps}{{\varepsilon}}
\newcommand{\vepsb}{\bs{\veps}}
\newcommand{\Hb}{\mb{H}}
\newcommand{\Hbt}{\Hb^t}
\newcommand{\Eb}{\mb{E}}
\newcommand{\Ebt}{\Eb^t}
\newcommand{\hb}{\mb{h}}
\newcommand{\eb}{\mb{e}}
\newcommand{\Hbh}{\hat{\mb{H}}}
\newcommand{\Ebh}{\hat{\mb{E}}}
\newcommand{\Hbs}{{\mb{H}^*}}
\newcommand{\Ebs}{{\mb{E}^*}}
\newcommand{\Ebht}{\Ebh^t}
\newcommand{\Hbht}{\Hbh^t}
\newcommand{\Lb}{\mb{L}}
\newcommand{\Gb}{\mb{G}}
\newcommand{\Lbh}{\hat{\Lb}}
\newcommand{\Lbs}{\Lb^*}
\newcommand{\Ib}{\mb{I}}
\newcommand{\Sigb}{\bs{\Sigma}}
\newcommand{\Piu}{\Pi_\u}
\newcommand{\Pisigb}{\Pi_\sigb}
\renewcommand{\Pr}{\mathbb{P}}
\newcommand{\Z}{Z}
\newcommand{\Y}{Y}
\newcommand{\rhou}{\rho\u}
\newcommand{\rhov}{\rho\v}
\newcommand{\E}{E}
\newcommand{\Escript}{\mathsf{\E}}
\renewcommand{\P}{P}
\newcommand{\h}{H}
\newcommand{\vel}{\bs{\vartheta}}
\newcommand{\vels}{\vel^*}
\newcommand{\vele}{\vel^e}
\newcommand{\velh}{\hat{\vel}}
\newcommand{\phis}{\phi^*}
\newcommand{\phie}{\phi^e}
\newcommand{\phih}{\hat{\phi}}
\newcommand{\phio}{\phi^0}
\newcommand{\uo}{\u^0}
\newcommand{\vo}{\v^0}
\newcommand{\vs}{\v^*}
\newcommand{\vh}{\hat{\v}}
\newcommand{\Ic}{\mc{I}}
\newcommand{\ue}{{\u^e}}
\newcommand{\ube}{{\ub^e}}
\newcommand{\ve}{{\v^e}}
\renewcommand{\b}{{b}}

\newcommand{\mygreen}[1]{{\color[rgb]{0,.65,0} #1}}
\newcommand{\mywhite}[1]{{\color[rgb]{1.0,1.0,1.0} #1}}
\newcommand{\myred}[1]{{\color[rgb]{0.65,0.0,0.0} #1}}
\newcommand{\myblue}[1]{{\color[rgb]{0.0,0.0,1.0} #1}}
\newcommand{\mygray}[1]{{\color[rgb]{.15,.35,.60} #1}}
\newcommand{\mythemec}[1]{{\color[RGB]{51,108,121} #1}}
\newcommand{\mydarkred}[1]{{\color[rgb]{.6,.1,.1} #1}}
\newcommand{\mydarkblue}[1]{{\color[rgb]{.1,.1,.9} #1}}
\newcommand{\mygreenback}[1]{{\color[rgb]{.75,1,.75} #1}}
\newcommand{\myorange}[1]{{\color[rgb]{.76,.39,.13} #1}}
\newcommand{\mygrass}[1]{{\color[rgb]{.19,.64,.13} #1}}
\newcommand{\mysierp}[1]{{\color[RGB]{209,28,209} #1}}
 
\newcommand{\ephih}{{\veps^h_\phi}}
\newcommand{\ephiI}{{\veps^I_\phi}}
\newcommand{\ephihh}{{\veps^h_{\phih}}}
\newcommand{\ephihI}{{\veps^I_{\phih}}}
\newcommand{\evelh}{{\bs{\veps}^h_{\vel}}}
\newcommand{\evelhh}{{\bs{\veps}^h_{\Ubh}}}
\newcommand{\evelI}{{\bs{\veps}^I_{\vel}}}
\newcommand{\evelhI}{{\bs{\veps}^I_{\Ubh}}}
\newcommand{\Proj}{{\mathbb{P}}}

\newcommand{\Uhat}{\hat{\U}}
\newcommand{\Vhat}{\hat{\V}}
\newcommand{\What}{\hat{\W}}
\newcommand{\pOmegah}{{\pOmega_h}}
\newcommand{\Omegah}{{\Omega_h}}
\newcommand{\Vscript}{\mathscr{V}}
\newcommand{\vtest}{{\bf v}}
\newcommand{\Np} {\ensuremath{{N_\text{p}}}}
\newcommand{\qbh}{\hat{\mb{\q}}}
\newcommand{\tila}{\tilde{a}}
\newcommand{\tilb}{\tilde{b}}
\newcommand{\tilc}{\tilde{c}}
\newcommand{\xb}{{\bf x}}
\newcommand{\hm}{h^-}
\newcommand{\hp}{h^+}
\newcommand{\s}{s}
\newcommand{\GammaD}{\Gamma_D}
\newcommand{\cGammaD}{\overline{\Gamma}_D}
\newcommand{\GammaN}{\Gamma_N}
\newcommand{\cGammaN}{\overline{\Gamma}_N}
\newcommand{\Fb}{{\bf F}}
\newcommand{\Ub}{{\bf U}}
\newcommand{\Ubh}{\hat{\Ub}}
\newcommand{\Ube}{\Ub^e}
\newcommand{\nb}{{\bf n}}
\newcommand{\U}{U}
\newcommand{\W}{W}
\newcommand{\Fcal}{\mathcal{F}}
\newcommand{\Fcalh}{\hat{\mathcal{F}}}
\newcommand{\Fcals}{\mathcal{F}^*}
\newcommand{\Acal}{\mathcal{A}}
\newcommand{\rvec}{{ \bf \hat{r}}}
\newcommand{\dtt}{\triangle t}
\newcommand{\Qb}{{\bf Q}}
\newcommand{\Qbi}{{\bf Q}^{(i)}}
\newcommand{\Qbip}{{\bf Q}^{(i)}_+}
\newcommand{\Qbim}{{\bf Q}^{(i)}_-}
\newcommand{\Qbj}{{\bf Q}^{(j)}}
\newcommand{\Qbh}{\hat{{\bf Q}}} 
\newcommand{\Qbhi}{\hat{{\bf Q}}^{(i)}}
\newcommand{\Qbhj}{\hat{{\bf Q}}^{(j)}}
\newcommand{\dQbh}{\delta\hat{{\bf Q}}} 
\newcommand{\dQb}{{\delta\bf Q}}
\newcommand{\Lcal}{\mathcal{L}}
\newcommand{\NLcal}{\mathcal{NL}}
\newcommand{\MassMatrix}{{ \bf M}}
\newcommand{\InvMassMatrix}{{ \bf M}^{-1}}
\newcommand{\ab}{{\bf a}}
\newcommand{\bb}{{\bf b}}
\newcommand{\Res}{{\mathcal Res}}
\newcommand{\Flx}{{\mathcal Flx}}
\newcommand{\tauh}{\hat{\tau}}
\newcommand{\ulon}{\u_\lambda}
\newcommand{\ulat}{\u_\theta}
\newcommand{\uinf}{\u_\infty}
\newcommand{\mass}{\text{mass}}
\newcommand{\energy}{\text{energy}}
\newcommand{\Uhatb}{\hat{\mb{U}}}

\graphicspath{{./}}

\maketitle

\begin{abstract}
   We propose IMEX HDG-DG schemes for planar and spherical shallow
   water systems. Of interest is subcritical flow, where the speed of
   the gravity wave is faster than that of nonlinear advection.
   In order to simulate these flows efficiently, we split the
   governing system into a stiff part describing the gravity wave and
   a non-stiff part associated with nonlinear advection. The former is
   discretized implicitly with the HDG method while an explicit Runge-Kutta
   DG discretization is employed for the latter. The proposed IMEX
   HDG-DG framework: 1) facilitates high-order solutions both in time
   and space; 2) avoids overly small time-step sizes; 3) requires only
   one linear system solve per time stage; 4) relative to DG generates smaller and
   sparser linear systems while promoting further parallelism.
   Numerical results for various test cases
   demonstrate that our methods are comparable to explicit Runge-Kutta
   DG schemes in terms of accuracy while allowing for much larger time
   step sizes.
\end{abstract}



\pagestyle{myheadings} \thispagestyle{plain}
\markboth{S. KANG, F.X. GIRALDO, AND T. BUI-THANH}{IMEX HDG-DG SCHEME}

\section{Introduction}

The shallow water equations describe the motion of a thin layer of
incompressible and inviscid fluid.  Because it captures  essential
dynamical characteristics such as nonlinear advection and gravity
waves in geophysical flows, it is widely used in oceanography and atmospheric sciences. 
For the modeling of geophysical flows, spatial discretizations using high order discontinuous
Galerkin (DG) finite element methods have been of considerable 
interest
\cite{blaise2015stabilization,nair2005discontinuous,
  Bui-Thanh16a, Bui-Thanh15, LiLiu01,AizingerDawson02,
  GiraldoWarburton08, GiraldoRestelli10} due to their flexibility in
dealing with complex geometries, high order accuracy, compact stencil, upwind stabilization, etc.
\cite{cockburn1998runge,
riviere2000discontinuous}.
However, DG methods have an important drawback, that is, they have
many degrees of freedoms (DOFs) 
since, by construction, DOFs on interfaces between elements are
duplicated.  Consequently  DG is in general more expensive than other existing
numerical methods, especially for steady-state or time-dependent nonlinear problems.

To tackle the aforementioned problem, 
Cockburn, coauthors, and others have introduced
hybridized (also known as hybridizable) discontinuous Galerkin (HDG)
methods for various types of PDEs including the Poisson equation
\cite{CockburnGopalakrishnanLazarov09,
KirbySherwinCockburn12}, 
convection-diffusion equation
\cite{NguyenPeraireCockburn09a, CockburnDongGuzmanEtAl09}, 
 Stokes equation \cite{CockburnGopalakrishnan09,
  NguyenPeraireCockburn10}, Euler and Navier-Stokes equations
\cite{NguyenPeraireCockburn11, MoroNguyenPeraire11}, Maxwell's equations
\cite{
 LiLanteriPerrrussel13}, acoustics and
elastodynamics equations \cite{NguyenPeraireCockburn11a}, Helmholtz equation
\cite{GriesmaierMonk11}
, and eigenvalue problems
\cite{CockburnLiNguyenEtAl10}, to name a few. 
At the heart of HDG methods is the introduction of trace unknowns on the mesh
skeleton, i.e. the faces, to hybridize the DG method. Once they are computed, the
usual DG (volume) unknowns can be recovered in an element-by-element fashion,
completely independent of each other.
The beauty of HDG
methods is that they reduce the number of coupled unknowns substantially
while retaining all other attractive properties of the DG method.  
Our recent attempt in developing HDG methods for both nonlinear and linearized shallow water systems
has been promising \cite{Bui-Thanh15, Bui-Thanh16a}. 

To fully discretize a time-dependent partial differential equation (PDE),
temporal discretization is also necessary. Explicit time integrators
such as Runge-Kutta methods are popular due to their simplicity and 
ease in computer implementation. However, fast waves, such as acoustic/gravity waves, limit the time-step size
severely for high-order DG methods (see, e.g.,
\cite{GiraldoWarburton08}). For long time integration, which is not
uncommon in geophysical fluid dynamics, this can lead to an excessive number of time steps,
and hence substantially taxing computing and storage resources. On the
other hand, fully-implicit methods could be expensive, especially for
nonlinear PDEs for which Newton-like methods are
required. Semi-implicit time-integrators have been designed to balance
the time-step size restriction due to fast waves and the computational
expense required by nonlinearities
\cite{ascher1997implicit,Kennedy2003additive,pareschi2005implicit}. In
the context of low-speed fluid flows, including Euler, Navier-Stokes, and
shallow water equations, IMEX DG methods have been proposed and proven
to be much more advantageous than either explicit or fully-implicit DG
methods \cite{
  FeistauerDolejsiKucera07, RestelliGiraldo09}. The common feature of these methods is that they
employ implicit time-stepping schemes for the linear(ized) part of the
PDE under consideration that contains the fastest waves, and explicit
time-integrators for the (resulting) nonlinear part for which the
fastest waves are removed. {\em Unlike standard operator splitting methods,
this class of IMEX schemes facilitate high-order solutions both in time
and space}. In particular, they provide the flexibility in employing
separate high-order discretization methods for the fast
linear and for the slow nonlinear operators.

The main goal of this paper is to construct a coupled HDG-DG scheme
under an IMEX framework to overcome the computational burden of the
pure DG IMEX scheme. We start by briefly discussing a class of
implicit-explicit Runge-Kutta (IMEX-RK) time integrators in section
\secref{IMEXRK}.  In section \secref{governingEqn}, we present shallow
water systems for planar and spherical surfaces. Of importance is the
introduction of a linear-nonlinear splitting of the flux tensor to
separate the fast wave. This is done by a linearization of the
flux tensor around the ``lake at rest'' condition (to be
defined). Using an energy method we show that the linearized PDEs are
well-posed, e.g., the total energy is a non-increasing function in
time. Next we present, in detail, a coupled HDG-DG spatial
discretization for the split system in section
\secref{spatialDiscretization}. The well-posedness of the
semi-discrete HDG system and its rigorous convergence analysis are
simultaneously shown for both planar and spherical geometries.
In section \secref{ARK}, we present an IMEX Runge-Kutta method for the
semi-discrete HDG-DG system as well as the procedure for solving the
implicit HDG part. Various numerical
results for the shallow water systems for both planar and spherical flows  will be presented in section
\secref{numericalResults} to confirm the accuracy and efficiency of the proposed IMEX HDG-DG scheme. Finally we conclude the paper and discuss future research
directions in section \secref{conclusion}.

\section{Implicit-Explicit (IMEX) Runge-Kutta methods}
\seclab{IMEXRK}

In this section, we briefly describe the key ideas behind a class of IMEX
Runge-Kutta (IMEX-RK) methods. The readers are referred to
\cite{ascher1997implicit,pareschi2005implicit,Kennedy2003additive,
  weller2013runge} for more details. 
We employ standard letters for scalars, boldface letters for vectors and calligraphic letters for tensors.
Let us begin by considering the following system of ordinary differential  equations
  \begin{align}
    \eqnlab{splitted_ode}
    \DD{\qb}{t} = \fb(\qb) + \gb(\qb), \quad t\in\LRp{0,T},
  \end{align}
with the initial condition $\qb(0)=\qb_0$.
The functions $\fb$ and $\gb$ correspond to the non-stiff (slow time-varying) and the stiff (fast time-varying)
parts, respectively. Note that they could be the result of applying
two different spatial discretizations (e.g. DG and HDG methods as in
this paper) for two differential operators associated with slow and
fast waves. Here, we employ explicit Runge-Kutta methods for the
temporal evolution corresponding to $\fb(\qb)$ and diagonally implicit Runge-Kutta
(DIRK) methods with $s$ stages for the temporal evolution corresponding to
$\gb(\qb)$. Combining these temporal discretizations into one formula gives the
IMEX-RK scheme at the $i$th stage \cite{ascher1997implicit,pareschi2005implicit,Kennedy2003additive}:
\begin{subequations}
   \eqnlab{IMEXRKi}
  \begin{align}
   \eqnlab{IMEXRK_qi}
     \Qb^{(i)} &= \qb^n
       + \dtt\sum_{j=1}^{i-1}a_{ij} \fb_j
       + \dtt\sum_{j=1}^{i}\tilde{a}_{ij} \gb_j,\quad i=1,\hdots,s,\\
   \eqnlab{IMEXRK_qn}
     \qb^{n+1} &= \qb^n
       + \dtt\sum_{i=1}^{s}b_{i} \fb_i
       + \dtt\sum_{i=1}^{s}\tilde{b}_{i} \gb_i,
  \end{align}
\end{subequations}
where $\fb_i = \fb\LRp{t^n+c_i\dtt, \Qb^{(i)}}$, $\gb_i =
\gb\LRp{t^n+\tilde{c}_i\dtt,\Qb^{(i)}}$, $\qb^n = \qb(t^n)$ and $\Qb^{(i)}$
is the $i$th intermediate state; here $\dtt$ is the time-step size.
The scalar coefficients $a_{ij}$, $\tila_{ij}$, $b_i$, $\tilb_i$,
$c_i$ and $\tilc_i$ determine all the properties of a given IMEX-RK
scheme.
The actual forms of the non-stiff term $\fb(\qb)$ and stiff term
$\gb(\qb)$ for our proposed coupled HDG-DG discretization for shallow
water systems will be described in section \secref{ARK}.

\section{Governing Equation}
\seclab{governingEqn}
The homogenous shallow water system in conservative form can be
written as follows
  \begin{subequations}
    \eqnlab{governingeq}
    \begin{align}
      \frac{\partial H}{\partial t}    + \nabla\cdot\LRp{H\ub}&=0,\\
      \frac{\partial \LRp{H\ub}}{\partial t} + \nabla\cdot\LRp{H\ub\otimes\ub + \frac{gH^2}{2}\mathcal{I}_d}&={\bf 0},
    \end{align}
  \end{subequations}
  where
  $H$ is the total water depth,
  $\ub$  the horizontal velocity, 
  $d$  the dimension,
  $\mathcal{I}_d$  the $d\times d$ identity matrix, and 
  $g$ the gravitational acceleration.

  We can rewrite \eqnref{governingeq} as 
  \begin{subequations}
    \eqnlab{governingeq2}
    \begin{align}
      \frac{\partial \phi}{\partial t}    + \nabla\cdot{\Ub}&=0,\\
      \frac{\partial \Ub}{\partial t} + \nabla\cdot\LRp{\frac{\Ub\otimes\Ub}{\phi} + \frac{\phi^2}{2}\mathcal{I}_d}&=0,
    \end{align}
  \end{subequations}
  where $\phi=gH$ is the geopotential height and $\Ub = \phi\ub$. 

  The nonlinear shallow water system \eqnref{governingeq2} has two characteristic time
  scales: nonlinear advection and gravity waves with corresponding
  speeds $\snor{\bf u}$ and $\sqrt{\phi}$, respectively. {\em
    In this paper, we consider subcritical flow ($\snor{\ub}
    <\sqrt{\phi}$), i.e., the differential operator associated with gravity
    waves is stiff.}

\subsection{Planar shallow water equations}
\seclab{SWEplane}
We first consider the two-dimensional shallow water equations on a plane.
We split the total water column $H$ into $\eta$ and $B$ such that $H=\eta + B$,
where $\eta$ is the free surface elevation over a reference plane (positive upward), and
$B$ is the water depth under the reference plane (positive downward), which is assumed to be constant in time. 
Following \cite{GiraldoRestelli10}, the governing equation \eqnref{governingeq2} can be rewritten as 
\begin{subequations}
  \eqnlab{psweGov}
  \begin{align}
    \frac{\partial \qb}{\partial t}+\nabla\cdot{\Fcal}&=\sb & \text{ in } \Omega,\\
    \qb &= \gb_D & \text{ on } \GammaD,\\
    \Fcal \cdot \nb &= \gb_N & \text{ on } \GammaN,
  \end{align}
\end{subequations}
where $\Omega \subset \R^2$ is a {\em planar domain},
$\pOmega=\cGammaD \cup \cGammaN$ is the boundary, $\n=(n_x,n_y)$ 
is outward unit normal vector on $\pOmega$, and $\qb=\LRp{\phi_{\eta},\Ub}^T := \LRp{g\eta,\LRp{\U,\V}}^T$ are the conservative variables. 
Here, $\Fcal=(\Fb_x,\Fb_y)$ defined by
\begin{align}
  \eqnlab{psweFluxTensor}
\Fb_x = 
\LRp{\begin{array}{c}
\U\\
\frac{\U\U}{\phi} + \frac{1}{2}\phi_\eta^2 + \phi_\eta\phi_B\\
\frac{\U\V}{\phi} 
\end{array}},\quad
\Fb_y = 
\LRp{\begin{array}{c}
\V\\
\frac{\V\U}{\phi} \\
\frac{\V\V}{\phi} + \frac{1}{2}\phi_\eta^2 + \phi_\eta\phi_B\\
\end{array}},
\end{align}
is  the flux tensor,
$\sb=
\LRp{0,
\phi_{\eta}\frac{\partial\phi_{B}}{\partial x},
\phi_{\eta}\frac{\partial\phi_{B}}{\partial y} }^T$
the source vector, and
$\phi_{B}=gB$ the reference geopotential height.

We can extract the fast gravity wave term (stiff
  operator), by linearizing the flux tensor \eqnref{psweFluxTensor}
  around the ``lake at rest'' condition, i.e., $\eta=0$ and ${\bf
    u}={\bf 0}$, to obtain the linearized flux $\Fcal_L$ corresponding
  to the fast gravity wave  \cite{RestelliGiraldo09,
    GiraldoRestelli10} as
\begin{align}
  \eqnlab{psweLinearFluxTensor}
     \Fcal_L=
     \LRp{\begin{array}{cc}
         U & V\\
         \phi_{\eta}\phi_{B} & 0\\
         0 & \phi_{\eta}\phi_{B}
       \end{array}
     }.
\end{align}

\subsection{Shallow water equations on a sphere}

In this paper, we are also interested in the shallow water equations on the
Earth surface, and for that reason, we consider the two-dimensional shallow
water equations on the sphere with the Earth radius $a=6.371 \times
10^6 \text{m}$. 
We adopt the Lagrange
multiplier approach
\cite{cote1988lagrange,gravel1994mass,giraldo2002nodal,lauter2008discontinuous}, 
 i.e., we embed the two-dimensional flow on the spherical manifold into
the three-dimensional space $\R^3$. 
The shallow water equation \eqnref{governingeq2}
on the spherical manifold can be cast into the following PDE in $\R^3$
  \begin{align}
    \eqnlab{ssweGov}
    \frac{\partial \qb}{\partial t}+\nabla\cdot{\Fcal}&=\sb & \text{ in } \Omega,
  \end{align}
where $\Omega $ is still the original surface of the sphere but now is
considered 
a subset of $\R^3$,  $\qb:= \LRp{\phi,\Ub}^T := \LRp{\phi,\U,\V,\W}^T$  are the
conservative variables, and 
\begin{align}
  \eqnlab{ssweFluxTensor}
\Fcal=
\LRp{
  \begin{array}{ccc}
    \U & \V & \W \\
    \frac{\U\U}{\phi} + \frac{1}{2}\phi^2 &
    \frac{\V\U}{\phi} &
    \frac{\W\U}{\phi} 
    \\
    \frac{\U\V}{\phi} &
    \frac{\V\V}{\phi} + \frac{1}{2}\phi^2 &
    \frac{\W\V}{\phi} 
    \\
    \frac{\U\W}{\phi} &
    \frac{\V\W}{\phi} &
    \frac{\W\W}{\phi} + \frac{1}{2}\phi^2 
    \\
  \end{array}
}
\end{align}
 is the flux tensor. Here
$\sb = \LRp{0,\sb_\Ub(\qb)}^T$, 
where $\sb_\Ub(\qb) = -\phi\nabla\phi_s - f\rvec\times\Ub + \mu \rb$, 
is the source vector, 
$\f=2\Omega\sin\theta$ is the Coriolis parameter,
 $\Omega$ is the Earth's angular velocity, $\theta$ is the latitude coordinate, $\rb=(x,y,z)$ is the position
vector on the sphere, $\rvec=\rb a^{-1}$ is the unit normal vector on the sphere,
$\phi_{s}$ is the surface topography, and $\mu$ is the Lagrange multiplier.
In this approach, the tangential velocity on the sphere is denoted by $\ub=(u,v,w)$ in the Cartesian
coordinate system. Clearly, the additional degree of freedom allows fluid
particles to depart from the spherical surface.
 One way to avoid this
undesirable effect is to introduce a fictitious force via a Lagrange
multiplier, which is chosen such that the velocity has no radial component on the sphere, i.e. $ \ub\cdot \rb = 0$ \cite{giraldo2002nodal}.
 By taking a dot product of $\rb$ and the momentum equation in \eqnref{ssweGov}, we have
 \begin{align}
   \rb\cdot \dd{\Ub}{t} = \rb\cdot R_\Ub + \mu \rb\cdot\rb,
 \end{align}
where $R_\Ub=-\nabla \cdot \LRp{ \frac{\Ub \otimes \Ub}{\phi} +\frac{\phi^2}{2} \mathcal{I}_3 } -\phi\nabla\phi_s - f\rvec\times\Ub$. 
Using the conditions  $ \ub\cdot \rb = 0$ and $\dd{\rb}{t}=0$, we obtain the Lagrange multiplier
$ \mu = - \frac{\rb \cdot R_\Ub}{a^2}.$
Substituting $\mu$ into the momentum equation yields 
 \begin{align}
   \dd{\Ub}{t} = \LRp{\mathcal{I}_3 - \rvec\rvec^T} R_\Ub,
 \end{align}
which maps the momentum equation onto the local tangential plane.
Note that $\rvec\rvec^T$ is the orthogonal projector that takes vectors to the direction normal to the sphere and, consequently, $(\mathcal{I}_3 - \rvec\rvec^T)$ is the complementary projector which takes all vectors along the tangent to the spherical surface.

  Similar to section \secref{SWEplane}, we extract the fast gravity
  wave by linearizing the flux tensor \eqnref{ssweFluxTensor} around
  the lake at rest condition, i.e., constant background geopotential height $\phi_B$ and zero horizontal velocity. We obtain the linearized flux  $\Fcal_L$  containing the fast gravity waves:
\begin{align}
  \eqnlab{ssweLinearFluxTensor}
     \Fcal_L=
     \LRp{\begin{array}{ccc}
         \U & \V & \W\\
         \phi_B \phi & 0 & 0\\
         0 & \phi_B \phi & 0 \\
         0 & 0 & \phi_B \phi
       \end{array}
     }.
\end{align}

We now show that the dynamics corresponding to the linearized
differential operator (associated with the fast waves) either in
\eqnref{psweLinearFluxTensor} or \eqnref{ssweLinearFluxTensor} is
well-defined.
\begin{lemma}[Stability]
Consider the following linear system of PDEs:
\begin{equation}
\eqnlab{linearizedPDE}
\frac{\partial \qb}{\partial t} + \Div\Fcal_L = 0, \text{ in } \Omega,
\end{equation}
where $\Fcal_L$ is either from \eqnref{psweLinearFluxTensor} or
\eqnref{ssweLinearFluxTensor}.  Suppose \eqnref{linearizedPDE} is
equipped with either wall boundary conditions, i.e. $\Ub\cdot\n = 0$ on
$\pOmega$ where $\n$ is the unit outward normal vector, or periodic boundary conditions, then it is well-defined in
the following sense
\begin{equation}
  \eqnlab{energyConservation}
\frac{\partial {E}}{\partial t} = 0,
\end{equation}
where the energy ${E}$ is defined as ${E} =
\int_\Omega\phi_\eta^2\,d\Omega + \int_\Omega\phi_B^{-1}\Ub\cdot\Ub\,d\Omega$.
\end{lemma}
\begin{proof}
We proceed by an energy approach. Specifically, taking the $L^2$-inner
product of the mass conservation equation with $\phi_\eta$ and the momentum equation with $\phi_B^{-1}\Ub$, and then adding the resulting equations together we have
\[
\half\pp{{E}}{t} + \int_\Omega\phi_\eta\Div\Ub\,d\Omega + \int_\Omega\Ub \cdot \Grad\phi_\eta\,d\Omega = 0,
\]
which yields \eqnref{energyConservation} after integrating the second
term by parts and applying the boundary conditions. That is, the
energy of the linearized shallow water system \eqnref{linearizedPDE}
remains constant over time.
\end{proof}

\section{Spatial Discretization}
\seclab{spatialDiscretization}

\subsection{Finite element definitions and notations}
Let $\Omega$ be either a plane or the  surface of the earth.
We denote by
$\Omega_h := \cup_{i=1}^\Nel \K_i$ the mesh containing a finite
collection of non-overlapping elements, $\K_i$, that partition
$\Omega$.  Here, $h$ is defined as $h := \max_{j\in
  \LRc{1,\hdots,\Nel}}diam\LRp{\Kj}$. Let $\pOmega_h := \LRc{\pK:\K
  \in \Omega_h}$ be the collection of the faces of all elements. Let
us define $\Gh:= \LRc{\e: \e \in \Gho \cup \Ghb }$ as the skeleton of
the mesh which consists of the set of all uniquely defined faces,
where $\Ghb$ is the set of all boundary faces on $\pOmega$, and
$\Gho:=\Gh \setminus \Ghb$ is the set of all interior interfaces. For
two neighboring elements $\Kp$ and $\Km$ that share an interior
interface $\e = \Kp \cap \Km$, we denote by $q^\pm$ the trace of their
solutions on $\e$.  We define $\nm$ as the unit outward normal vector on
the boundary $\pK^-$ of element $\Km$, and $\np = -\nm$ the unit outward
normal of a neighboring element $\Kp$.  On the interior interfaces $\e
\in \Gho$, we define the mean/average operator $\average{\bf v}$, where $\bf v$ is
either a scalar or a vector quantify, as
$\average{{\bf v}}:=\LRp{{\bf v}^- + {\bf v}^+}/2$, 
and the jump operator $\jump{\bf v} := 2  \average{\bf v}$.
 On the boundary faces $\e \in \Ghb$, we define the mean and jump operators as 
$\average{{\bf v}}:={\bf v}, \quad \jump{{\bf v}} :={\bf v}$.

Let $\Poly^\p\LRp{D}$ denote the space of polynomials of degree at
most $\p$ on a domain $D$. Next, we introduce discontinuous
piecewise polynomial spaces for scalars and vectors as
\begin{align*}
\Vh\LRp{\Omega_h} &:= \LRc{v \in L^2\LRp{\Omega_h}:
  \eval{v}_{\K} \in \Poly^\p\LRp{\K}, \forall \K \in \Omega_h}, \\
\Lamh\LRp{\Gh} &:= \LRc{\lambda \in \Lte:
  \eval{\lambda}_{\e} \in \Poly^\p\LRp{\e}, \forall \e \in \Gh},\\
\Vbh\LRp{\Omega_h} &:= \LRc{{\bf v} \in \LRs{L^2\LRp{\Omega_h}}^m:
  \eval{{\bf v}}_{\K} \in \LRs{\Poly^\p\LRp{\K}}^m, \forall \K \in \Omega_h},\\
\Lambh\LRp{\Gh} &:= \LRc{\lambdab \in \LRs{\Lte}^m:
  \eval{\lambdab}_{\e} \in \LRs{\Poly^\p\LRp{\e}}^m, \forall \e \in \Gh}.
\end{align*}
and similar spaces $\VhK$, $\Lamhe$, $\VbhK$, and $\Lambhe$ by
replacing $\Omega_h$ with $\K$ and $\Gh$ with $\e$. Here, $m$ is the
number of components of the vector under consideration.

We define $\LRp{\cdot,\cdot}_\K$ as the $L^2$-inner product on an
element $\K \in \R^d$, and $\LRa{\cdot,\cdot}_{\pK}$ as the
$L^2$-inner product on the element boundary $\pK \in
\R^{d-1}$.  We also define the broken inner products as
$\LRp{\cdot,\cdot}_\Omega := \LRp{\cdot,\cdot}_{\Omega_h} :=
\sum_{\K\in \Omega_h}\LRp{\cdot,\cdot}_\K$ and
$\LRa{\cdot,\cdot}_{\pOmega} := \LRa{\cdot,\cdot}_{\pOmega_h} :=
\sum_{\pK\in \pOmega_h}\LRa{\cdot,\cdot}_\pK$, and on the mesh
skeleton as $\LRa{\cdot,\cdot}_\Gh := \sum_{\e\in
  \Gh}\LRa{\cdot,\cdot}_\e$.

\subsection{DG and HDG spatial discretization}
\seclab{DGHDG}

The DG discretization \cite{LiLiu01,hesthaven2007nodal,GiraldoWarburton08} for either \eqnref{psweGov} or
\eqnref{ssweGov} can be written in the following form:
seek $\qb \in \VbhK$ such that the weak formulation
  \begin{align}
  \eqnlab{WeakForm_DG}
  \LRp{\dd{\qb}{t}, \vtest}_\K
  -\LRp{\Fcal\LRp{\qb}, \Grad{\vtest}}_\K
   + \LRa{\Fcal^*\LRp{\qb^\pm}\cdot\n,\vtest}_\pK
    = \LRp{\sb,\vtest}_\K,
  \end{align}
holds for each element $\K \in \Omega_h$, where $\Fcal^*\LRp{\qb^\pm}$
is a numerical flux \cite{leveque2002finite} such as the  Lax-Friedrichs (i.e.,  Russanov) \cite{rusanov1962calculation} or Roe \cite{Roe81} flux. 
Note that the standard numerical flux $\Fcal^*\LRp{\qb^\pm}$ is
a function of the solution traces $\qb^\pm$ from both sides of $\pK$.
For
convenience, we have ignored the fact that \eqnref{WeakForm_DG} must
hold for all test functions $\vtest \in \VbhK$; throughout this paper,
this should be implicitly understood.

The key idea of the HDG framework is to introduce a new single-valued numerical trace
 $\qbh$ on the mesh skeleton \cite{NguyenPeraireCockburn09a,CockburnGopalakrishnanLazarov09,
MoroNguyenPeraire11,Bui-Thanh15} so that the numerical flux is now the function of the solution in element $\K$ and $\qbh$. In particular, the weak
formulation for the HDG discretization (compared with the DG discretization in
\eqnref{WeakForm_DG}) reads
  \begin{align}
  \eqnlab{WeakForm_HDG}
  \LRp{\dd{\qb}{t}, \vtest}_\K
  -\LRp{\Fcal\LRp{\qb}, \Grad\vtest}_\K
   + \LRa{\Fcalh\LRp{\qb,\qbh}\cdot\n,\vtest}_\pK
    = \LRp{\sb,\vtest}_\K,
  \end{align}
where $\Fcalh$ is a hybridization of the numerical flux
$\Fcal^*\LRp{\qb^\pm}$ in \eqnref{WeakForm_DG}, and $\qbh$
approximates $\qb$ on $\Gh$.  In other words, we have hybridized the DG formulation \eqnref{WeakForm_DG} to obtain the HDG formulation \eqnref{WeakForm_HDG}. Since we introduce a new variable, $\qbh$, we need
one more equation to close the system. To that end, we note that for the
HDG discretization \eqnref{WeakForm_HDG} to be conservative the HDG
flux $\Fcalh$ needs to be continuous across the mesh skeleton. Thus, a
natural equation (a sufficient condition for conservation)
is a weak continuity of the HDG normal flux on each interface $\e
\in \pK$, i.e.,
\begin{equation}
\eqnlab{HDGcont}  \LRa{\jump{\Fcalh\LRp{\qb,\qbh} \cdot \n},\mub}_\e = \mb{0},
\end{equation}
for all $\mub \in \Lambh(e)$.  By summing \eqnref{WeakForm_HDG} over
all elements and \eqnref{HDGcont} over the mesh skeleton we obtain the
complete HDG discretization: find the approximate solution $(\qb,\qbh) \in
\Vbh(\Omegah) \times \Lambh(\Gh)$ such that
\begin{subequations}
  \eqnlab{HDGsystem_all}
  \begin{align}
  \eqnlab{HDGsystem_all_local}
  \LRp{\dd{\qb}{t}, \vtest}_\Omegah
  -\LRp{\Fcal\LRp{\qb}, \Grad\vtest}_\Omegah
   + \LRa{\Fcalh\LRp{\qb,\qbh}\cdot\n,\vtest}_\pOmegah
    = \LRp{\sb,\vtest}_\Omegah,\\
  \eqnlab{HDGsystem_all_global}
    \LRa{\jump{\Fcalh\LRp{\qb,\qbh} \cdot \n},\mub}_\Gh = \mb{0}, 
  \end{align}
\end{subequations}
for all $({\bf v},\mub)\in \Vbh(\Omegah) \times \Lambh(\Gh)$, where
the numerical flux $\Fcalh$ can be defined as \cite{CockburnGopalakrishnan09,NguyenPeraireCockburn09,Bui-Thanh15}
\begin{equation}
\eqnlab{RiemannFluxBoth}
\Fcalh\LRp{\qb,\qbh} = \Fcal\LRp{\qb} + \tau \LRp{\qb - \qbh}\otimes \n,
\end{equation}
with $\tau$ as the stabilization parameter (to be described in detail later).

\subsection{Coupled HDG-DG spatial discretization}

As discussed in section \secref{governingEqn} we decompose the nonlinear
differential operator associated with the shallow water equations into
a linear (stiff) part $\Div{\Fcal_L}$ and a nonlinear (non-stiff)
part $\Div{\LRp{\Fcal - \Fcal_L}}$. 
Unlike most of the existing literature, our
decomposition is on the continuous level instead of the discrete
one. The advantage of this strategy is that it allows one to
employ two separate spatial discretizations for the stiff and
non-stiff parts, respectively. In this paper, we choose HDG for the
former and DG for the latter. Clearly, we can choose DG
\cite{blaise2015stabilization,nair2005discontinuous,
  Bui-Thanh16a, Bui-Thanh15, LiLiu01,AizingerDawson02,
  GiraldoWarburton08, GiraldoRestelli10}
 for the former
as well but, as will be shown, HDG provides several advantages over DG
including lower storage and more efficiency. 
The coupled HDG-DG discretization (see section \secref{DGHDG}) of the decomposed system reads:
   seek  $(\qb,\qbh) \in
\Vbh(\Omegah) \times \Lambh(\Gh)$ such that
  \begin{subequations}
    \eqnlab{SemiDiscretizedForm_ode}
\begin{align}
      \LRp{\pp{\qb}{t},\vtest}_\Omegah 
       &= \NLcal(\qb) + \Lcal(\qb,\qbh), \\
\eqnlab{SemiDiscretizedForm_odeJ}
\LRa{\jump{\Fcalh_L\LRp{\qb,\qbh} \cdot \n},\mub}_\Gh &= 0,
\end{align}
  \end{subequations}
  where 
  \begin{align*}
    \NLcal(\qb) &= \LRp{\Fcal_{NL}\LRp{\qb}, \Grad{\vtest}}_\Omegah
                  - \LRa{\Fcals_{NL}\LRp{\qb^\pm} \cdot \n,\vtest}_\pOmegah,\\
    \Lcal(\qb)  &= \LRp{\Fcal_L\LRp{\qb}, \Grad{\vtest}}_\Omegah
                  + \LRp{\sb(\qb),\vtest}_\Omegah
                  - \LRa{\Fcalh_L\LRp{\qb,\qbh} \cdot \n,\vtest}_\pOmegah.
  \end{align*}
Here, $\Fcal_{NL}:=\Fcal - \Fcal_L$, $\Fcal_{NL}^*:=\Fcal^* -
\Fcal_L^*$ is a nonlinear DG numerical flux, and $\Fcalh_L$ is a
linear HDG numerical flux. We now present a choice for these numerical
fluxes.

 For the two-dimensional shallow water equations on a plane, we choose the Lax-Friedrichs 
 numerical flux \cite{rusanov1962calculation,toro1992riemann} for the DG
 discretization and the upwind HDG flux \cite{Bui-Thanh16a}:
  \begin{subequations}
  \eqnlab{NumericalFluxCombination1}
  \begin{align}
    \eqnlab{Full_russanov}
      \Fcal^*\LRp{\qb} = \average{\Fcal(\qb)} + \frac{\tau^*}{2}\jump{\qb\otimes \n},\\
    \eqnlab{Linear_russanov}
      \Fcal_L^*\LRp{\qb} = \average{\Fcal_L(\qb)} + \frac{\tau_L^*}{2}\jump{\qb\otimes \n},\\
    \eqnlab{Linear_upwindhdg3}
      \Fcalh_L\LRp{\qb,\qbh} = \Fcal_L(\qb) + \tauh(\qb-\qbh)\otimes \n,
  \end{align}
  \end{subequations}
where
 $\tau^* = \max\LRp{  \LRp{|\ub\cdot\n|+\sqrt{\phi}}^+, \LRp{|\ub\cdot\n|+\sqrt{\phi}}^- }$,
$\tau_L^* = \max\LRp{\sqrt{\phi_B^+},\sqrt{\phi_B^-} }$, and 
$\tauh = \tau_L^*$.
Note that $\tau^*$ is the (advection + gravity) wave speed of the
shallow water equations, while $\tau_L^*$ is the (gravity) wave speed of the stiff term.
Here, a hybridized Lax-Friedrichs flux\footnote{Note that for polygonal domain $\Omega$, \eqnref{Linear_russanov} and \eqnref{Linear_upwindhdg3} are the same \cite{Bui-Thanh16a}, and hence there is no splitting error. Otherwise, the splitting error is of order $\mc{O}\LRp{h^{p+\half}}$, which is the same as the convergence order, and thus not affecting the convergence rate of the whole scheme.}, \eqnref{Linear_upwindhdg3}, 
is defined \cite{Bui-Thanh16a} as
  \begin{align}
    \eqnlab{linear_hdgflux3}
    \n\cdot\Fcalh_L\LRp{\qb,\qbh} =
      \begin{pmatrix}
        n_x \U + n_y \V + \sqrt{\phi_B}\LRp{\phi_\eta - \phih_\eta}\\ 
        n_x \phi_B \phi_\eta + \sqrt{\phi_B}\LRp{\U - \Uhat}\\
        n_y \phi_B \phi_\eta + \sqrt{\phi_B}\LRp{\V - \Vhat}
      \end{pmatrix}
      .
  \end{align}

 For the two-dimensional shallow water equations on a sphere, 
 the Lax-Friedrichs flux for DG methods has the same form as \eqnref{Full_russanov} and \eqnref{Linear_russanov}, while
the hybridized Lax-Friedrichs flux is defined as
  \begin{align}
    \eqnlab{linear_hdgflux4}
    \n\cdot\Fcalh_L\LRp{\qb,\qbh} =
      \begin{pmatrix}
        n_x \U + n_y \V + n_z \W + \sqrt{\phi_B}\LRp{\phi - \phih}\\ 
        n_x \phi_B \phi+ \sqrt{\phi_B}\LRp{\U - \Uhat}\\
        n_y \phi_B \phi+ \sqrt{\phi_B}\LRp{\V - \Vhat}\\
        n_z \phi_B \phi+ \sqrt{\phi_B}\LRp{\W - \What}
      \end{pmatrix}.
  \end{align}

  We next show that the HDG discretization with hybridized
Lax-Friedrichs flux is stable. Without loss of generality, we can
ignore the source term $\sb$.  For periodic boundary condition (or
similarly no boundary in  spherical cases), all faces are
interior faces, and hence no special treatment is needed. To enforce the wall boundary condition, we use a
reflection principle. In
particular, for an element $\Km$ that is adjacent to the domain
boundary, i.e. $\pK^- \cap \pOmega \ne \emptyset$, we assume that there is an
imaginary neighbor element $\Kp$ whose state $\qb^+ = \LRp{\phi^+,
  \Ub^+}^T$ is determined as
\begin{subequations}
\eqnlab{reflection}
\begin{align}
\phi^+ &= \phi^-, \\
\Ub^+ &= \Ub^- - 2\LRp{\Ub^-\cdot\n^-}\n^-,
\end{align}
\end{subequations}
which, together with the conservation condition
\eqnref{SemiDiscretizedForm_odeJ} and the HDG flux
\eqnref{linear_hdgflux3} (or \eqnref{linear_hdgflux4}) on boundary
faces $\e\in \pK^- \cap \pOmega$, leads to
\begin{subequations}
\eqnlab{reflectionBC}
  \begin{align}
        \LRa{\Ub\cdot\n+ \sqrt{\phi_B}\LRp{\phi - \phih},\mu}_{\pK^- \cap \pOmega} &= 0,\\ 
        \LRa{\sqrt{\phi_B}\LRp{\Ub^t - \hat{\Ub}^t},\bs{\mu}^t}_{\pK^- \cap \pOmega} &= 0, \quad \hat{\Ub}\cdot\n = 0,
\end{align}
\end{subequations}
where the superscript ``t'' denotes the tangential part. In what follows, 
we adapt the energy analysis in
\cite{Bui-Thanh16a} to prove stability and convergence
of the HDG discretization with hybridized Lax-Friedrichs fluxes.

\begin{lemma}[Semi-discrete stability]
  \lemlab{sdStability}
    Consider the following semi-discrete system for the linear part using the HDG discretization
  \begin{subequations}
    \eqnlab{sdHDG}
    \begin{align}
      \eqnlab{sdHDGlocal}
      \LRp{\pp{\qb}{t},\vtest}_\Omegah 
       &= \LRp{\Fcal_L\LRp{\qb}, \Grad{\vtest}}_\Omegah
      - \LRa{\Fcalh_L\LRp{\qb,\qbh} \cdot \n,\vtest}_\pOmegah, \\
      \eqnlab{sdHDGglobal}
\LRa{\jump{\Fcalh_L\LRp{\qb,\qbh} \cdot \n},\mub}_\Gho &= 0,
\end{align}
  \end{subequations}
  where $\Fcal_L$ and $\Fcalh_L$ are defined in
  \eqnref{psweLinearFluxTensor} and \eqnref{linear_hdgflux3} for
  planar flow or in \eqnref{ssweLinearFluxTensor} and
  \eqnref{linear_hdgflux4} for spherical flow. The system
  \eqnref{sdHDG} is stable in the sense that the discrete
  total energy $E^h := \nor{\phi}^2_\Omegah +
  \nor{\Ub}^2_{\Omegah,\phi_B^{-1}}$ is non-increasing over time, i.e.,
  \[
  \frac{\partial E^h}{\partial t} \le 0.
  \]
  \end{lemma}
  \begin{proof}
We  take $\vtest = \LRp{\phi,\phi_B^{-1}\Ub}$, and  integrate by parts the first
term on the right hand side  of the mass conservation part of \eqnref{sdHDGlocal}, adding the resulting equations
together, and summing over all elements we obtain
\[
\half\frac{\partial {E}^h}{\partial t} = -\LRa{\Ub\cdot\n + \phi_B^\half\LRp{\phi-\phih},\phi}_\pOmegah
-\LRa{\phi_B^{-\half}\LRp{\Ub - \hat{\Ub}},\Ub}_\pOmegah,
\]
which, together with the boundary condition \eqnref{reflectionBC}, leads to
\begin{multline}
\half\frac{\partial {E^h}}{\partial t} = -\LRa{\Ub\cdot\n + \phi_B^\half\LRp{\phi-\phih},\phi}_{\pOmegah\cap\Gho}
-\LRa{\phi_B^{-\half}\LRp{\Ub - \hat{\Ub}},\Ub}_{\pOmegah\cap\Gho} \\
-\nor{\LRp{\Ub-\hat{\Ub}}\cdot\n}^2_{\Ghb,\phi_B^{-\half}}.
\eqnlab{localTotal}
\end{multline}
On the other hand, taking $\bs{\mu} = \LRp{\phih,\phi_B^{-1}\hat{\Ub}}$ and summing over all the interior faces $\e$ in the mesh skeleton, i.e. $\e \in \Gho$, yields
\begin{equation}
  \eqnlab{globalTotal}
 0 = \LRa{\Ub\cdot\n + \phi_B^\half\LRp{\phi-\phih},\phih}_{\pOmegah\cap\Gho}
+\LRa{\phi\n + \phi_B^{-\half}\LRp{\Ub - \hat{\Ub}},\hat{\Ub}}_{\pOmegah\cap\Gho}.
\end{equation}
Now adding \eqnref{localTotal} and \eqnref{globalTotal} and using the
fact that $\LRa{\phih,\hat{\Ub}\cdot\n}_{\pOmegah\cap\Gho} = 0$ we arrive at
\begin{multline*}
\half\frac{\partial {E^h}}{\partial t} = -\nor{\phi-\phih}^2_{\pOmegah\cap\Gho,\phi_B^\half} -\nor{\Ub-\hat{\Ub}}^2_{\pOmegah\cap\Gho,\phi_B^{-\half}} \\- \LRa{\phi-\phih,\LRp{\Ub - \hat{\Ub}}\cdot\n}_{\pOmegah\cap\Gho} - \nor{\LRp{\Ub-\hat{\Ub}}\cdot\n}^2_{\Ghb,\phi_B^{-\half}},
\end{multline*}
which, after using the Cauchy-Schwarz inequality for the third term on the
right hand side, ends the proof, i.e.,
\begin{multline*}
\half\frac{\partial {E^h}}{\partial t} \le -\half\nor{\phi-\phih}^2_{\pOmegah\cap\Gho,\phi_B^\half} -\half\nor{\Ub-\hat{\Ub}}^2_{\pOmegah\cap\Gho,\phi_B^{-\half}} \\ - \nor{\LRp{\Ub-\hat{\Ub}}\cdot\n}^2_{\Ghb,\phi_B^{-\half}} \le 0.
\end{multline*}
  \end{proof}
\begin{corollary}[Well-posedness]
At any point in time, the HDG
scheme \eqnref{sdHDG} is well-posed. In particular, there exists a unique HDG solution. 
\end{corollary}
\begin{proof}
Since the HDG solution $\qb = \LRp{\phi,\Ub}$ resides in a finite element space with finite dimensions, well-posedness is equivalent to
uniqueness. Furthermore, it is sufficient to show that HDG solutions vanish for
zero initial condition. Integrating the last inequality in the proof of Lemma \lemref{sdStability}
from $0$ to $t$ we have
\begin{multline*}
  {E}^h\LRp{t} \le -\half\int_{0}^t\nor{\phi-\phih}^2_{\pOmegah\cap\Gho,\phi_B^\half}\,dt 
  -\half\int_{0}^t\nor{\Ub-\hat{\Ub}}^2_{\pOmegah\cap\Gho,\phi_B^{-\half}}\,dt \\
  -\int_{0}^t \nor{\LRp{\Ub-\hat{\Ub}}\cdot\n}^2_{\Ghb,\phi_B^{-\half}}\,dt,
\end{multline*}
whose left hand side is non-negative and right hand side is
non-positive. This can only be true if both vanish,
i.e. ${E}^h\LRp{t} = 0$ and $\nor{\phi-\phih}^2_{\pOmegah\cap\Gho,\phi_B^\half} = \nor{\Ub-\hat{\Ub}}^2_{\pOmegah\cap\Gho,\phi_B^{-\half}} = \nor{\LRp{\Ub-\hat{\Ub}}\cdot\n}^2_{\Ghb,\phi_B^{-\half}} = 0$.
Combining this result and  the boundary condition \eqnref{reflectionBC}
we conclude $\phi = 0$, $\Ub =
\mb{0}$, $\phih = 0$, and $\hat{\Ub} = \mb{0}$, and hence demonstrate uniqueness.
\end{proof}

We are now in the position to prove the convergence of the
semi-discrete HDG discretization. To that end, let us denote by
$\Proj$ and $\Pi$ the local $L^2$-projections on an element and an
edge, respectively. The following errors between the $L^2$-projection
of the exact solution and the HDG solution (and the exact solution respectively) are useful for our error analysis:
\begin{align*}
\ephih &:= \Proj\phie - \phi, \quad \ephiI := \phie - \Proj\phie, \\
\ephihh &:= \Pi\phie - \phih, \quad \ephihI := \phie - \Pi\phie, \\
\evelh &:= \Proj\Ub^e - \Ub, \quad \evelI := \Ub^e - \Proj\Ub^e, \\
\evelhh &:= \Pi\Ube - \Ubh, \quad \evelhI := \Ube - \Pi\Ube, \\
\Sigma &:= \nor{\ephih}_{\Omega_h}^2 + \nor{\evelh}_{\Omega_h,\phi_B^{-1}}^2.
\end{align*}

\begin{theorem}[Convergence]
  Assume $\eval{\qb^e}_\K = \eval{\LRp{\phi^e,\vel^e}}_\K \in
\LRs{H^{s}\LRp{\K}}^3, s \ge 3/2$. There exists a constant $c$ that depends only on the angle
condition of $\K$, $s$, and on $\phi_B$ such that
\begin{equation}
\eqnlab{noForcingEstimateF}
\Sigma\LRp{t} \le c \frac{h^{2\sigma - 1}}{\p^{2s-1}} t \max_{\theta
  \in \LRs{0,t}}{E}^e\LRp{\theta},
\end{equation}
with $\sigma = \min\LRc{\p+1,s}$ and
\[
{E}^e\LRp{t} := \sum_\K \nor{\phi^e\LRp{t}}_{H^{s}\LRp{\K}}^2 + \nor{\vel^e\LRp{t}}_{H^{s}\LRp{\K}}^2.
\]
  \end{theorem}
\begin{proof}
Using the fact that the exact solution satisfies the shallow water
system we can rewrite the HDG system \eqnref{sdHDG} in terms of the errors as
\begin{subequations}
  \eqnlab{sdError}
  \begin{align}
    \eqnlab{sdErrorLocal}
    &\LRp{
      \pp{}{t}
      \LRp{
        \begin{array}{c}
          \ephih \\
          \evelh
        \end{array}
      }
      ,\vtest}_\Omegah 
    = \LRp{
      \Fcal_L\LRp{
        \begin{array}{c}
          \ephih \\
          \evelh
        \end{array}
      }
      , \Grad{\vtest}}_\Omegah \\
    & - \LRa{\LRp{
      \begin{array}{c}
        \evelh\cdot\n + \sqrt{\phi_B}\LRp{\ephih - \ephihh }\\
        \phi_B\ephih \n + \sqrt{\phi_B}\LRp{\evelh - \evelhh}
      \end{array}
      }
      ,\vtest}_\pOmegah 
    - \LRa{
        \LRp{
          \begin{array}{c}
            \evelI\cdot\n + \sqrt{\phi_B}\ephiI \\
            \phi_B\ephiI \n + \sqrt{\phi_B}\evelI
          \end{array}
        } 
       ,\vtest}_\pOmegah,\nonumber \\
      \eqnlab{sdErrorGlobal}
&      \LRa{\jump{
          \begin{array}{c}
            \evelh\cdot\n + \sqrt{\phi_B}\LRp{\ephih - \ephihh }\\
            \phi_B\ephih \n + \sqrt{\phi_B}\LRp{\evelh - \evelhh}
          \end{array}
      },\mub}_\Gho = -
\LRa{\jump{
\begin{array}{c}
  \evelI\cdot\n + \sqrt{\phi_B}\ephiI \\
  \phi_B\ephiI \n + \sqrt{\phi_B}\evelI
\end{array} 
  }
,\mub}_\Gho.
\end{align}
  \end{subequations}
Now first taking $\vtest = \LRp{\ephih,\phi_B^{-1}\evelh}$ in
\eqnref{sdErrorLocal} and $\bs{\mu} = \LRp{\ephihh,
  \phi_B^{-1}\evelhh}$ in \eqnref{sdErrorGlobal}, and then using a similar energy argument as in the
  proof of Lemma \lemref{sdStability} we obtain
  \begin{multline*}
    \half\frac{\partial {\Sigma}}{\partial t} = -\nor{\ephih-\ephihh}^2_{\pOmegah\cap\Gho,\phi_B^\half} -\nor{\evelh-\evelhh}^2_{\pOmegah\cap\Gho,\phi_B^{-\half}} \\- \LRa{\ephih-\ephihh,\LRp{\evelh - \evelhh}\cdot\n}_{\pOmegah\cap\Gho} - \nor{\evelh\cdot\n}^2_{\Ghb,\phi_B^{-\half}} \\
    -\LRa{\evelI\cdot\n + \sqrt{\phi_B}\ephiI, \ephih - \ephihh}_{\pOmegah\cap\Gho} 
-\LRa{\phi_B\ephiI \n + \sqrt{\phi_B}\evelI, \evelh - \evelhh}_{\pOmegah\cap\Gho} \\
-\LRa{\ephiI + \phi_B^{-\half}\LRp{\evelI + \evelhI}\cdot\n, \evelh\cdot\n}_{\Ghb},
\end{multline*}
which, after applying Cauchy-Schwarz for the third term on the right hand side, leads to
  \begin{multline*}
    \half\frac{\partial {\Sigma}}{\partial t} \le -\half\nor{\ephih-\ephihh}^2_{\pOmegah\cap\Gho,\phi_B^\half} -\half\nor{\evelh-\evelhh}^2_{\pOmegah\cap\Gho,\phi_B^{-\half}}  - \nor{\evelh\cdot\n}^2_{\Ghb,\phi_B^{-\half}} \\
    -\LRa{\evelI\cdot\n + \sqrt{\phi_B}\ephiI, \ephih - \ephihh}_{\pOmegah\cap\Gho} 
-\LRa{\ephiI \n + \phi_B^{-\half}\evelI, \evelh - \evelhh}_{\pOmegah\cap\Gho} \\
-\LRa{\ephiI + \phi_B^{-\half}\LRp{\evelI + \evelhI}\cdot\n, \evelh\cdot\n}_{\Ghb},
\end{multline*}
which, in turn, becomes
  \begin{multline}
\eqnlab{conv}
    \half\frac{\partial {\Sigma}}{\partial t} \le \half\nor{\phi_B^{-1/4}\evelI\cdot\n + \phi_B^{1/4}\ephiI}^2_{\pOmegah\cap\Gho} + \half\nor{\phi_B^{1/4}\ephiI \n + \phi_B^{-1/4}\evelI}^2_{\pOmegah\cap\Gho} \\ \frac{1}{4} \nor{\phi_B^{1/4}\ephiI + \phi_B^{-1/4}\LRp{\evelI + \evelhI}\cdot\n}^2_{\Ghb}
\end{multline}
after completing squares and ignoring negative square terms on the
right hand side. We observe that the right hand side of \eqnref{conv}
involves the projection errors of the exact solution on the mesh
skeleton. Using interpolation/projection error analysis from
\cite{BabuskaSuri87,
BabuskaSuri94} we conclude that
there exists a positive  constant $c$ depending only on the angle condition of
$\K$, $s$, and on $\phi_B$ such that
\[
\frac{\partial {\Sigma}\LRp{t}}{\partial t} \le c \frac{h^{2\sigma - 1}}{\p^{2s-1}} \max_{\theta
  \in \LRs{0,t}}{E}^e\LRp{\theta},
\]
which ends the proof.
\end{proof}

\section{Temporal Discretization}
\seclab{ARK}
   
In this section, we adapt the general IMEX-RK idea in section
\secref{IMEXRK} to the semi-discrete system
\eqnref{SemiDiscretizedForm_ode}. In particular, the $i$th stage
IMEX-RK stated in \eqnref{IMEXRKi}, when specified to \eqnref{SemiDiscretizedForm_ode}, reads
\begin{subequations}
  \begin{align}
   \eqnlab{hdgIMEXRK_qi}
     \Qb^{(i)} &= \qb^n
       + \dtt\sum_{j=1}^{i-1}a_{ij} \InvMassMatrix\NLcal_j
       + \dtt\sum_{j=1}^{i}\tilde{a}_{ij} \InvMassMatrix\Lcal_j, \quad i=1,\hdots,s,\\
   \eqnlab{hdgIMEXRK_qn}
     \qb^{n+1} &= \qb^n
       + \dtt\sum_{i=1}^{s}b_{i} \InvMassMatrix\NLcal_i
       + \dtt\sum_{i=1}^{s}\tilde{b}_{i} \InvMassMatrix\Lcal_i,
  \end{align}
\end{subequations}
where $\NLcal_i:=\NLcal\LRp{\Qbi}$ and
$\Lcal_i:=\Lcal\LRp{\Qbi,\Qbhi}$.  Due to the last term on the right-hand
 side, the $i$th stage equation \eqnref{hdgIMEXRK_qi} is implicit
in both $\Qbi$ and $\Qbhi$.  They can be solved for by combining
\eqnref{hdgIMEXRK_qi} and \eqnref{SemiDiscretizedForm_odeJ}. Since $\Lcal_j$ is a result of the HDG discretization, this combination is nothing more than
an HDG discretization with the local equation and the conservation condition defined as
  \begin{subequations}
  \eqnlab{IMEXRKDiscretizedForm}
  \begin{align}
    \eqnlab{IMEXRKDiscretizedForm_localSolver}
      \Qbi -\dtt \tilde{a}_{ii} \InvMassMatrix \Lcal_i = \Res_0, \\
     \eqnlab{IMEXRKDiscertizedForm_globalSolver}
       \LRa{\jump{\Fcalh_L\LRp{\Qbi,\Qbhi} \cdot \n},\mub}_\Gh = \mb{0}, 
  \end{align}
  \end{subequations}
  where $\Res_0= \qb^n + \dtt \InvMassMatrix
                        \sum_{j=1}^{i-1}\LRp{ a_{ij} \NLcal_j
                                    + \tilde{a}_{ij} \Lcal_j }.$

To solve the
                        HDG system \eqnref{IMEXRKDiscretizedForm},
                        we note that both equations are linear in 
                        $\Qbi$ and $\Qbhi$, and can be written as a coupled linear system.
  We define 
  \begin{subequations}
  \eqnlab{linearizedHDGsystem_Res_Flx}
  \begin{align}
    \Res\LRp{\Qb,\Qbh} &= \Qb -\dtt \tilde{a}_{ii} \InvMassMatrix \Lcal\LRp{\Qb,\Qbh} - \Res_0, \\
    \Flx\LRp{\Qb,\Qbh} &= \LRa{\jump{\Fcalh_L\LRp{\Qb,\Qbh} \cdot \n},\mub}_\Gh.
  \end{align}
  \end{subequations}

The HDG system \eqnref{IMEXRKDiscretizedForm} can be written algebraically as

  \begin{align}
    \eqnlab{linearizedHDGsystem}
      \begin{pmatrix}
        \A & \B \\
        \C & \D
      \end{pmatrix}
      \begin{pmatrix}
        \Qbi\\
        \Qbhi
      \end{pmatrix}
      =
      \begin{pmatrix}
        \mb{R}_1\\
        \mb{R}_2
      \end{pmatrix},
  \end{align}
  where $\A=I-\triangle t \tilde{a}_{ii}\InvMassMatrix\dd{\Lcal}{\Qb}$, 
  $\B= -\triangle t \tilde{a}_{ii}\InvMassMatrix\dd{\Lcal}{\Qbh}$,
  $\C=\dd{\Flx}{\Qb}$,
  $\D=\dd{\Flx}{\Qbh}$,
  $\mb{R}_1=\Res_0$ and $\mb{R}_2=0$.
Note that the term $\dd{\Lcal}{\Qb}$ does not involve the computation of a Jacobian. Since $L$ is linear, $\dd{\Lcal}{\Qb}$ is a constant matrix.

  To solve \eqnref{linearizedHDGsystem}, we can first eliminate the
   volume unknowns $\Qbi$ 
  \begin{align}
    \eqnlab{recover_volume_unknowns}
    \Qbi = \A^{-1} \LRp{ \mb{R}_1 - \B\Qbhi}.
  \end{align}
Since $\A$ is block-diagonal (each block corresponding to one element in
the mesh), the inversion in \eqnref{recover_volume_unknowns} is
actually done in an element-by-element fashion, completely independent of each
other. This Schur complement step allows us to condense $\Qbi$ to
arrive at a much smaller linear system of equations in terms of $\Qbhi$:
  \begin{align}
    \eqnlab{linearizedHDGsystem_schur}
    \LRp{\D - \C\A^{-1}\B} \Qbhi= \mb{R}_2 -\C\A^{-1}\mb{R}_1.
  \end{align}

Once $\Qbhi$ is computed, the volume unknowns $\Qbi$ can be obtained
using \eqnref{recover_volume_unknowns}, in an element-by-element fashion. 
Compared to 
 IMEX DG schemes \cite{
  FeistauerDolejsiKucera07, RestelliGiraldo09,
  xu2004local},
  our IMEX HDG-DG scheme has a smaller number of coupled unknowns. On quadrilateral meshes with $n\times n$ elements and polynomial order $p$, for example, the number of coupled IMEX HDG-DG unknowns is $2n(n+1)(p+1)$, whereas that of the IMEX DG is $n^2(p+1)^2$. The ratio of the IMEX DG unknowns to the IMEX HDG-DG counterparts is $\frac{p+1}{2(1+1/n)}$. The IMEX HDG-DG schemes thus become beneficial in terms of the number of coupled degrees of freedom, and hence the size of the linear system, when the solution order $p\ge 1+2/n$.
In particular, IMEX HDG-DG becomes advantageous starting from second order approximations. A detailed complexity comparison between HDG and DG can be found in \cite{Bui-Thanh16a}.
  Once all the intermediate solutions are computed, the next time-step solution $\qb^{n+1}$ is determined through \eqnref{hdgIMEXRK_qn}. Algorithm \algref{HDG-IMEX} summarizes all the steps of our proposed IMEX scheme.

\begin{algorithm}
  \begin{algorithmic}[1]
    \ENSURE Given solution state $\q^n$, compute its next solution state $\q^{n+1}$.
      \FOR{$i=1$ to $s$}
        \IF {$\tilde{a}_{ii} = 0$}
          \STATE $Q^{(i)} \gets \qb^n$
          \STATE $\Lcal_i \gets \Lcal (\Qbi)$
        \ELSE
          \STATE $Res0 \gets q^n + \dtt\InvMassMatrix\sum_{j=1}^{i-1} \LRp{a_{ij} \NLcal_j + \tilde{a}_{ij} \Lcal_j}$
          \STATE Solve for $\Qbhi$ using \eqnref{linearizedHDGsystem_schur}
          \STATE Obtain the volume unknowns $\Qbi$ using \eqnref{recover_volume_unknowns}
          \STATE $\Lcal_i \gets \Lcal (\Qbi,\Qbhi)$ 
        \ENDIF

        \STATE $\NLcal_i \gets \NLcal\LRp{\Qbi}$
      \ENDFOR
      \STATE Update the solution $q^{n+1} \gets q^n + \dtt\InvMassMatrix\sum_{i=1}^s \LRp{b_i \NLcal_i + \tilde{b}_i \Lcal_i} $
  \end{algorithmic}
  \caption{IMEX HDG-DG scheme for $s$-stages.}
  \alglab{HDG-IMEX}
\end{algorithm}
The IMEX methods considered in this paper are the ARS2(2,3,2) and
ARS3(4,4,3) \cite{ascher1997implicit} methods, which have the singly diagonally implicit Runge-Kutta (SDIRK) property. (ARK methods
\cite{giraldo2013implicit, Kennedy2003additive} with the same order of
accuracy behave similarly and hence are not shown in the paper.)
Here, the triplet $(s,\sigma,\p)$ denotes the $s$ stages of the implicit scheme, $\sigma$ stages of the explicit scheme, and the order of accuracy of the scheme.

\section{Numerical Results}
\seclab{numericalResults}

In this section, we demonstrate the accuracy and efficiency of the
proposed coupled IMEX HDG-DG methods for the shallow water equations
through several numerical experiments.  For  planar shallow water
flow, two test cases are considered: the translating vortex test
case and the water height perturbation problem. For the former, in
which an exact solution exists, we present the numerical convergence
for both the spatial and temporal discretizations. For the latter, in which no analytical solution is available, we perform a
qualitative comparison with explicit schemes. For the spherical shallow water
equations, the well-known standard test cases proposed by 
\cite{williamson1992standard} and the barotropic instability phenomenon
\cite{galewsky2004initial} are chosen to verify the IMEX HDG-DG scheme.

\subsection{Moving vortex}

We consider the vortex translation test
\cite{Ricchiuto2009} in which the
initial condition in the domain $\Omega=[-2,2]\times[-2,2]$ is chosen in
such a way that the pressure gradient force and the centrifugal force
are balanced. This allows the initial vortex to translate across the domain without
changing its shape. The exact solution for the vortex at any time $t$ is given by
  \begin{subequations}
    \eqnlab{tc_vortex_ic}
    \begin{align}
      H &= H_\infty - \frac{\beta^2}{32\pi^2} e^{-2(r^2-1)},\\
      u &= u_\infty - \frac{\beta}{2\pi} e^{-(r^2-1)}y_t,\\
      v &= v_\infty + \frac{\beta}{2\pi} e^{-(r^2-1)}x_t,
    \end{align}
  \end{subequations}
  where $\beta$ is the vortex strength,  $(x_c,y_c)$ the
  center of the vortex, $(u_\infty,v_\infty)$ the reference horizontal velocity, $x_t=x
  - x_c - u_\infty t$, $y_t=y - y_c - v_\infty t$,  $r^2 = x_t^2 + y_t^2$, and $H_\infty$ the reference water depth. For the
  numerical results in this section, we choose $H_\infty=1$ and
  $(u_\infty,v_\infty)=(1,0)$, $\beta=5$, and $g=2$. We use the exact
  solution to impose the boundary condition. Initially the vortex is located
  at $(x_c,y_c)=(0,0)$. Figure
  \figref{movingvortex} shows numerical results for the free surface elevation $\eta:=H - H_\infty$ at different times. Here, the solution order is $p = 6$ and the
  results are computed on a uniform mesh with $32\times 32$ elements.

  \begin{figure}[h!t]
  \begin{centering}
    \includegraphics[trim=9.2cm 4cm 9.2cm 4cm,clip=true,width=0.3\textwidth]{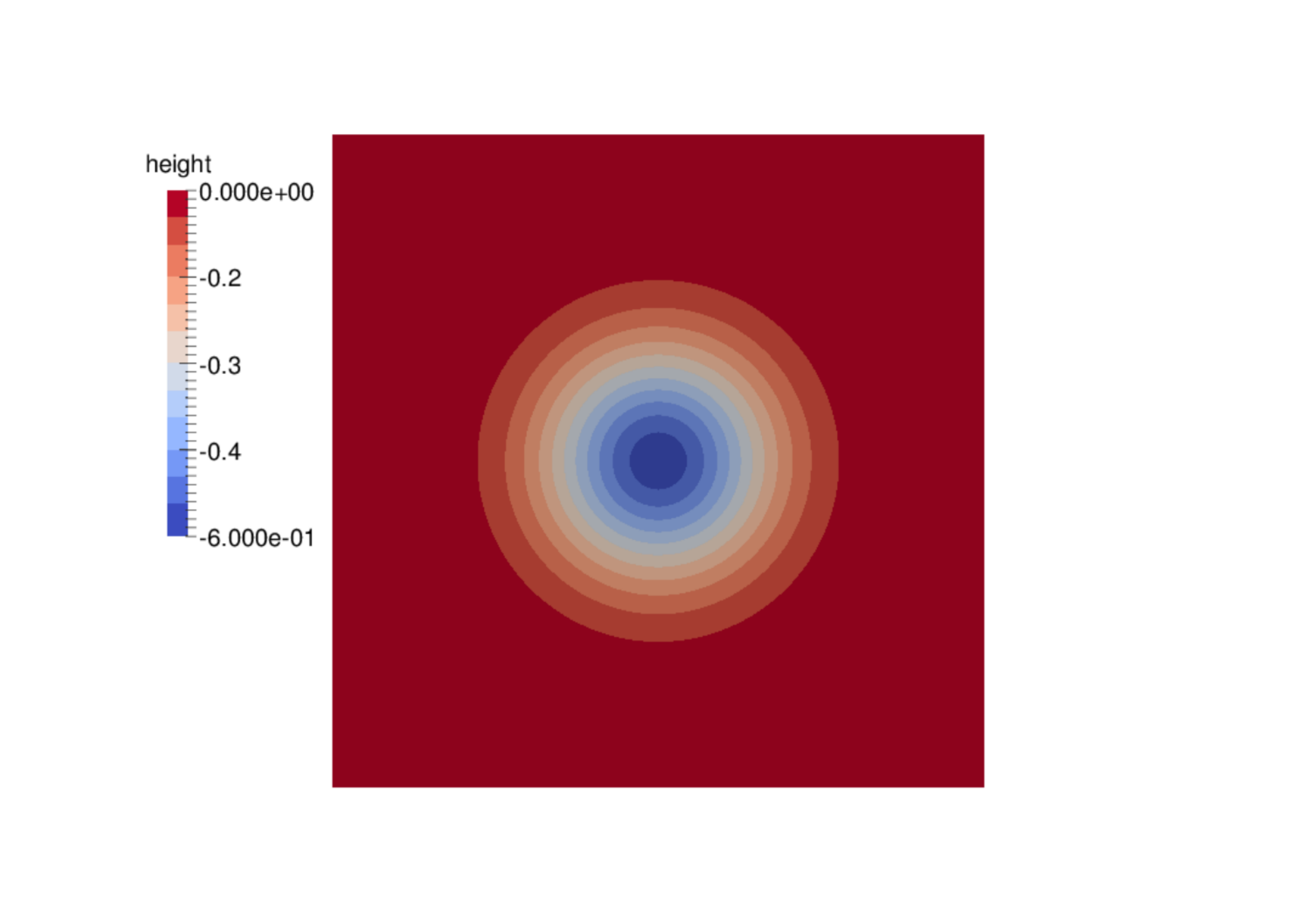}
    \includegraphics[trim=9.2cm 4cm 9.2cm 4cm,clip=true,width=0.3\textwidth]{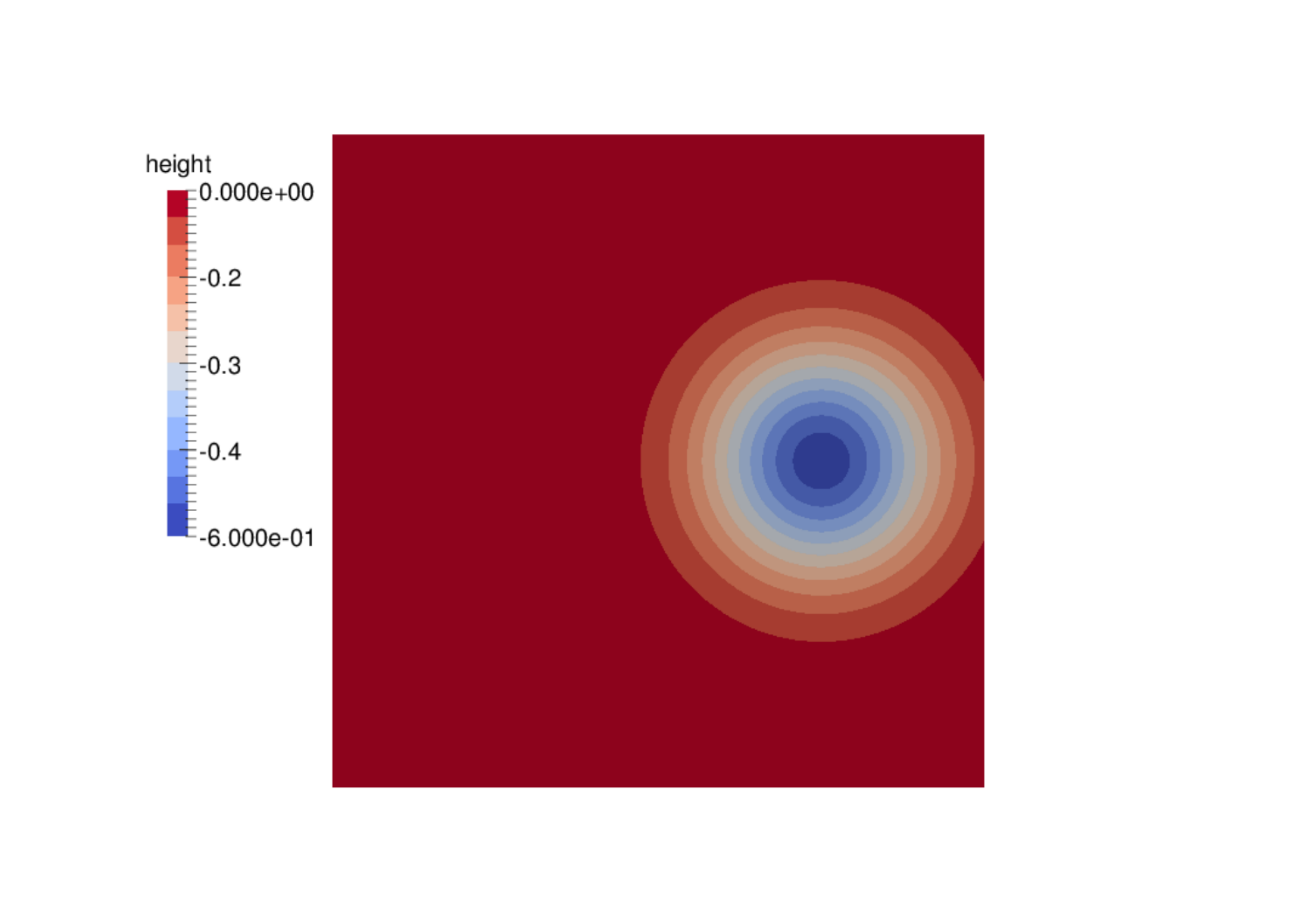}
    \includegraphics[trim=4.7cm 3.5cm 1.2cm 2.0cm,clip=true,width=0.355\textwidth]{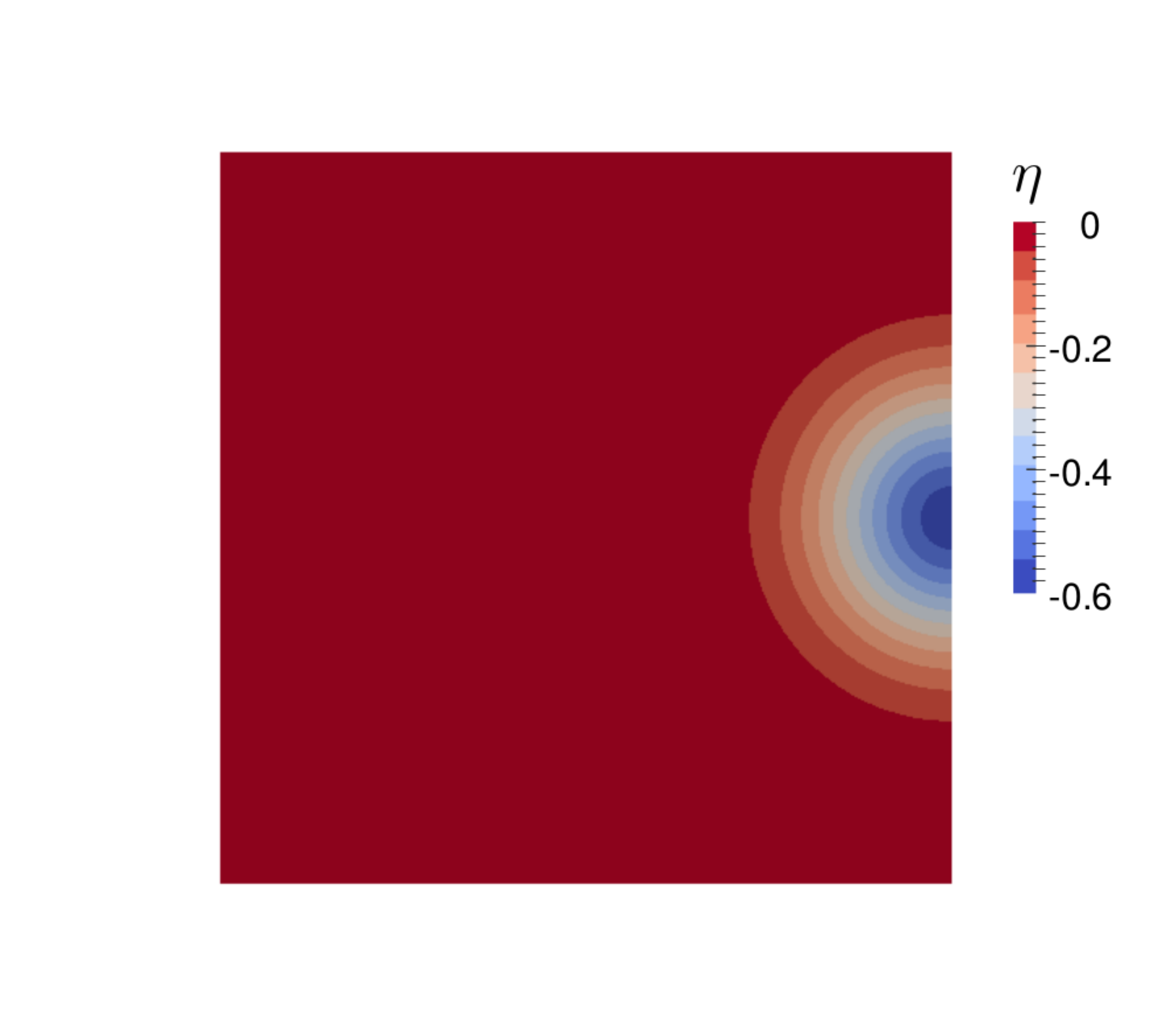}
    \caption{ The moving vortex test case: time evolution of the free
      surface elevation $\eta$ at times $t=0$, $t=1$ and
      $t=2$ computed using the ARS3 HDG-DG. The contour levels are from $-0.6$m to $0$m with the step-size of $0.05$m.} \figlab{movingvortex}
  \end{centering}
  \end{figure}

  We compute the errors of the free surface elevation and the velocity
  using the following $L_1$, $L_2$ and $L_\infty$ norms:
   \begin{align*}
      L_1(q) &:= \sum_{\K \in \Omegah} \int_K \abs{q - q_T} dK,\\
      L_2(q) &:= \sqrt{ \sum_{\K \in \Omegah} \int_K (q - q_T)^2 dK},\\
      L_\infty(q) &:= \max_{{ q\in \Omega_h}  } \abs{q - q_T}, 
    \end{align*}
  where $q_T$ is the exact solution at the final time $T$. 
  
  For the spatial convergence test, we use the ARS3 HDG-DG scheme and take $\dtt=0.000625$ for $p=5$, $\dtt=0.00125$ for $p=4$, $\dtt=0.0025$ for $p=3$ and $\dtt=0.005$ for $p=2$. The errors are computed at $T=2$ when the center of the vortex reaches the right boundary of the domain. Table \tabref{tc_mv_hconv_hdg3} shows the spatial convergence results using the $L_1,L_2$ and $L_\infty$ norms for the free surface elevation $\eta$.
As can be seen, the predicted convergence rate of $\LRp{p+1/2}$ is observed for all cases except for the case  $p=5$ in which  superconvergence is observed. In Table \tabref{tc_mv_l2_hconv_hdg3}, similar convergence rates are observed for 
 the $L_2$ errors of the horizontal velocity $(u,v)$ and the error in the energy norm $E:=\frac{1}{2}(\norm{\eta}^2_\Omegah +\norm{\ub}^2_\Omegah)$. 

\begin{table}[t]
\caption{Spatial convergence of $\eta$ for the traveling vortex test using ARS3 HDG-DG.}
\begin{center}
\begin{tabular}{c|c|ccc|ccc}
\multicolumn{2}{c}{ } & \multicolumn{3}{c}{Error} & \multicolumn{3}{c}{Order} \\
$p$ & $N_e$ &  $L_1$ & $L_2$ & $L_\infty$ & $L_1$ & $L_2$ & $L_\infty$ \\
     \hline\hline 
\multirow{4}{*}{2} & 4x4 & 2.036e-01 & 8.189e-02 & 3.372e-02 &   $-$ &   $-$ &   $-$  \tabularnewline
 & 8x8 & 4.455e-02 & 1.693e-02 & 8.623e-03 &  2.19 &  2.27 &  1.97  \tabularnewline
 & 16x16 & 6.561e-03 & 2.793e-03 & 2.083e-03 &  2.76 &  2.60 &  2.05  \tabularnewline
 & 32x32 & 1.057e-03 & 5.412e-04 & 4.269e-04 &  2.63 &  2.37 &  2.29  \tabularnewline
\hline
\multirow{4}{*}{3} & 4x4 & 8.991e-02 & 3.645e-02 & 2.167e-02 &   $-$ &   $-$ &   $-$  \tabularnewline
 & 8x8 & 1.001e-02 & 4.226e-03 & 2.494e-03 &  3.17 &  3.11 &  3.12  \tabularnewline
 & 16x16 & 1.139e-03 & 5.864e-04 & 7.896e-04 &  3.14 &  2.85 &  1.66  \tabularnewline
 & 32x32 & 6.117e-05 & 3.228e-05 & 3.495e-05 &  4.22 &  4.18 &  4.50  \tabularnewline
\hline
\multirow{4}{*}{4} & 4x4 & 1.999e-02 & 8.123e-03 & 6.767e-03 &   $-$ &   $-$ &   $-$  \tabularnewline
 & 8x8 & 1.287e-03 & 5.723e-04 & 6.756e-04 &  3.96 &  3.83 &  3.32  \tabularnewline
 & 16x16 & 5.053e-05 & 2.705e-05 & 3.020e-05 &  4.67 &  4.40 &  4.48  \tabularnewline
 & 32x32 & 2.244e-06 & 1.467e-06 & 1.273e-06 &  4.49 &  4.20 &  4.57  \tabularnewline
\hline
\multirow{4}{*}{5} & 4x4 & 7.790e-03 & 3.015e-03 & 2.428e-03 &   $-$ &   $-$ &   $-$  \tabularnewline
 & 8x8 & 2.856e-04 & 1.408e-04 & 1.604e-04 &  4.77 &  4.42 &  3.92  \tabularnewline
 & 16x16 & 6.040e-06 & 3.378e-06 & 3.676e-06 &  5.56 &  5.38 &  5.45  \tabularnewline
 & 32x32 & 6.822e-08 & 4.193e-08 & 1.739e-08 &  6.47 &  6.33 &  7.72  \tabularnewline
\hline\hline
\end{tabular}
\end{center}
\tablab{tc_mv_hconv_hdg3}
\end{table}

\begin{table}[t]
\caption{Spatial convergence of $\ub$ and $\sqrt{E}$ for the traveling vortex test using ARS3 HDG-DG.}
\begin{center}
\begin{tabular}{c|c|ccc|ccc}
\multicolumn{2}{c}{ } & \multicolumn{3}{c}{$L_2$ Error}  & \multicolumn{3}{c}{Order} \\
$p$ & $N_e$   & $u$ & $v$ & $\sqrt{E}$    & $u$ & $v$ & $\sqrt{E}$  \\
     \hline\hline
\multirow{4}{*}{2} & 4x4 & 1.654e-01 & 2.277e-01 & 2.072e-01 &   $-$ &   $-$ &   $-$  \tabularnewline
 & 8x8 & 4.993e-02 & 4.260e-02 & 4.793e-02 &  1.73 &  2.42 &  2.11  \tabularnewline
 & 16x16 & 1.196e-02 & 5.513e-03 & 9.517e-03 &  2.06 &  2.95 &  2.33  \tabularnewline
 & 32x32 & 2.534e-03 & 9.236e-04 & 1.945e-03 &  2.24 &  2.58 &  2.29  \tabularnewline
\hline
\multirow{4}{*}{3} & 4x4 & 1.665e-01 & 8.503e-02 & 1.347e-01 &   $-$ &   $-$ &   $-$  \tabularnewline
 & 8x8 & 9.142e-03 & 5.988e-03 & 8.285e-03 &  4.19 &  3.83 &  4.02  \tabularnewline
 & 16x16 & 9.278e-04 & 7.128e-04 & 9.254e-04 &  3.30 &  3.07 &  3.16  \tabularnewline
 & 32x32 & 5.166e-05 & 4.479e-05 & 5.347e-05 &  4.17 &  3.99 &  4.11  \tabularnewline
\hline
\multirow{4}{*}{4} & 4x4 & 1.734e-02 & 1.707e-02 & 1.814e-02 &   $-$ &   $-$ &   $-$  \tabularnewline
 & 8x8 & 1.199e-03 & 1.092e-03 & 1.216e-03 &  3.85 &  3.97 &  3.90  \tabularnewline
 & 16x16 & 5.942e-05 & 3.699e-05 & 5.306e-05 &  4.33 &  4.88 &  4.52  \tabularnewline
 & 32x32 & 3.593e-06 & 1.660e-06 & 2.985e-06 &  4.05 &  4.48 &  4.15  \tabularnewline
\hline
\multirow{4}{*}{5} & 4x4 & 1.403e-02 & 6.759e-03 & 1.121e-02 &   $-$ &   $-$ &   $-$  \tabularnewline
 & 8x8 & 2.666e-04 & 1.623e-04 & 2.421e-04 &  5.72 &  5.38 &  5.53  \tabularnewline
 & 16x16 & 5.734e-06 & 3.357e-06 & 5.270e-06 &  5.54 &  5.60 &  5.52  \tabularnewline
 & 32x32 & 6.808e-08 & 4.744e-08 & 6.574e-08 &  6.40 &  6.14 &  6.32  \tabularnewline
\hline\hline
\end{tabular}
\end{center}
\tablab{tc_mv_l2_hconv_hdg3}
\end{table}

To numerically compute the temporal convergence for ARS2 HDG-DG and
ARS3 HDG-DG, we simulate the translational vortex with a $6th$-order
solution on the $32\times 32$-element mesh. The time-step size $\dtt$
varies from $10^{-4}$ to $5\times 10^{-3}$, which corresponds to Courant numbers
from $0.11$ to $5.6$. We compute the error at $T=0.1$.  The mean water
depth $H_\infty$ is set to be $50$ so that the reference Froude
number, $Fr=\frac{u_\infty}{\sqrt{gH_\infty}}$, is $0.1$, that is,
the gravity wave dominates the convection.  In Figure
\figref{pswe_tc_mv_tconv_l2}, we observe the correct second-order and
third-order convergence in time for ARS2 HDG-DG and ARS3 HDG-DG,
respectively. To demonstrate the stability benefit of the IMEX HDG-DG scheme we perform simulations for a wide range of Courant numbers (Cr) from $0.28$ (the point over which the second order RKDG, denoted as RK2 DG, blows up)
to $5.6$.

Clearly, the IMEX HDG-DG approaches are more economical than our previous
work on IMEX DG \cite{GiraldoRestelli10,giraldo2013implicit}
due to the fewer number of coupled degrees of freedom. 
Compared to standard fully implicit methods, 
they are much more advantageous since only one linear solve
is needed for each stage per time-step. For this paper, our methods are in fact ``optimal''
in the sense that the HDG matrix in \eqnref{linearizedHDGsystem}, and
hence the matrices $\A$ in \eqnref{recover_volume_unknowns} and
$\LRp{\D - \C\A^{-1}\B}$ in \eqnref{linearizedHDGsystem_schur}, is the
same for any time-step and any stage (since $\tilde{a}_{ii}$ are the
same at any stage for the chosen schemes). Thus, we need to perform
the LU factorization 
(here we use UMFPACK \cite{davis2006direct}) of the
HDG-trace matrix $\LRp{\D - \C\A^{-1}\B}$ once, and the same LU
factors can be recycled (via a forward substitution followed by a
backward substitution) for all subsequent computations involving
\eqnref{linearizedHDGsystem_schur}.

Clearly, our approaches cannot compete with fully explicit methods
in terms of wallclock time since we still have to solve
\eqnref{recover_volume_unknowns} and
\eqnref{linearizedHDGsystem_schur} for each time-step. To demonstrate
this we plot in Figure \figref{pswe_tc_mv_tconv_wc} the $L^2$ error
of the free surface height against the wallclock time for ARS2 HDG-DG, ARS3 HDG-DG,
RK2 DG, and RK3 DG (the third order RKDG). 
To improve the wallclock time we can, for example, develop
iterative solvers for \eqnref{linearizedHDGsystem_schur} and solve
\eqnref{recover_volume_unknowns}; this will be presented in our
future work. Nevertheless, for applications in which fast time
marching to the solution is more important than an accurate solution,
our methods are more advantageous than the fully explicit counterparts
due to their ability to take (much) larger time-step sizes: we will
confirm this claim in section \secref{wdrop}.

  \begin{figure}[h!t]
    \centering
    \subfigure[Accuracy/Stability]{
      \includegraphics[trim=0.5cm 0.5cm 4cm 0cm, clip=true,width=0.45\textwidth]{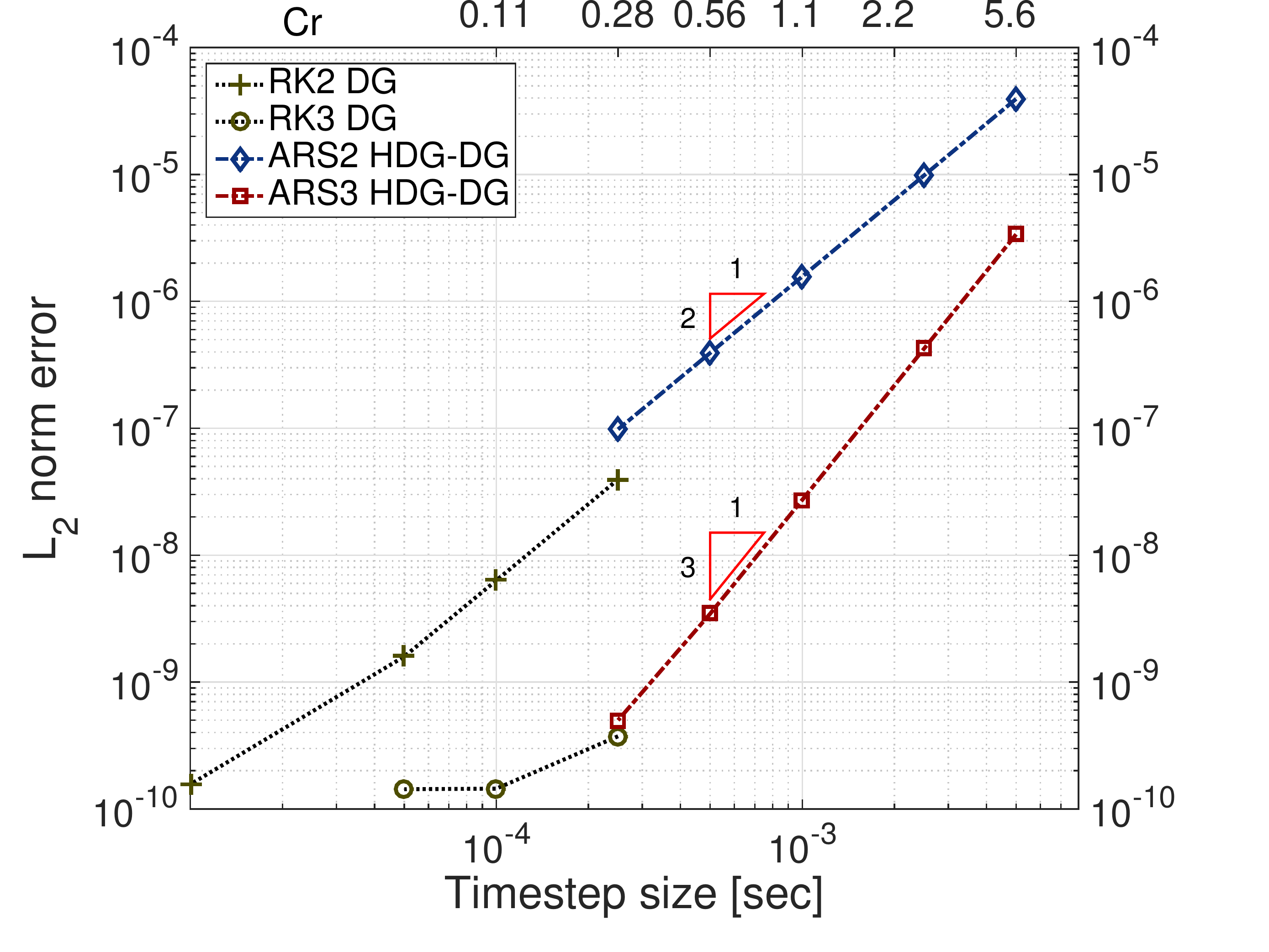}
      \figlab{pswe_tc_mv_tconv_l2}
    }
    \subfigure[Wallclock time]{
      \includegraphics[trim=0.5cm 0.5cm 4cm 0cm, clip=true,width=0.45\textwidth]{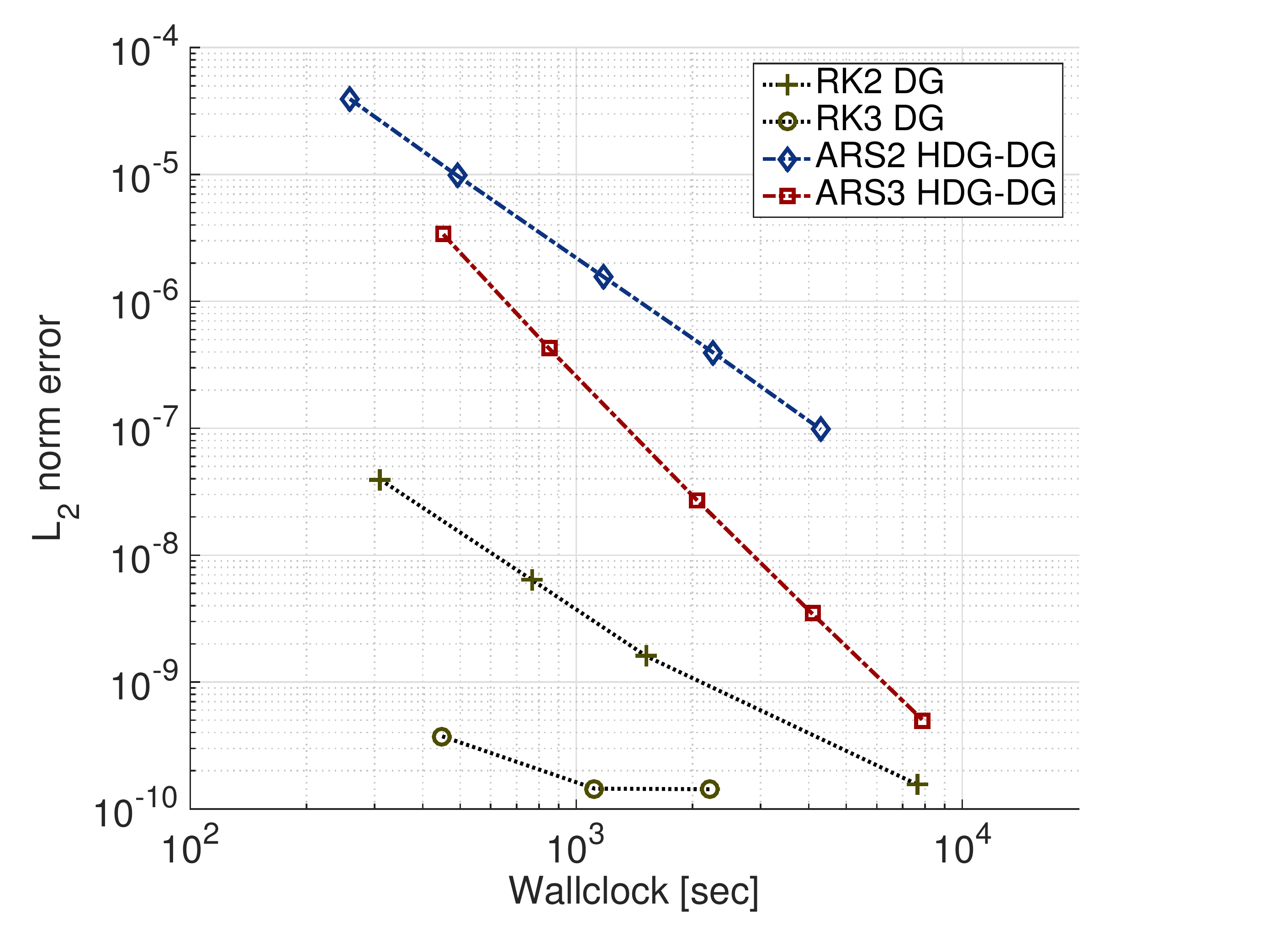}
      \figlab{pswe_tc_mv_tconv_wc}
    }
    \caption{Comparison between IMEX HDG-DG and RKDG for the moving vortex test case: (a) the accuracy/stability and (b) the wallclock time.}
    \figlab{pswe_tc_mv_tconv}
  \end{figure}

\subsection{Water height perturbation}
\seclab{wdrop}
In this section, we consider the propagation of smooth gravity waves \cite{Dumbser2013} over the domain $\Omega=[-1,1]\times[-1,1]$. The initial condition is given as
  \begin{align*}
    H = H_\infty + e^{-\frac{x^2+y^2}{2 \sigma^2}},\quad \text{ and } \quad
    u = v = 0,
  \end{align*}

  where $H_\infty = 100$. We set the gravitational acceleration $g$ to
  be unity. The domain is discretized with $20 \times 20$ finite
  elements and with $8th$ order polynomials. The time horizon is
  $T=0.16$.

  We choose different time-step sizes for RK2 DG and ARS2 HDG-DG.
  Since RK2 DG blows up after a few 
    iterations with $\triangle t$=0.0002 (see Figure \figref{waterdrop_blowup}), we take 
 $\triangle t=0.0001$.
    The time-step sizes of $\triangle t=0.002$ and $\triangle t=0.02$
    are chosen for ARS2 HDG-DG.
    Figure \figref{waterdrop_evolution} quantitatively shows a
    three-dimensional plot of the evolution of the free surface
    elevation using RK2 DG and ARS2 HDG-DG at times $t=0$, $t=0.02$,
    $t=0.06$, and $t=0.1$. We observe that ARS2 HDG-DG with
    $\triangle t=0.002$ is in good agreement with RK2 DG.  While 
    IMEX-RK methods allow us to increase the time-step size without being penalized by stability constraints 
    their accuracy is reduced due to the truncation error in the time discretization. This can be observed in the last column of Figure \figref{waterdrop_evolution} in which
    ARS2 HDG-DG with $\triangle t=0.02$ shows damped (inaccurate) solutions due to large truncation error.

\begin{figure}[h!t]
  \begin{centering}

    \subfigure[T=0.016]{
      \includegraphics[clip=true,width=0.3\textwidth]{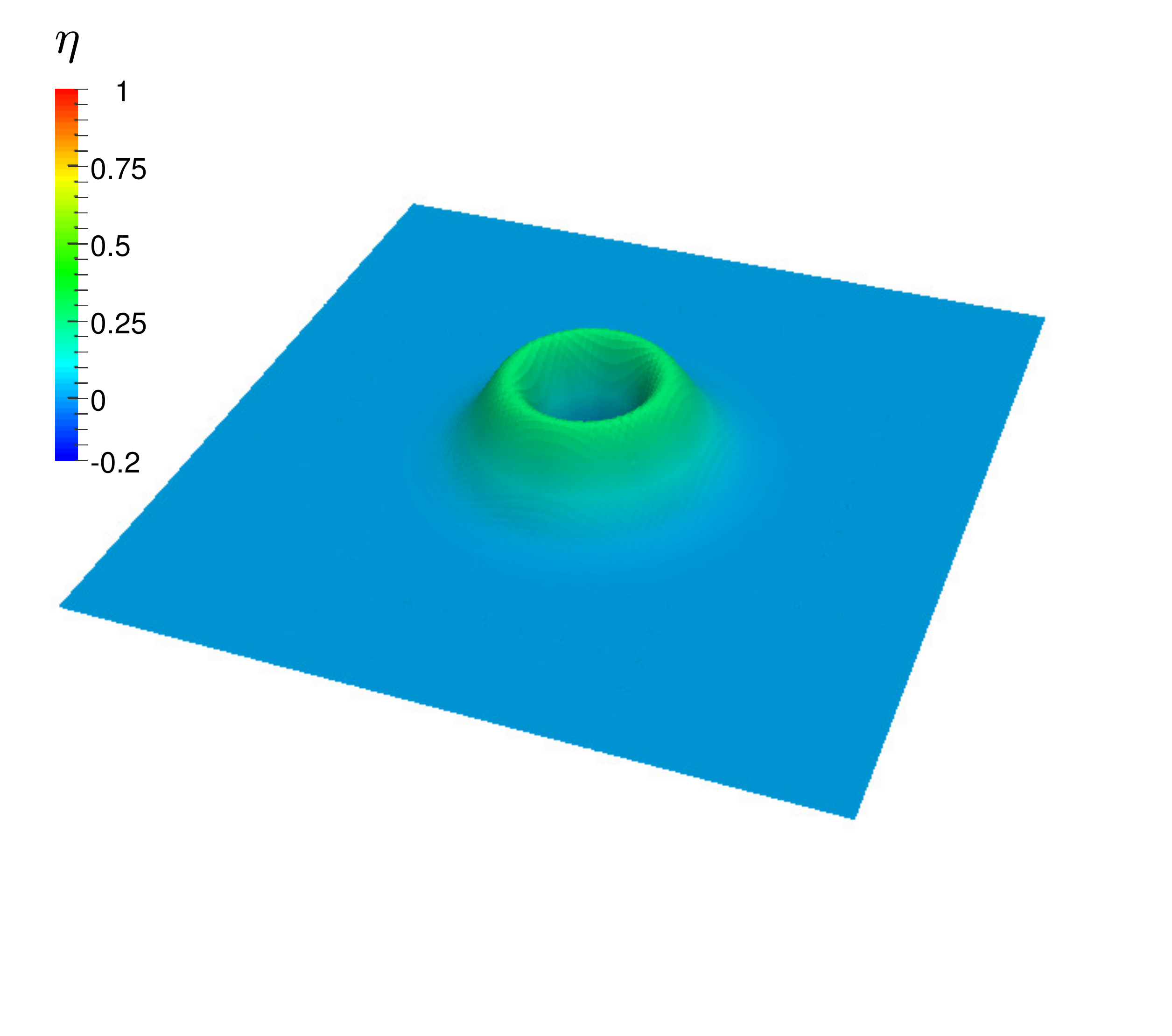}
    }
    \subfigure[T=0.017]{
      \includegraphics[clip=true,width=0.3\textwidth]{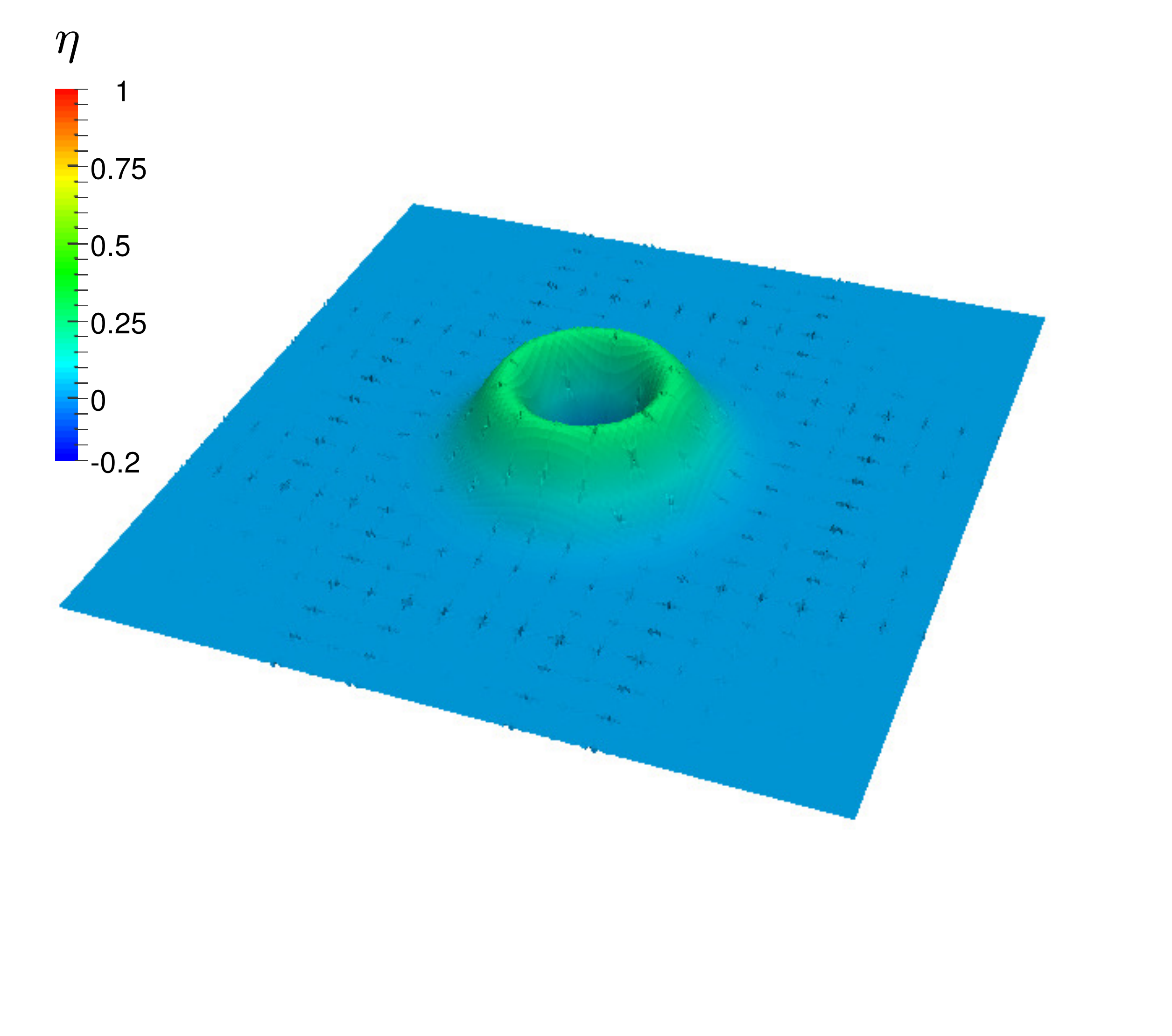}
    }
    \subfigure[T=0.018]{
      \includegraphics[clip=true,width=0.3\textwidth]{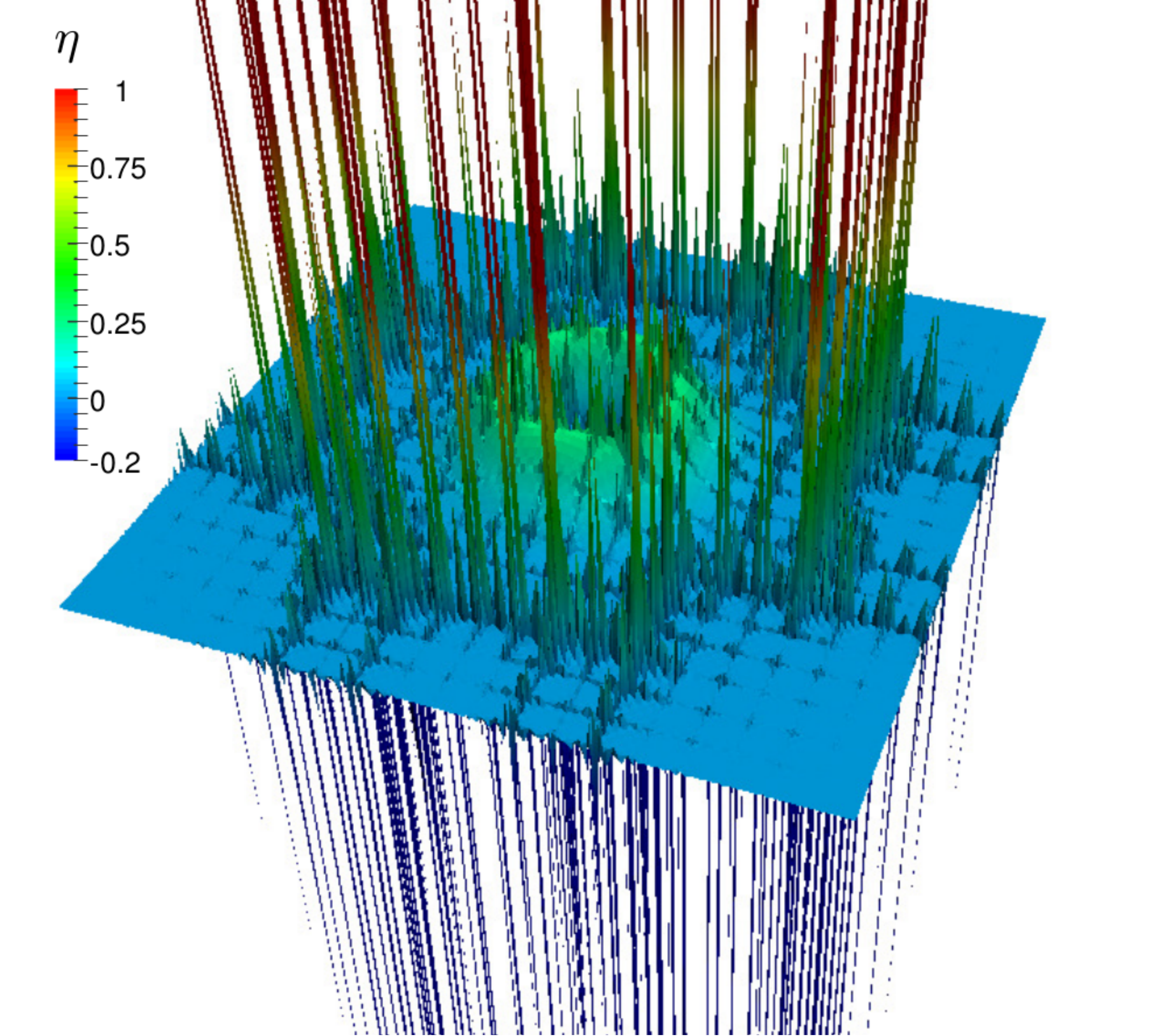}
    }
    \caption{ Water height perturbation test: free surface evolution for RK2 DG ($\triangle t$=0.0002) at $t=0.016$, $t=0.017$ and $t=0.018$. }
    \figlab{waterdrop_blowup}
  \end{centering}
  \end{figure}

  \begin{figure}[h!t]
  \begin{centering}
    \includegraphics[clip=true,width=0.3\textwidth]{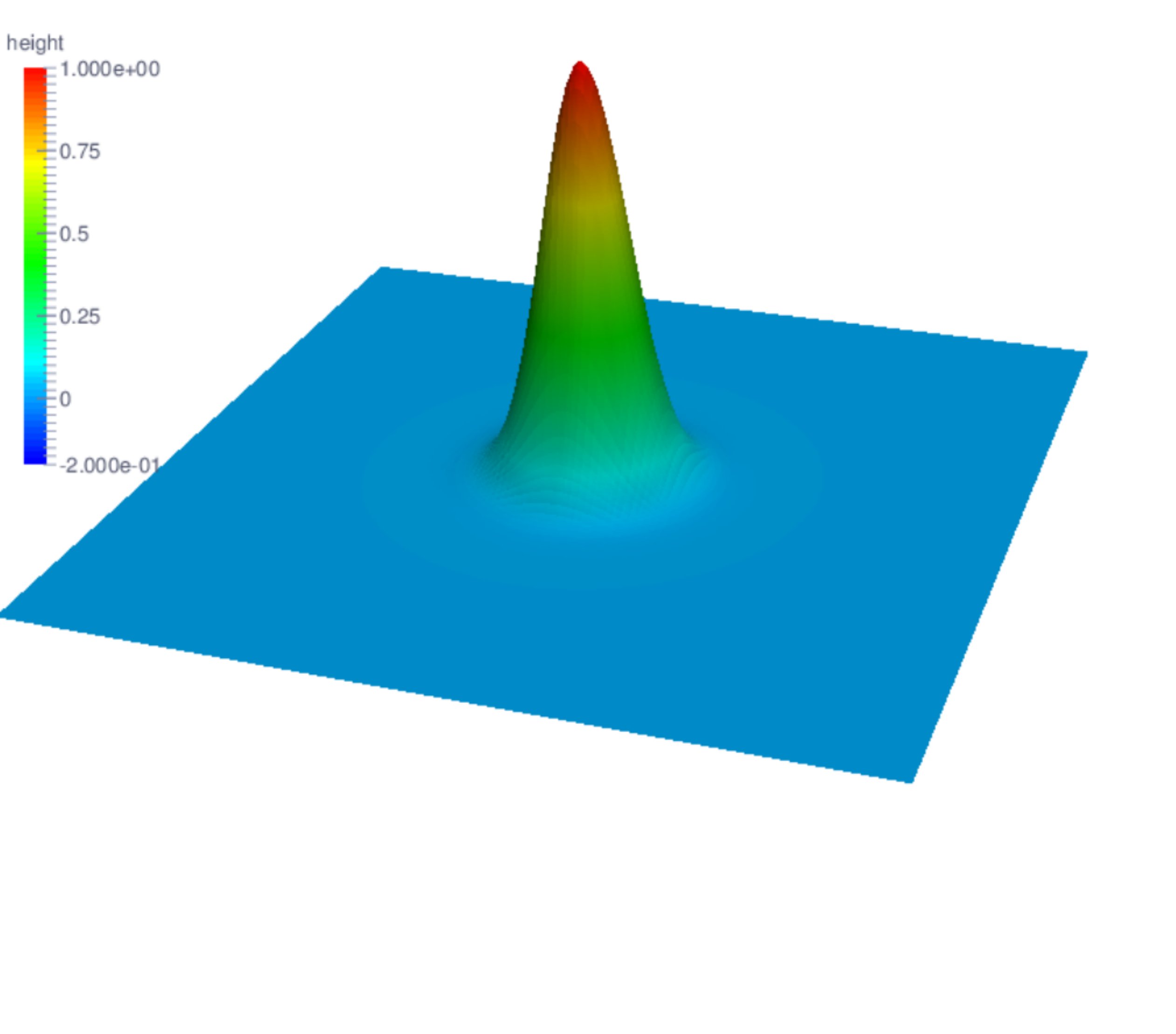}
    \includegraphics[clip=true,width=0.3\textwidth]{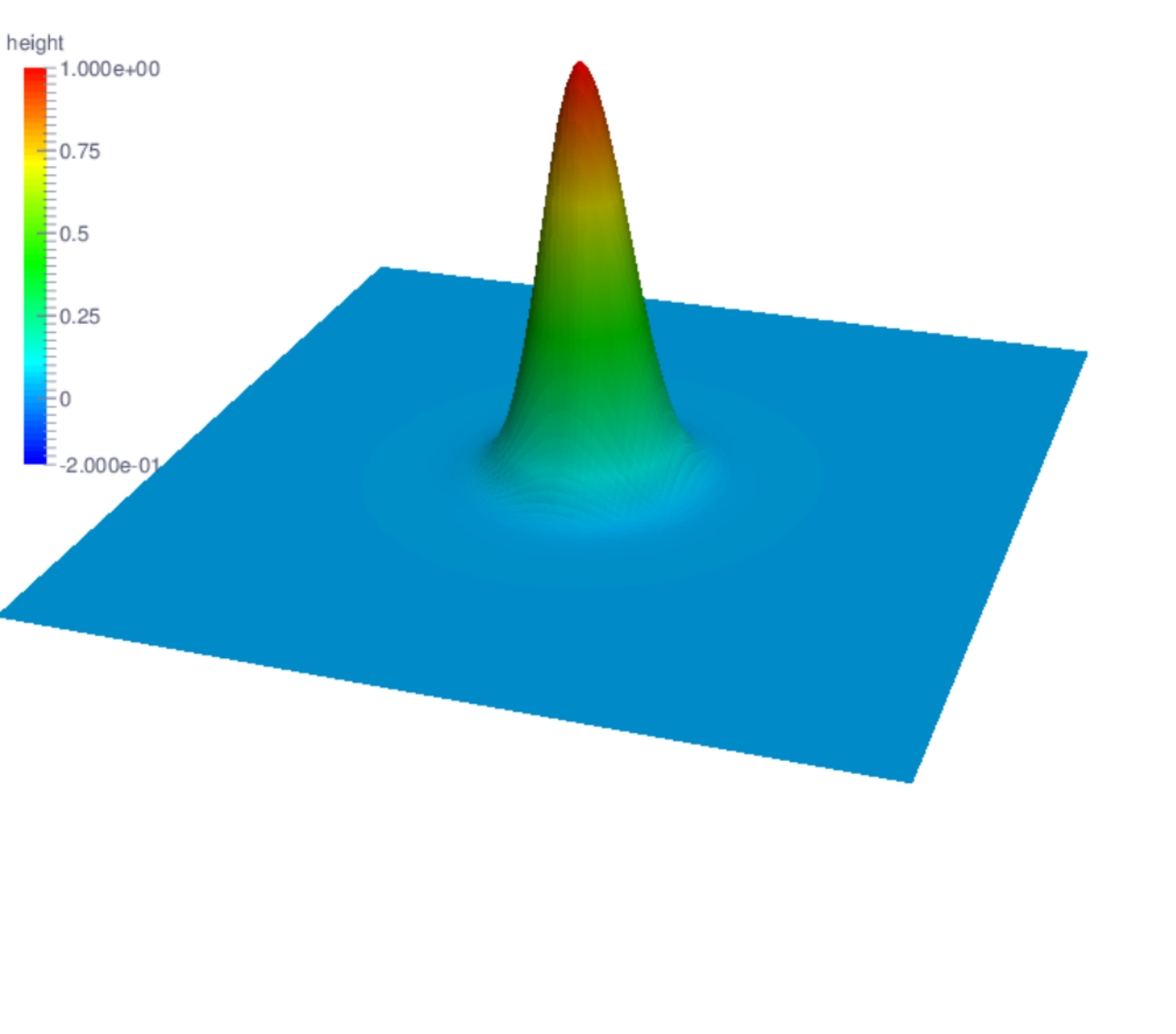}
    \includegraphics[clip=true,width=0.3\textwidth]{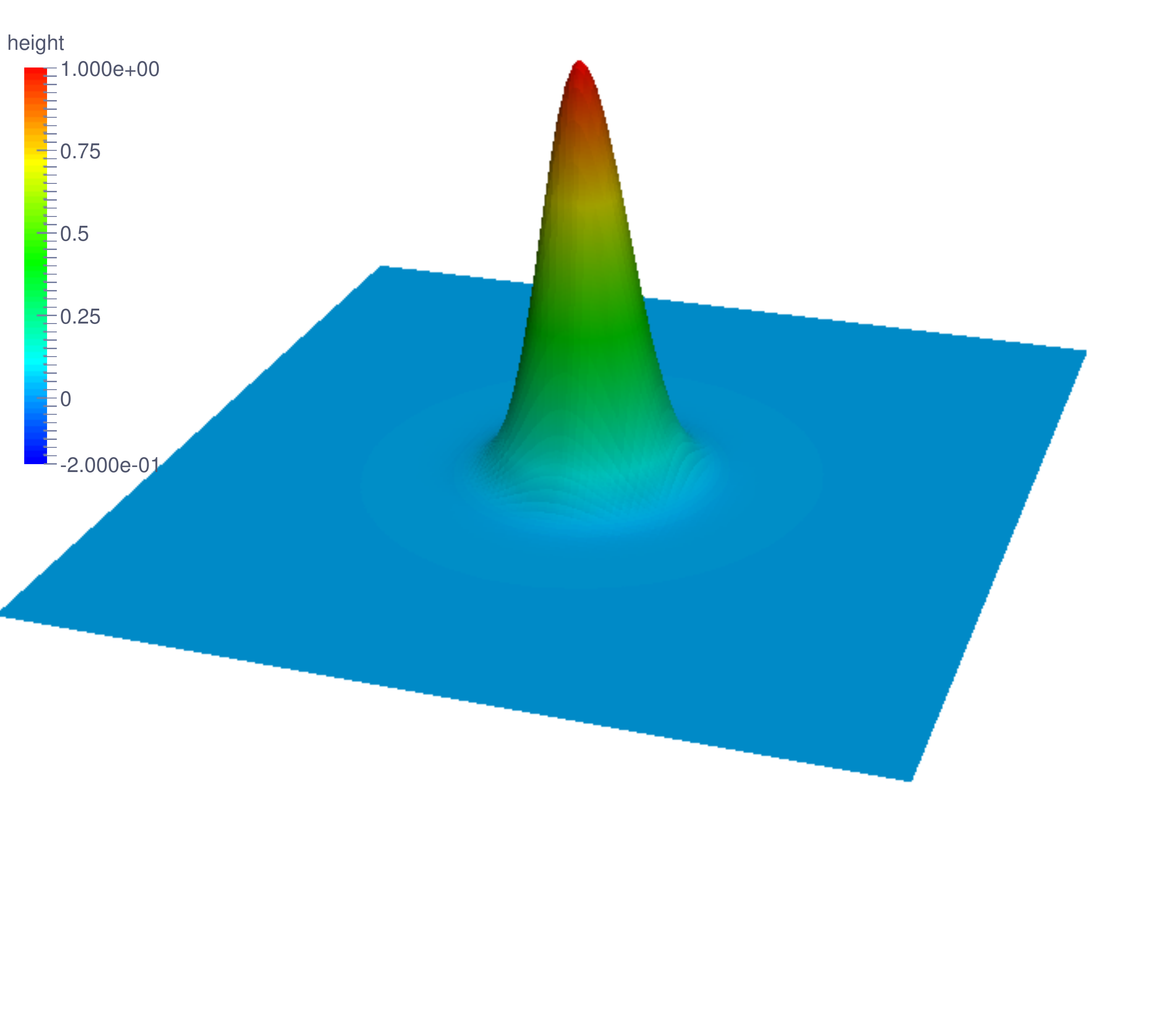}\\
    \vspace{-7mm}
    \includegraphics[clip=true,width=0.3\textwidth]{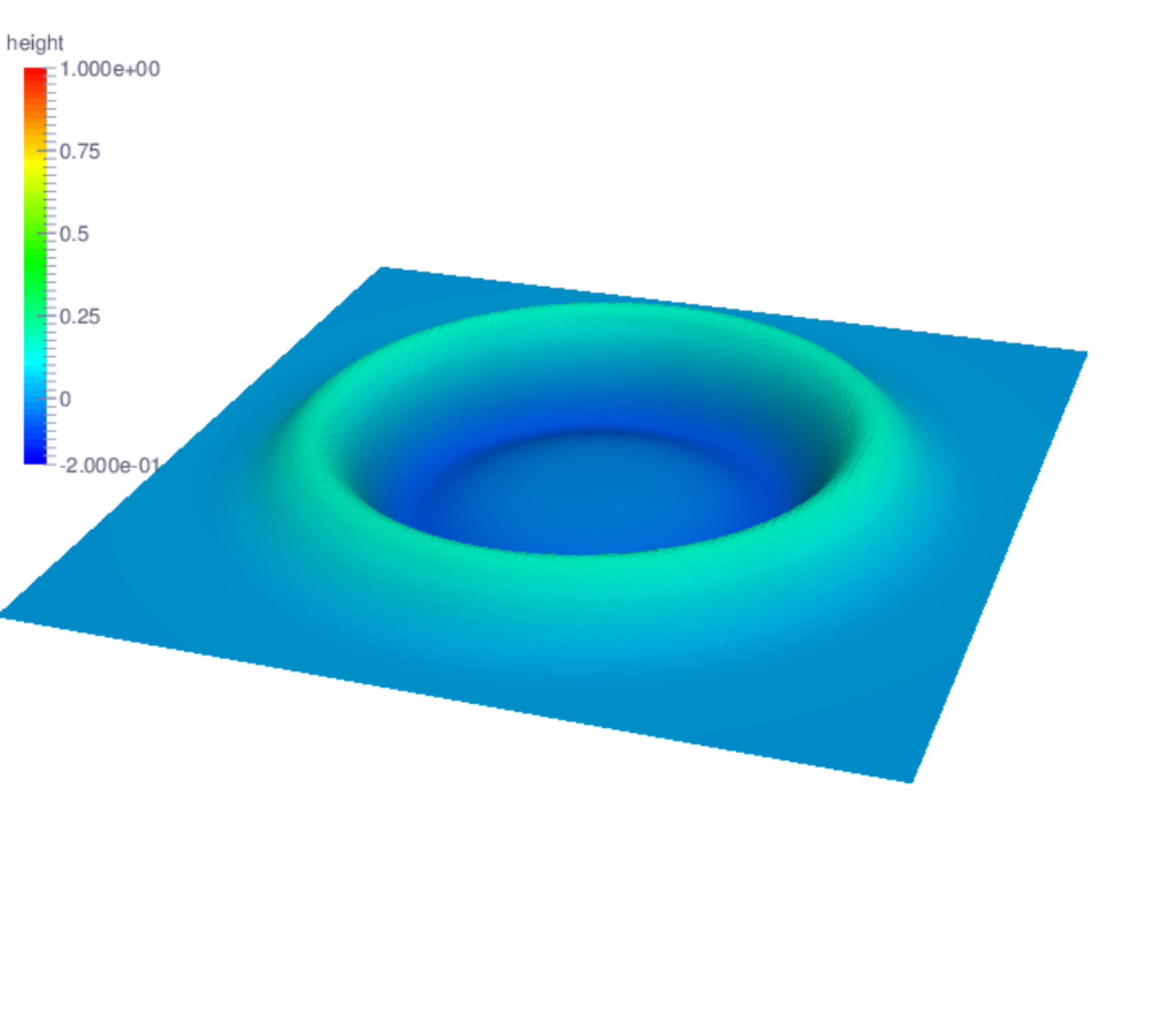}
    \includegraphics[clip=true,width=0.3\textwidth]{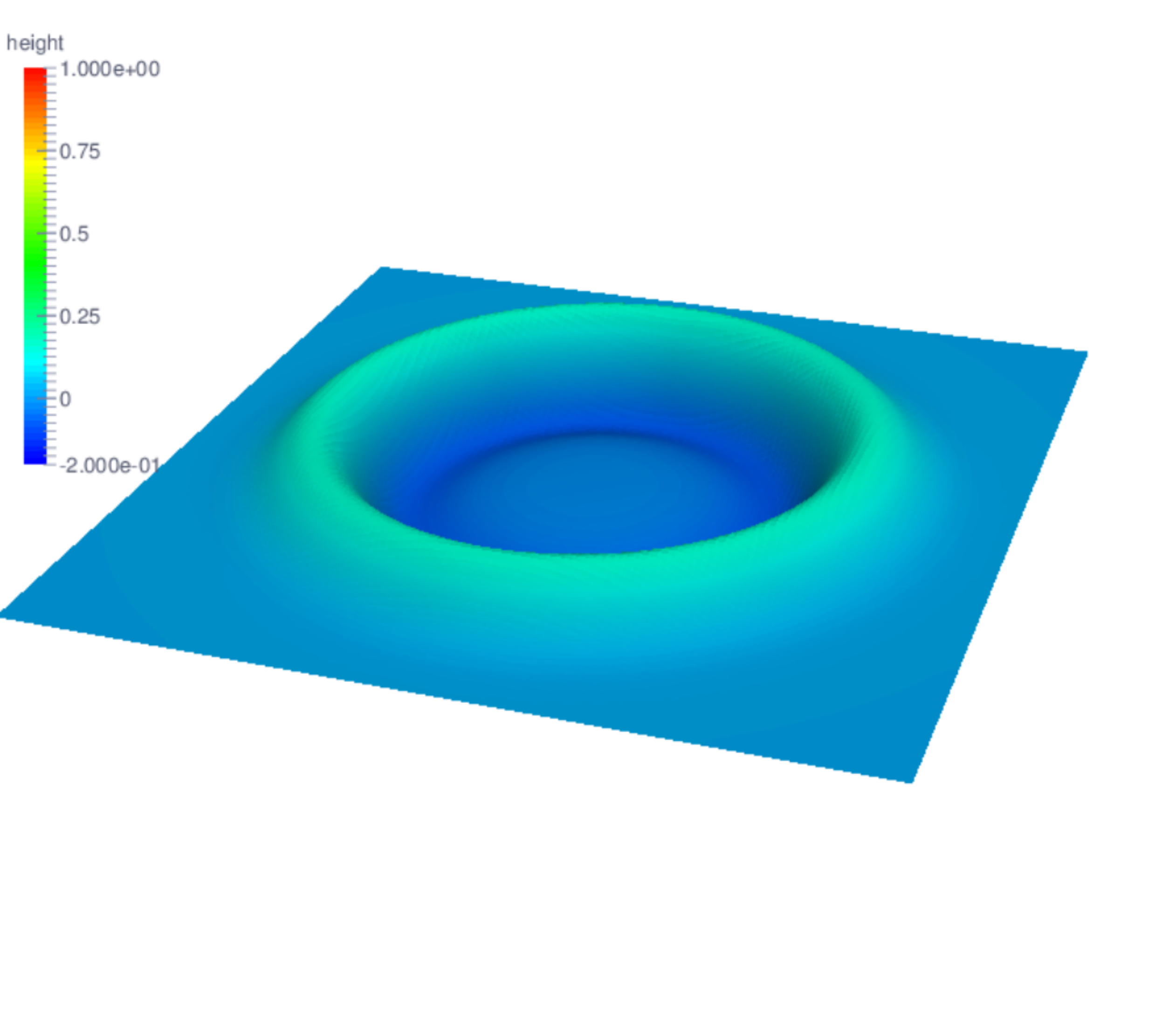}
    \includegraphics[clip=true,width=0.3\textwidth]{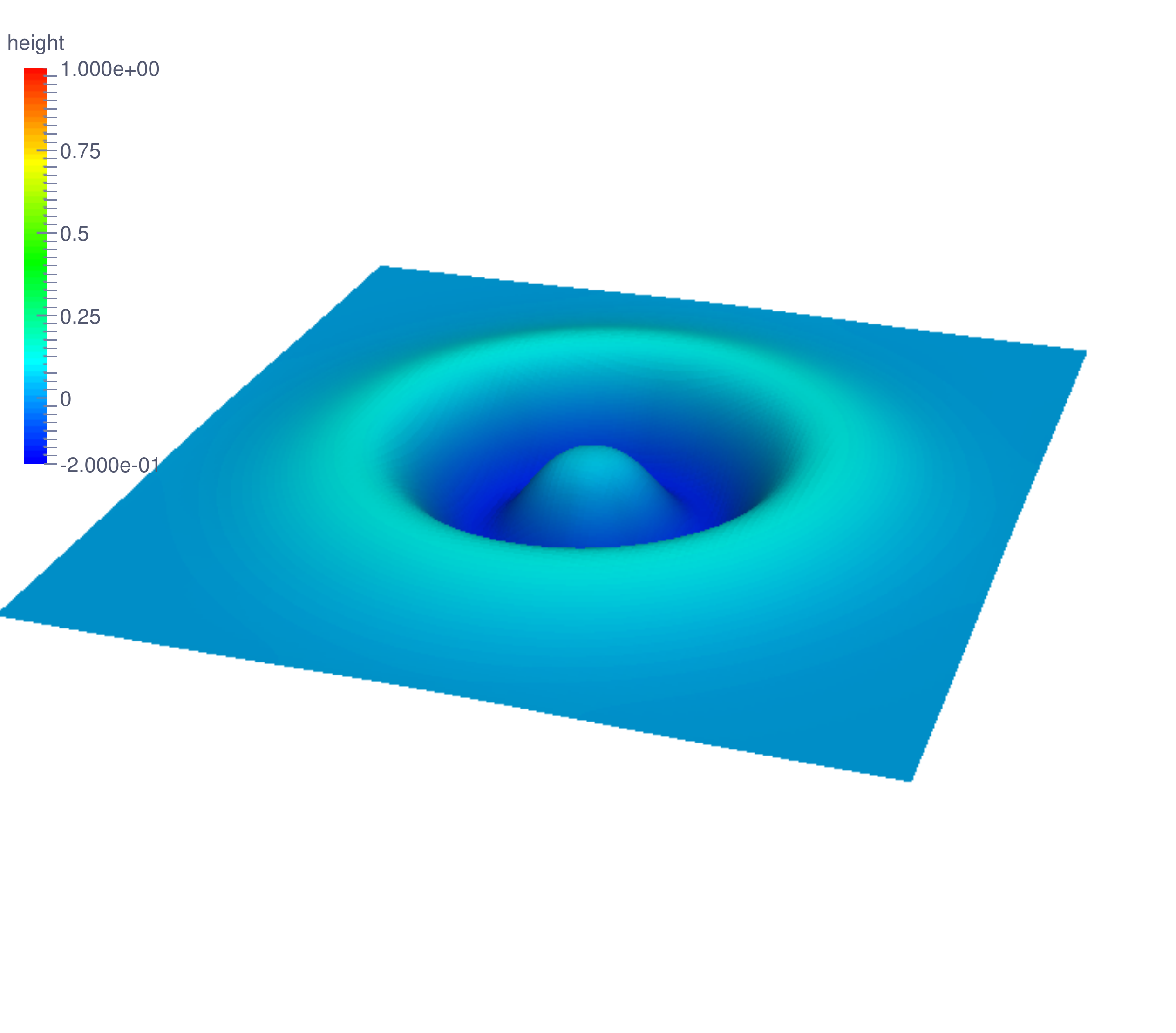}\\
    \vspace{-7mm}
    \includegraphics[clip=true,width=0.3\textwidth]{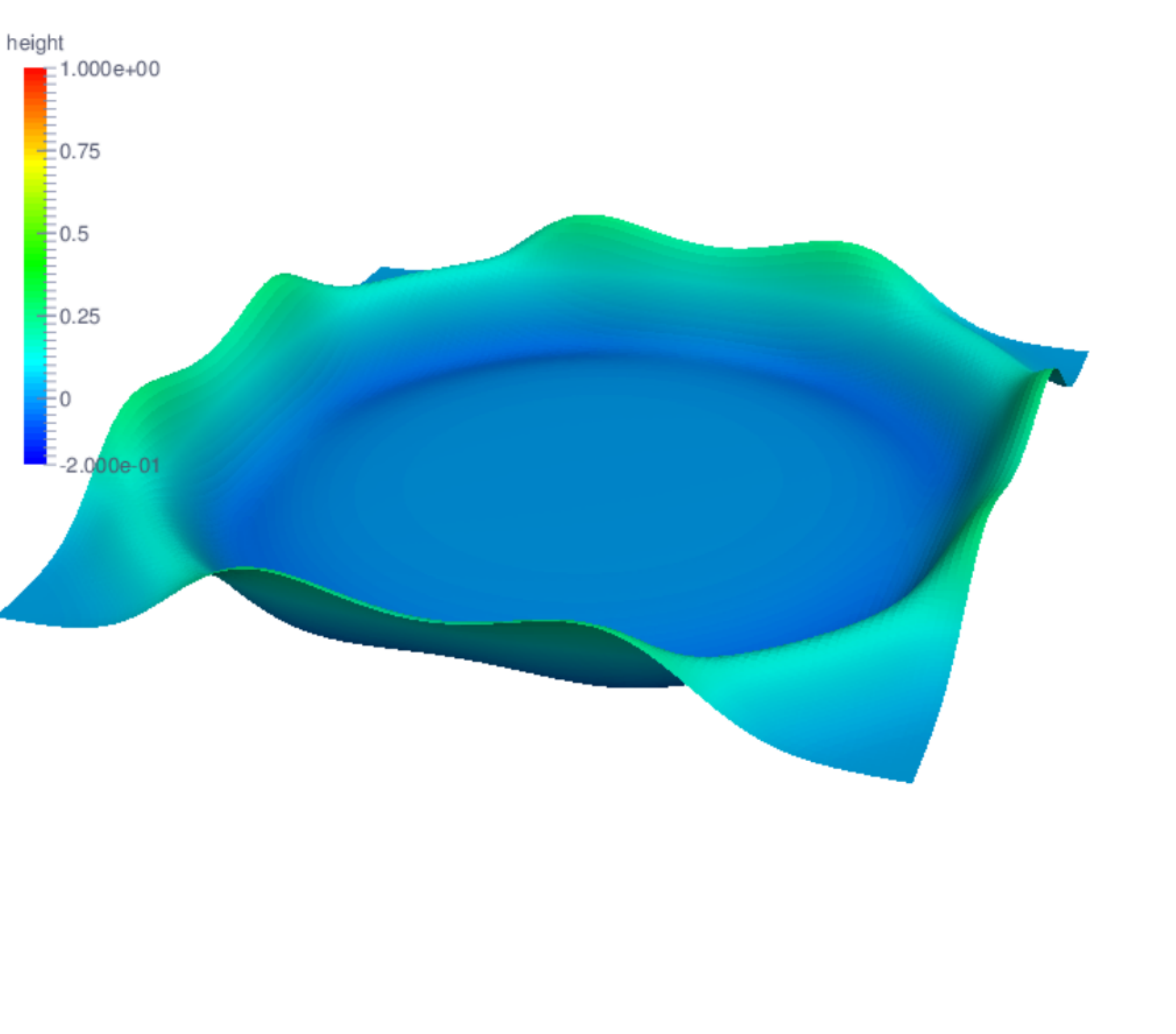}
    \includegraphics[clip=true,width=0.3\textwidth]{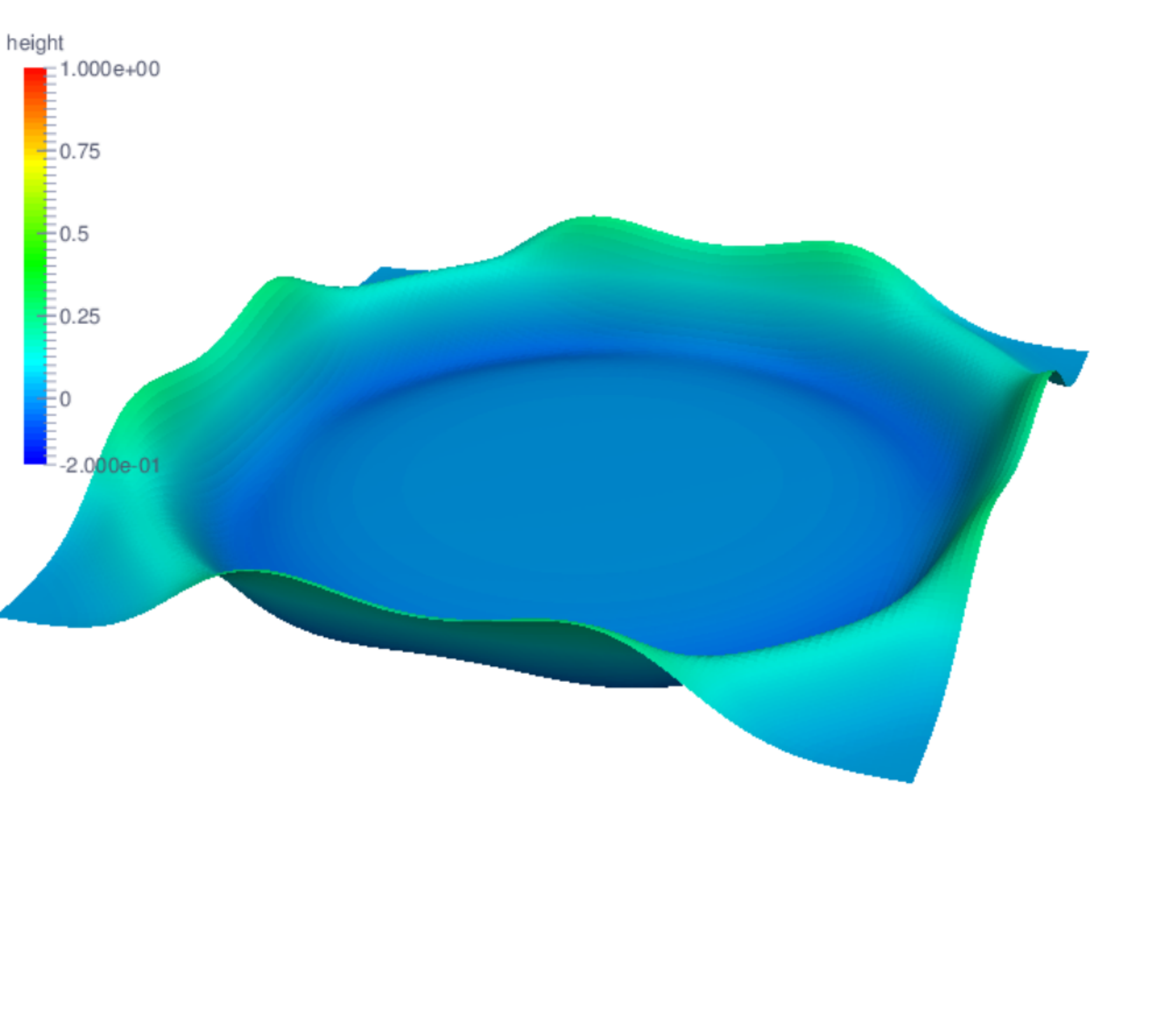}
    \includegraphics[clip=true,width=0.3\textwidth]{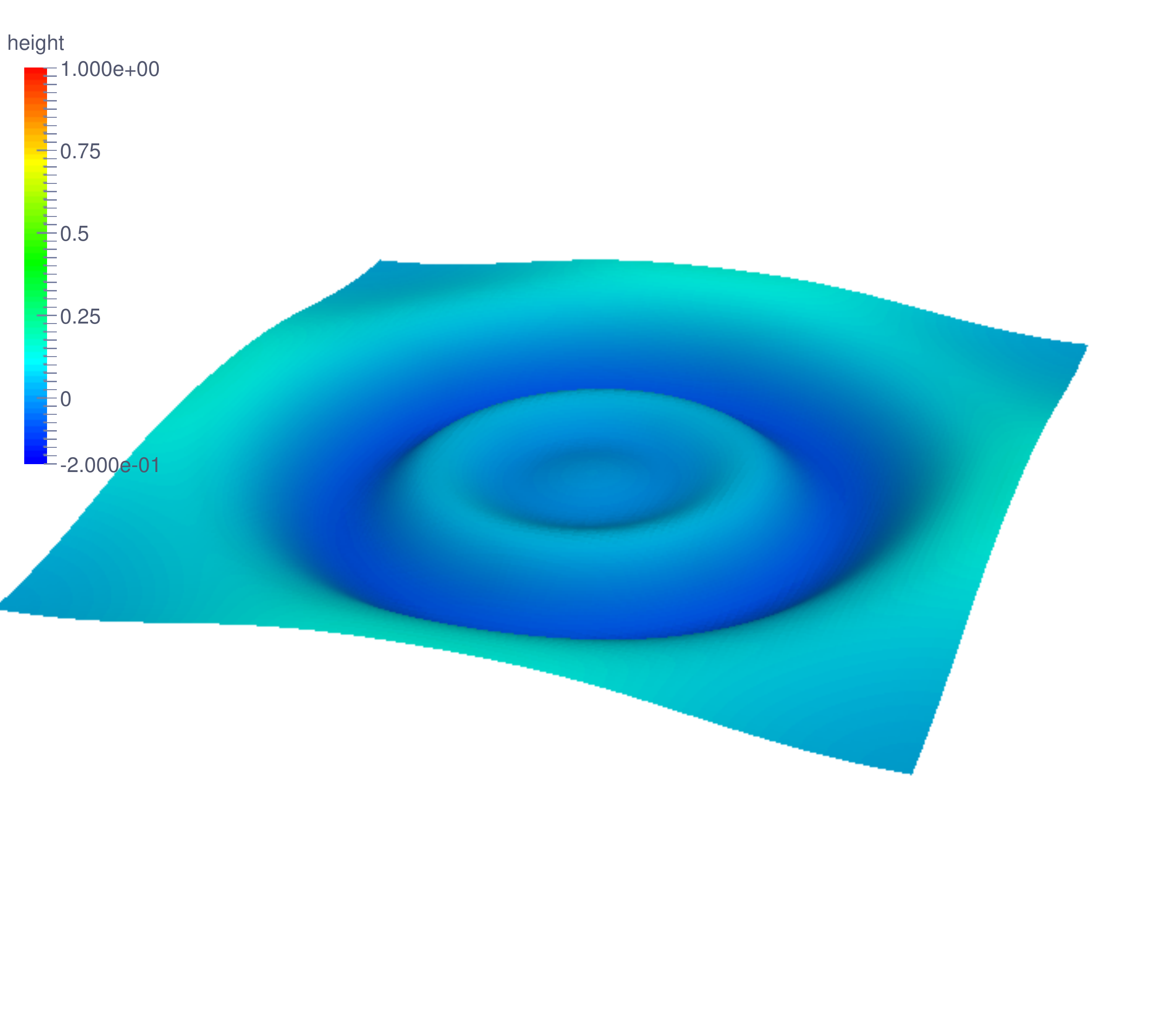}\\

    \caption{ Water height perturbation test: free surface evolution for RK2 DG ($\triangle t$=0.0001, left), ARS2 HDG-DG ($\triangle t$=0.002, center) and ARS2 HDG-DG ($\triangle t$=0.02, right) at times $t=0$, $t=0.06$, and $t=0.1$. }
    \figlab{waterdrop_evolution}
  \end{centering}
  \end{figure}

 In Table \tabref{waterdrop_wallclock}, we compare the wallclock times
 of ARS2 HDG-DG with those of RK2 DG. As can be seen, ARS2 HDG-DG with
 $\triangle t = 0.002$ is comparable to RK2 DG with $\triangle t = 0.0001$ in both
 wallclock time and in accuracy (see Figure
 \figref{waterdrop_evolution}). When $\triangle t = 0.02$ (two hundred times
 larger than RK2 DG stable time-step size), ARS2 HDG-DG outperforms
 RK2 DG in terms of computational cost (though with a less accurate solution).

  \begin{table}[t]
    \caption{Water height perturbation test: wallclock time comparison for ARS2 HDG-DG and RK2 DG where TI denotes time-integrator.}
    \begin{center}
      \begin{tabular}{c|c|c|c|c}
        TI &  $\triangle t$ & Cr  & Wallclock time \\
        \hline\hline
         RK2 DG    &   0.0001 &      0.2 & 18m 20s \\
         ARS2 HDG-DG&   0.0020 &      4.0 & 20m 41s  \\
         ARS2 HDG-DG&   0.0200 &     40.0 &  2m 28s \\
        \hline\hline
      \end{tabular}
    \end{center}
    \tablab{waterdrop_wallclock}
  \end{table}

\subsection{Steady-state geostrophic flow}
\seclab{SSgeotrophic}
We consider the steady-state geostrophic flow in
\cite{williamson1992standard} (a 
geostrophically balanced flow). This flow admits an analytical solution for the shallow water equations on the sphere
\cite{nair2005discontinuous}.
The initial condition is given by
  \begin{subequations}
  \begin{align}
    H &= H_\infty - \frac{1}{g}\LRp{a\Omega + \frac{u_\infty^2}{2} }\cos^2\theta,\\
    \ulon &= \uinf\cos\theta,\\
    \ulat &= 0,
  \end{align}
  \eqnlab{sswe_tc2_ic}
  \end{subequations}
where $(\ulon,\ulat)$ is the local tangential velocity in
latitude-longitude coordinates $(\lambda,\theta)$. We take
$gH_\infty=2.94\times 10^{4} m^2s^{-2}$, and $\uinf=38.61ms^{-1}$.
The numerical experiment is performed on a grid with $N_e=1536$ elements ($16 \times 16$ elements on each of the six faces of the cubed-sphere) and solution order $p=4$. 
The time-step size for ARS2 HDG-DG is $864s$.

 Figure \figref{sswe_tc2_h} shows the snapshot of the height
field from the ARS2 HDG-DG approach (Figure  \figref{sswe_tc2_h_hars2}) and the exact field  
(Figure \figref{sswe_tc2_h_exact}) after $12$ days.
The height field from ARS2 HDG-DG is almost the same as the exact solution.
Indeed, we show in Figure 
\figref{sswe_tc2_h_relerr} the relative error of the height field, $\vert \frac{H_{num} - H_{exact}}{H_{exact}} \vert$,  and  the maximum relative error of 
$\mc{O}\LRp{10^{-7}}$ is observed (see also Figure \figref{sswe_tc2_hars2_errhistory}).

  \begin{figure}[h!t!b!]
    \centering
    \subfigure[ARS2 HDG-DG]{
      \includegraphics[trim=3.7cm 0.4cm 3.3cm 1cm,clip=true,width=0.3\columnwidth]{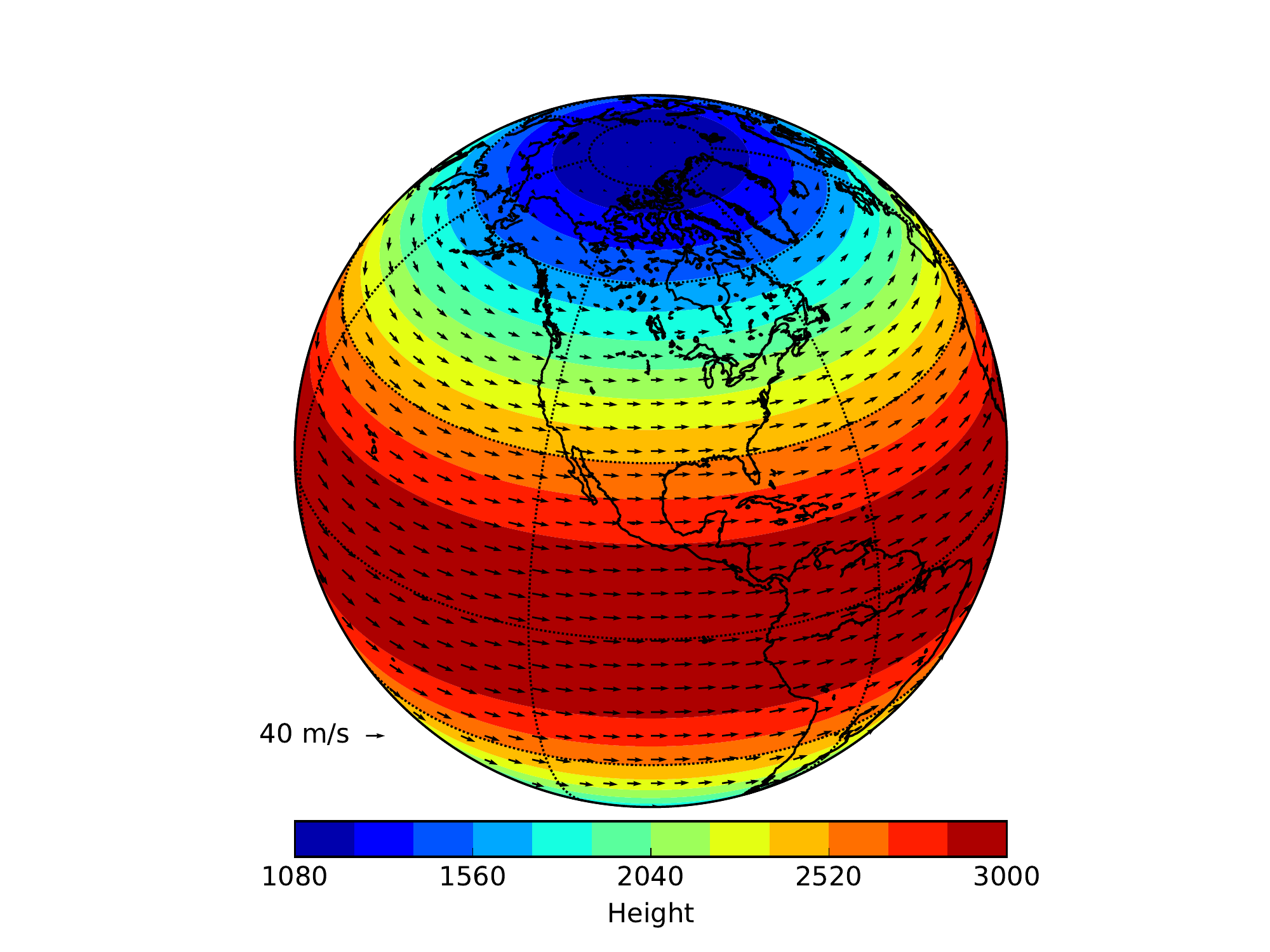}
      \figlab{sswe_tc2_h_hars2}
    }
    \subfigure[Exact solution]{
      \includegraphics[trim=3.7cm 0.4cm 3.3cm 1cm,clip=true,width=0.3\columnwidth]{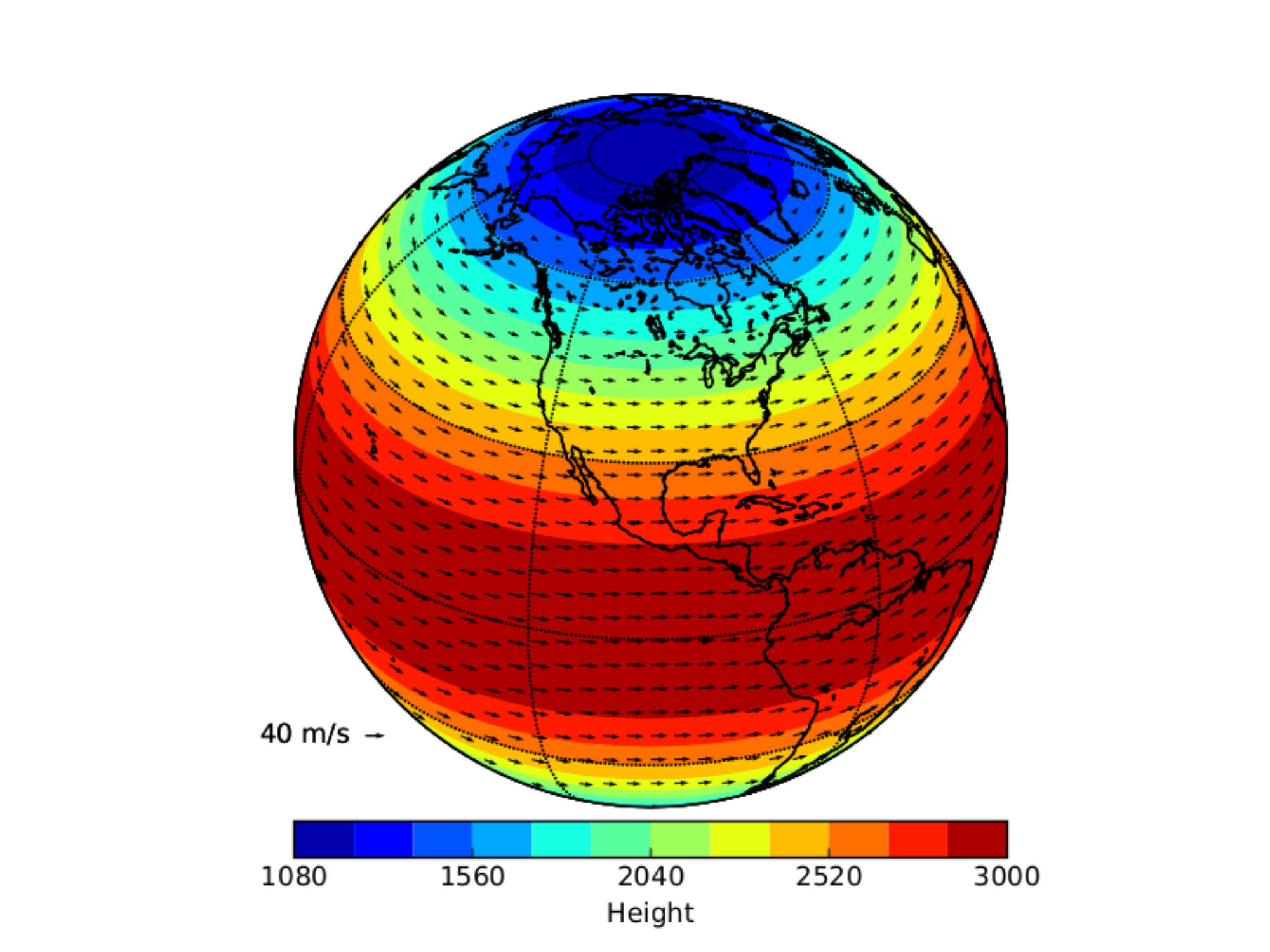}
      \figlab{sswe_tc2_h_exact}
    }
    \subfigure[Relative error]{
      \includegraphics[trim=3.7cm 0.4cm 3.3cm 1cm,clip=true,width=0.3\columnwidth]{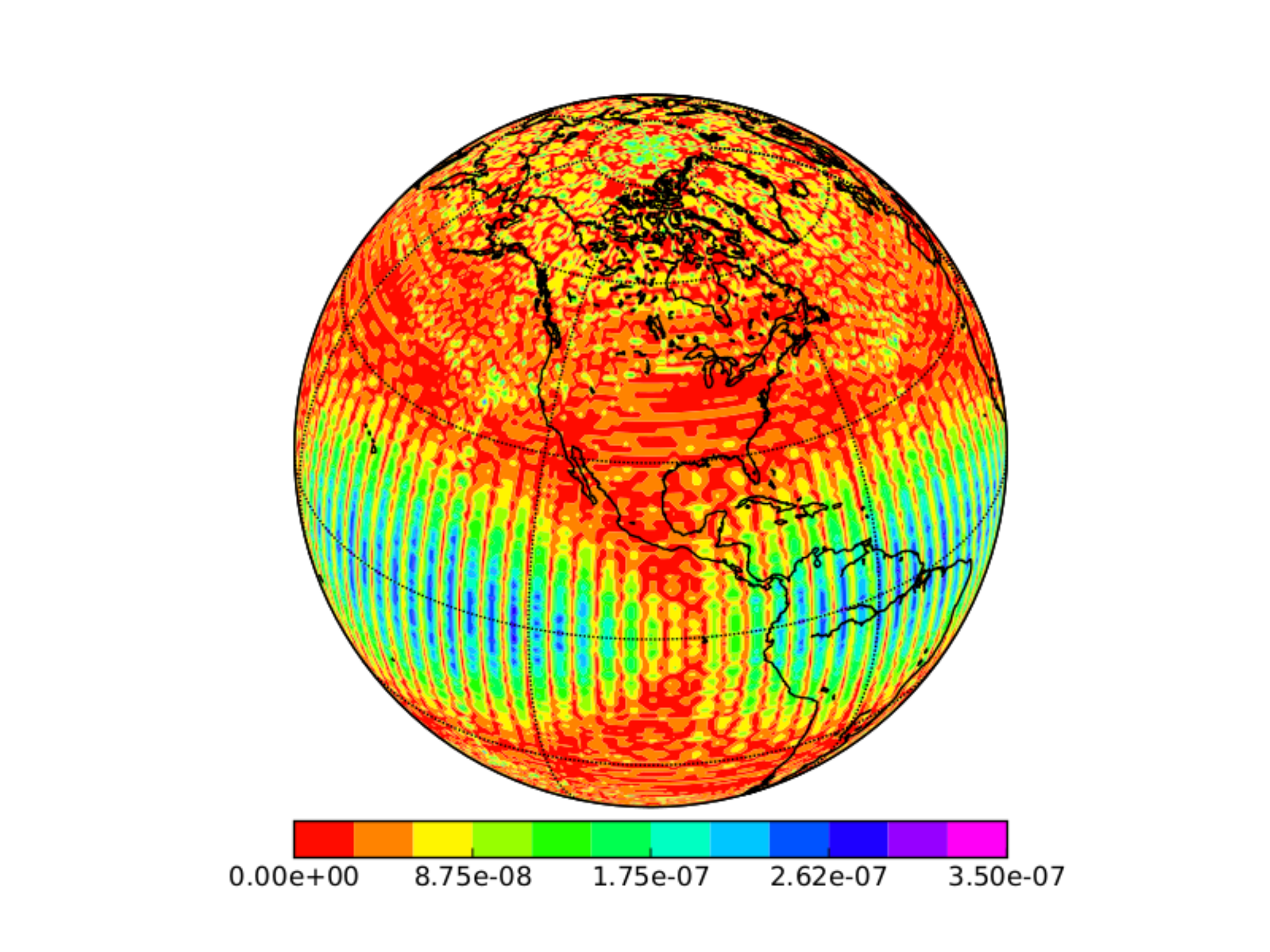}
      \figlab{sswe_tc2_h_relerr}
    }
    \caption{Geostrophic flow test case: (a) total height field from  ARS2 HDG-DG at day 12 with Cr $=1.36$, (b) the exact solution, and 
      (c) the relative error of the height field.
    }
    \figlab{sswe_tc2_h}
  \end{figure}

  Figure \figref{sswe_tc2_history} shows the time series of the height field error, mass, and energy loss in $L_1$, $L_2$ and $L_\infty$ norms.

\begin{figure}[h!t!b!]
  \centering
  \subfigure[Height errors in $L_1$, $L_2$ and $L_\infty$ norms]{
    \includegraphics[trim=0.8cm 0.5cm 2.5cm 0.5cm,clip=true,width=0.45\columnwidth]{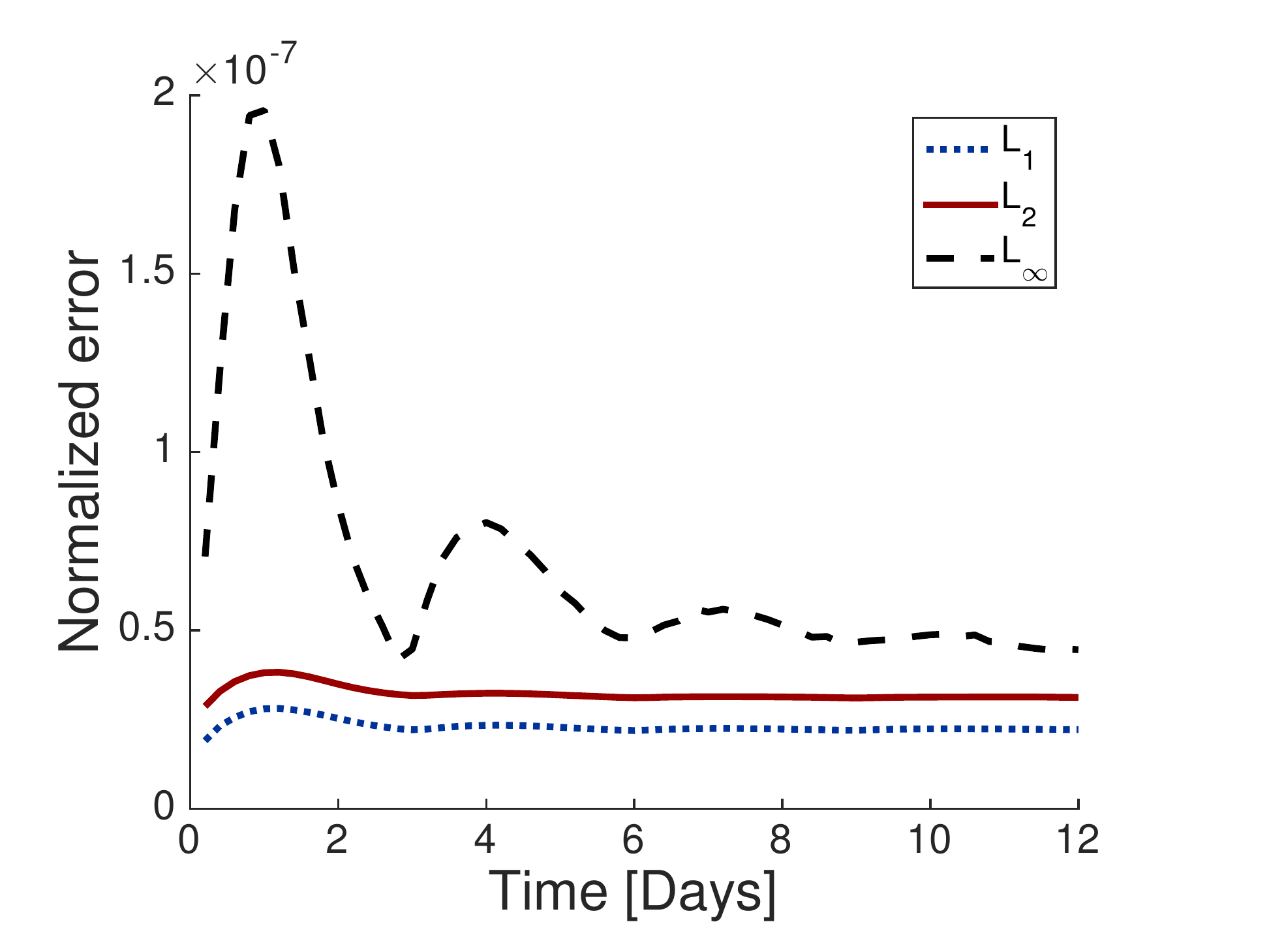}
    \figlab{sswe_tc2_hars2_errhistory}
  }
  \subfigure[Mass and energy loss]{
    \includegraphics[trim=0.8cm 0.5cm 2.5cm 0.5cm,clip=true,width=0.45\columnwidth]{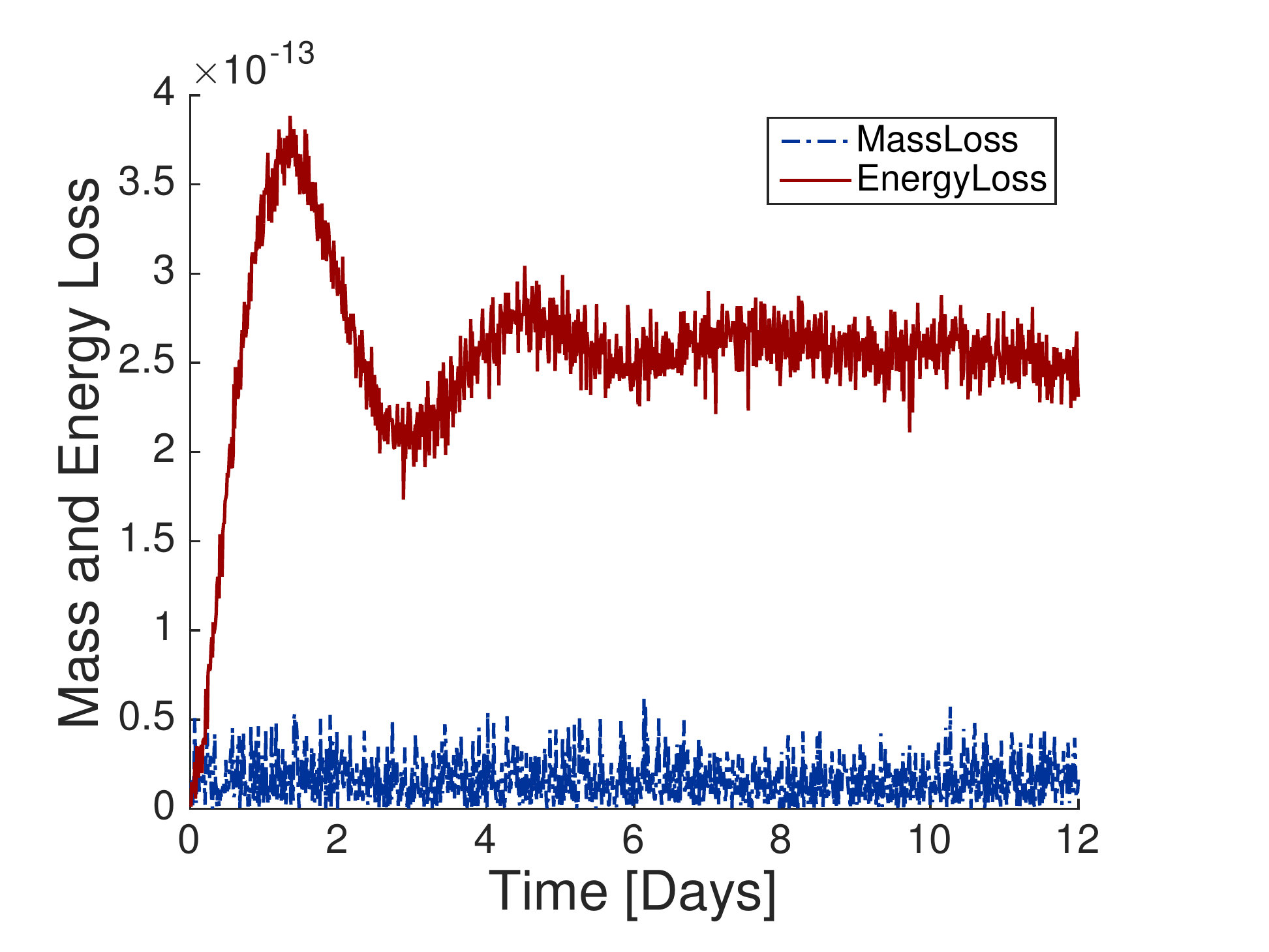}
    \figlab{sswe_tc2_hars2_masshistory}
  }
  \caption{Geostrophic flow test case: (a) time evolution of height field error in  $L_1$, $L_2$ and $L_\infty$ norms, and (b) mass and energy loss of the ARS2 HDG-DG solution with time-step size of $864 \text{s}$ (i.e. Cr $=1.36$).}
  \figlab{sswe_tc2_history}
 \end{figure}

 Here,
   the mass and energy losses are defined as 
\begin{align*}
  \text{ mass loss }   = \abs{ \frac{\mass(t)-\mass(0)}{\mass(0)} },
  \text{ energy loss } = \abs{ \frac{\energy(t)-\energy(0)}{\energy(0)} },
\end{align*}

where $\mass:=\norm{H}^2_\Omegah$ and
$\energy:=\norm{H\ub\cdot\ub+gH^2}^2_\Omegah$.
 We observe the energy and mass are conserved
and this is a direct consequence of the fact
that both DG and HDG are conservative discretizations.

\subsection{Steady-state geostrophic flow with compact support}

This case is similar to the steady-state  geostrophic flow in section \secref{SSgeotrophic}. The difference is that it is equipped 
with a compactly supported wind field, 
considered as a high latitude jet in the northern hemisphere.
The initial condition is given as

  \begin{subequations}
  \begin{align}
    H &= H_\infty - \frac{a}{g}\int_{-\pi/2}^\theta 
         \LRp{f + \frac{u(\tau)\tan\tau}{a} }u(\tau) d\tau,\\
    \ulon &= u_\infty b(x)b(x_e - x)e^{4/x_e},\\
    \ulat &= 0,
  \end{align}
  \eqnlab{sswe_tc3_ic}
  \end{subequations}
  where 
  $\theta_b=-\pi/6$, $\theta_e=\pi/2$, $x_e=0.3$,
  $x = x_e \frac{\theta - \theta_b}{\theta_e - \theta_b},$ and 
  $ 
    b(x) = 
      \begin{cases}
        0 & \text{for  } x\le 0,\\
        e^{-1/x} & \text{for  } 0 < x.
      \end{cases}
  $

The  total simulation time is 12 days.
  As shown in Figure \figref{sswe_tc3_h}, the height field of the ARS2 HDG-DG solution is similar to that of the exact solution: indeed Figure \figref{sswe_tc3_h_relerr}  shows that the relative error is of order $\mc{O}\LRp{10^{-5}}$. 

  \begin{figure}[h!t!b!]
    \centering
    \subfigure[ARS2 HDG-DG]{
      \includegraphics[trim=3.7cm 0.4cm 3.3cm 1cm,clip=true,width=0.3\columnwidth]{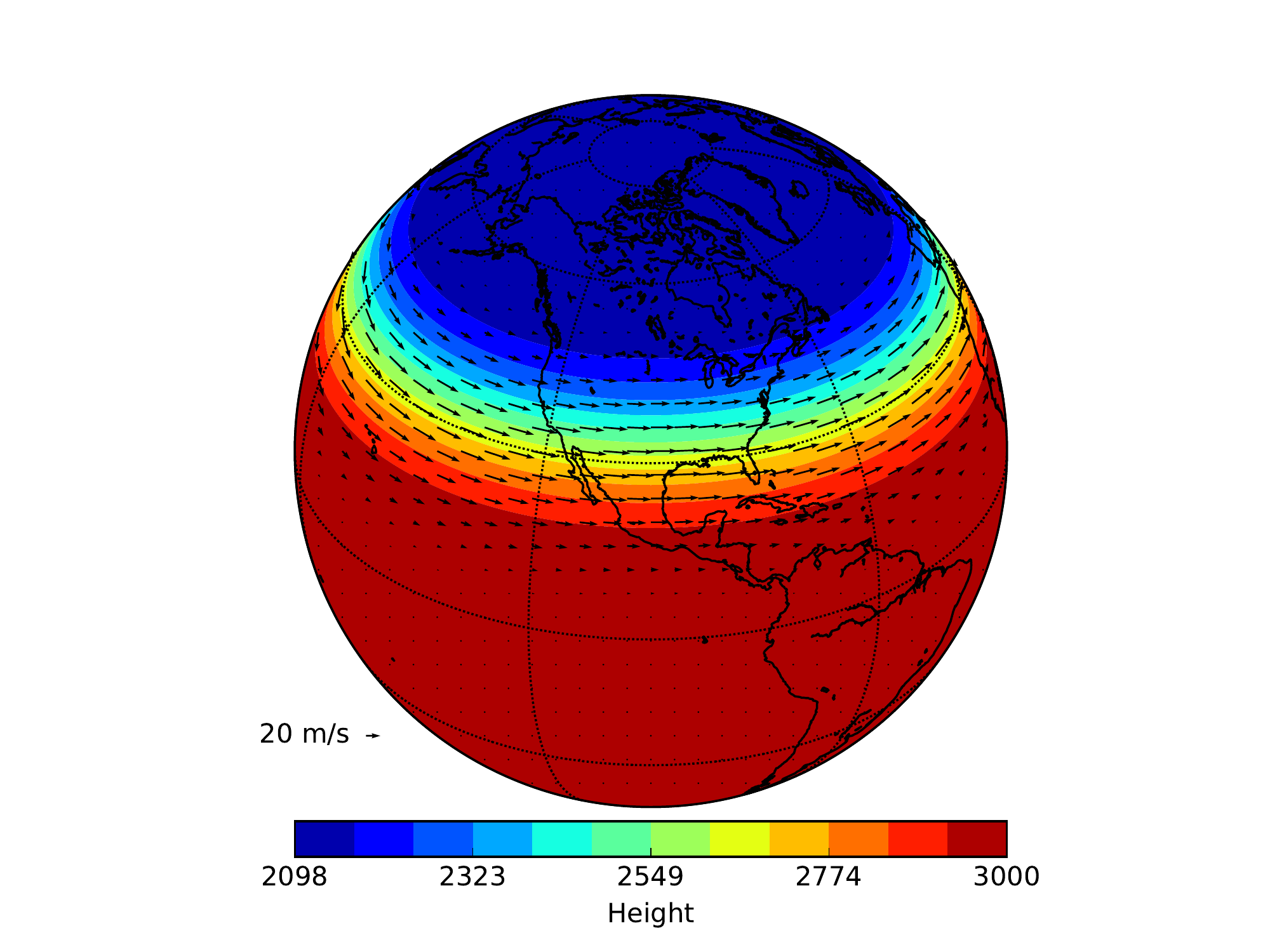}
      \figlab{sswe_tc3_h_hars2}
    }
    \subfigure[Exact solution]{
      \includegraphics[trim=3.7cm 0.4cm 3.3cm 1cm,clip=true,width=0.3\columnwidth]{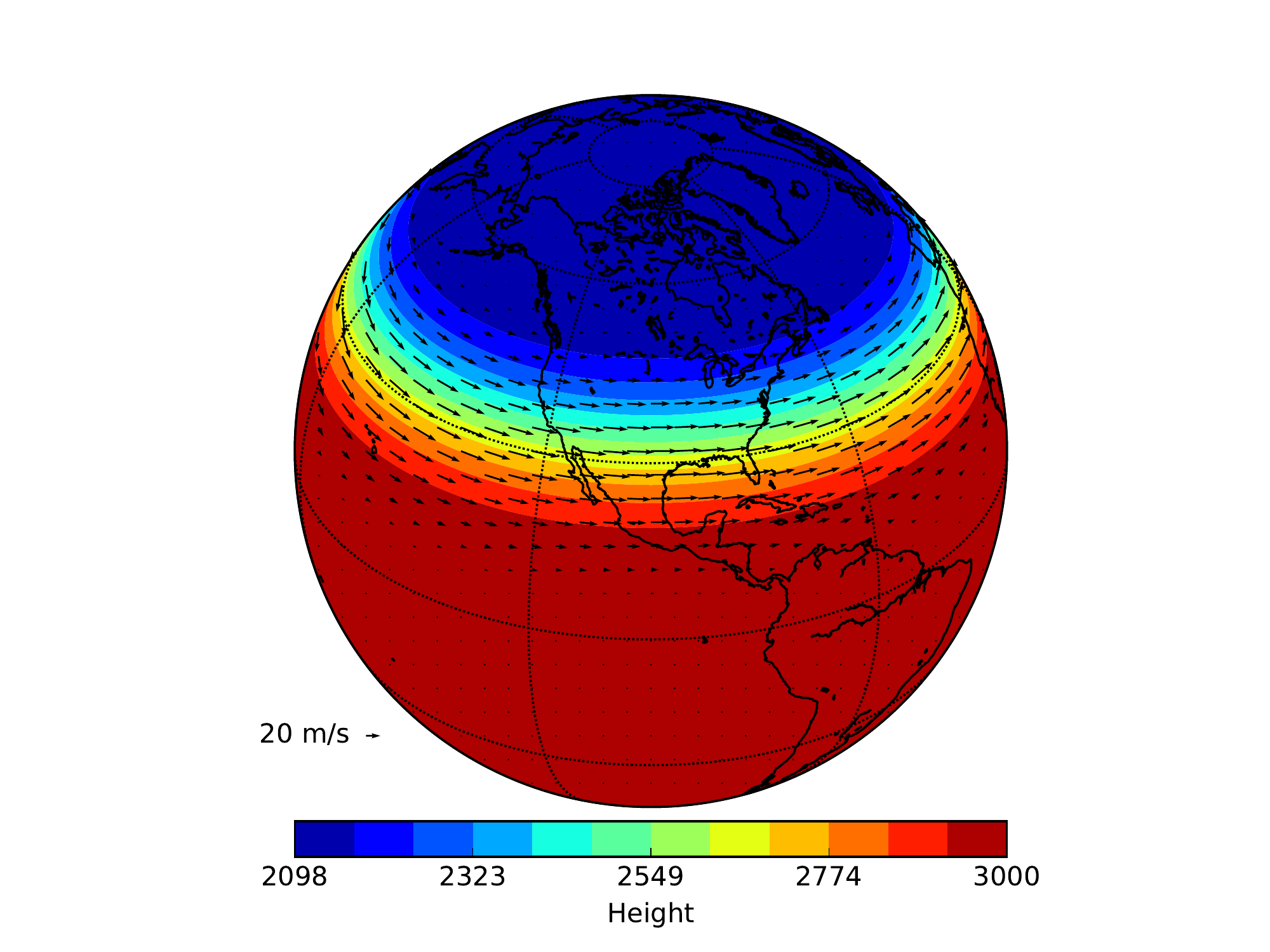}
      \figlab{sswe_tc3_h_exact}
    }
    \subfigure[Relative error]{
      \includegraphics[trim=3.7cm 0.4cm 3.3cm 1cm,clip=true,width=0.3\columnwidth]{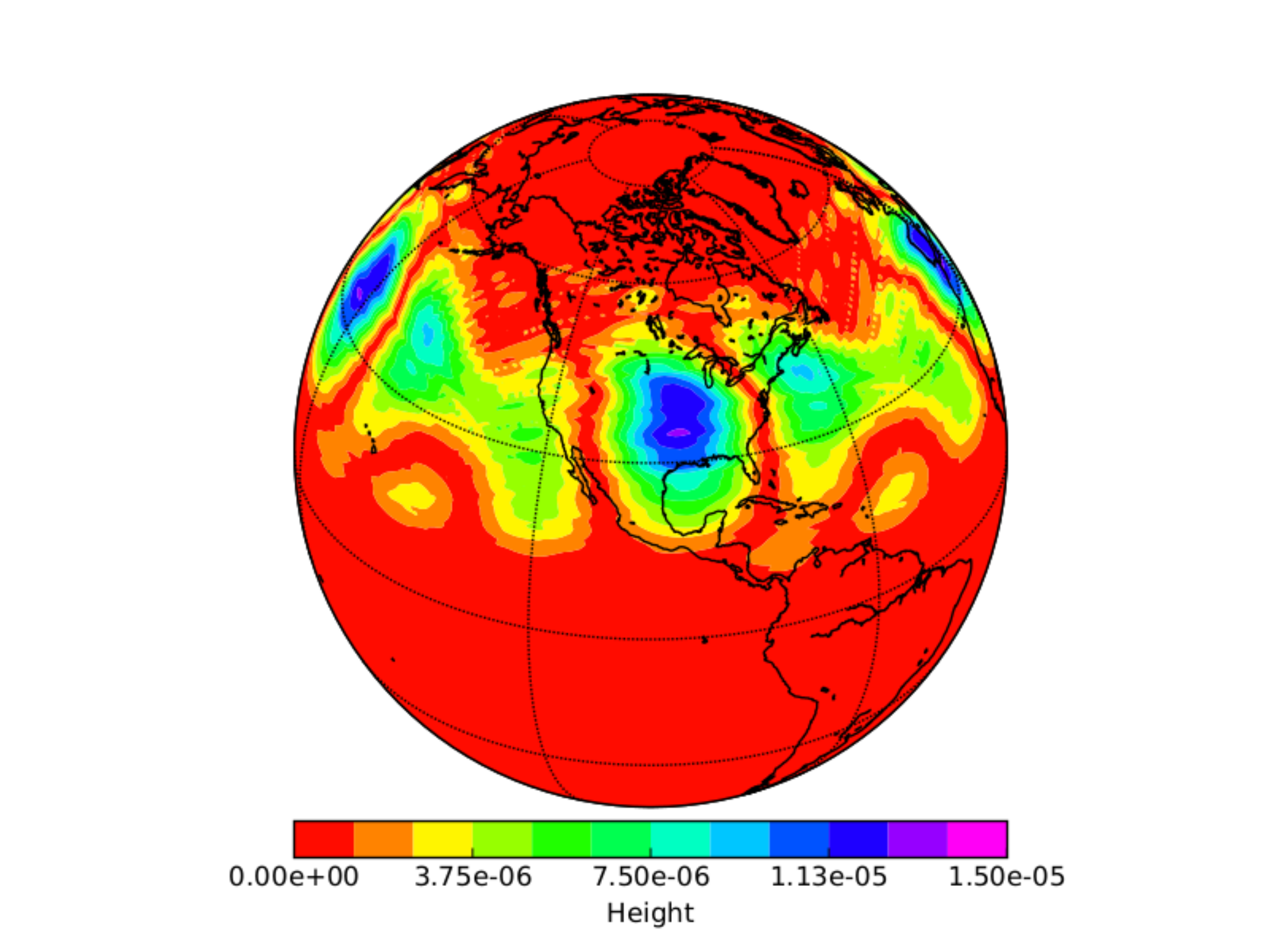}
      \figlab{sswe_tc3_h_relerr}
    }
    \caption{
Steady-state geostrophic flow with compact support: (a) total height field from  ARS2 HDG-DG at day 12 with Cr $=1.2$, (b) the exact solution, and 
      (c) the relative error of the height field.
    }
    \figlab{sswe_tc3_h}
  \end{figure}

 For the spatial convergence test, we conduct both $h$-convergence and
  $p$-convergence studies in Figure \figref{sswe_tc3_spatial_conv}.

  \begin{figure}[h!t!b!]
    \centering
    \subfigure[h-convergence]{
      \includegraphics[trim=0.6cm 0.5cm 3cm 0.3cm,clip=true,width=0.45\columnwidth]{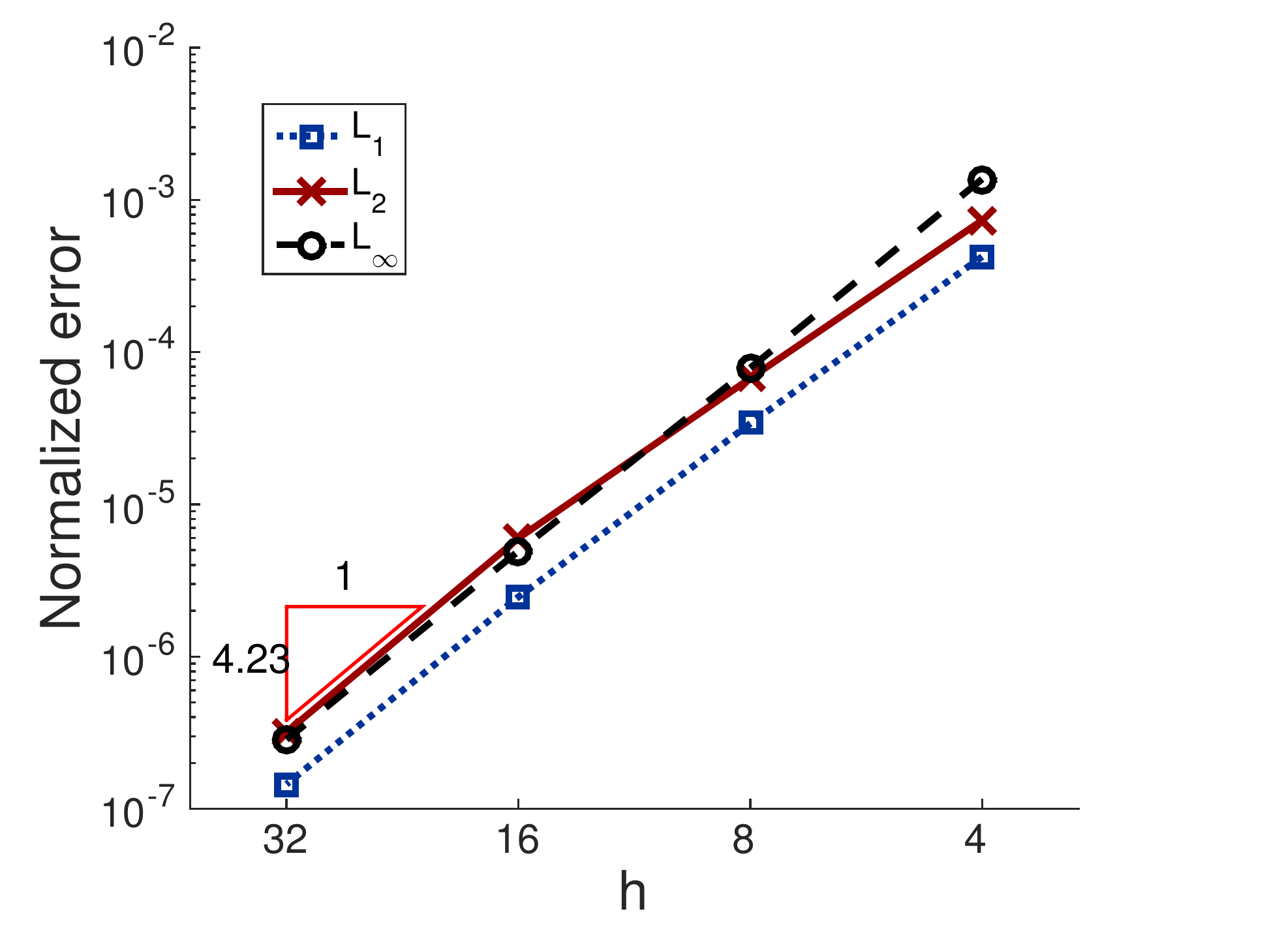}
      \figlab{sswe_tc3_h_conv_hars2}
    }
    \subfigure[p-convergence]{
      \includegraphics[trim=0.6cm 0.5cm 3cm 0.3cm,clip=true,width=0.45\columnwidth]{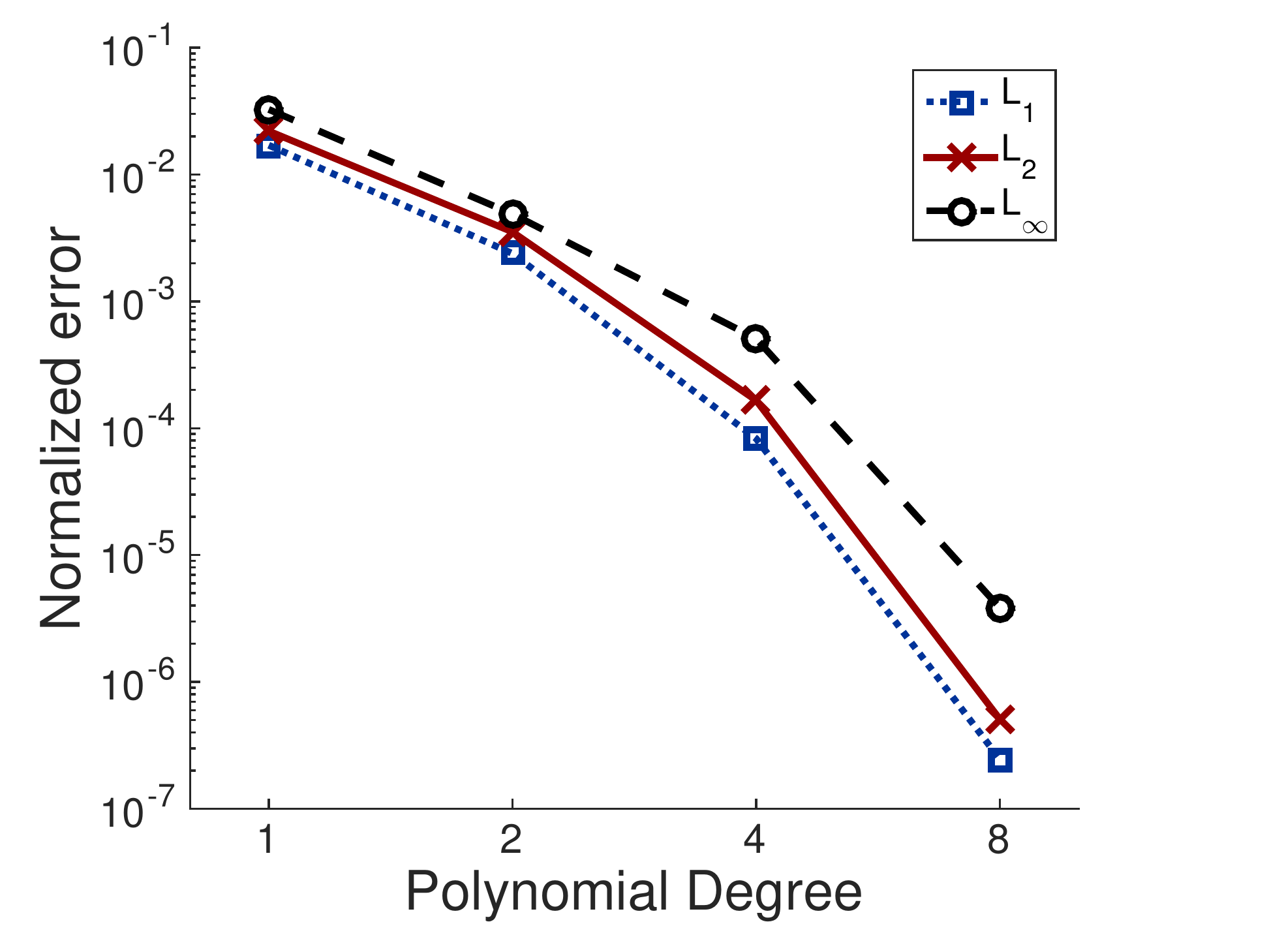}
      \figlab{sswe_tc3_p_conv_hars2}
    }
    \caption{Convergence studies for the ARS2 HDG-DG scheme when applied to the steady-state
      geostrophic flow with compact support with Cr $=0.7$: 
      (a) $\mathrm{h}$-convergence with $p=3$ and (b) $p$-convergence with $N_e=6144$.}
    \figlab{sswe_tc3_spatial_conv}
  \end{figure}

 The errors are measured at $T=0.4$  with  Cr $=0.7$.
 For $h$-convergence, the height field error in $L_1$, $L_2$ and $L_\infty$ norms
 are computed for $p=3$ and the total number of elements is given by $N_e=6{\mathrm{n}}^2$,
 where ${\mathrm{n}} =\LRc{4, 8, 16, 32}$. As can be seen in Figure  \figref{sswe_tc3_h_conv_hars2}, the convergence rate is $p+1$.  For $p$-convergence, an exponential rate is observed in Figure
 \figref{sswe_tc3_p_conv_hars2}.

\subsection{Zonal flow over an isolated mountain}

We consider the zonal flow over an isolated mountain test proposed in
\cite{williamson1992standard}. 
The height and wind fields are similar to those of the steady-state 
  geostrophic flow, but now $H_\infty=5960 \text{m}$ and $u_\infty=20 \text{ms}^{-1}$.
A mountain with height $H_s = 2000 (1-r/r_s) \text{m}$, located at 
$(\lambda_c,\theta_c) = (3\pi/2,\pi/6)$, is introduced in the flow, 
 where $r_s = \frac{\pi}{9}$
and $r^2 = \min(r^2,(\lambda-\lambda_c)^2 + (\theta-\theta_c)^2)$.

We plot the height field at days 5, 10 and 15 in Figure
\figref{sswe_tc5_hars2} on a grid with $N_e=384$ elements ($8 \times 8 \times 6$ elements on the cubed-sphere) and solution order $p=8$.
  The time-step size of 432 seconds is taken.
 As can be seen, the height fields are smooth and
comparable to the corresponding results in
\cite{nair2005discontinuous,ullrich2010high} (note
that this problem has no analytical solution).

  \begin{figure}[h!t!b!]
    \centering
    \subfigure[Day 5]{
      \includegraphics[trim=3.7cm 0.4cm 3.3cm 1cm,clip=true,width=0.3\columnwidth]{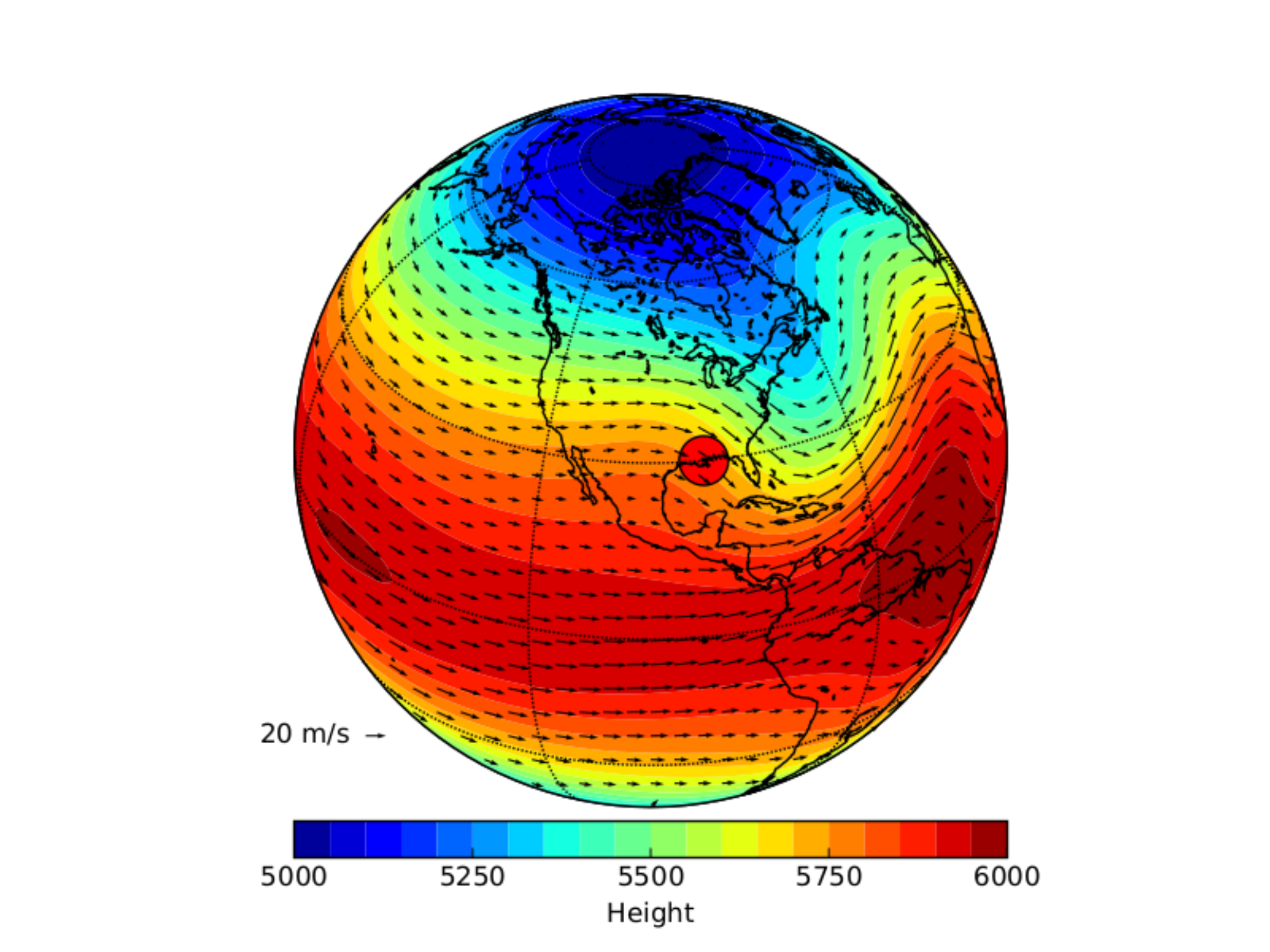}
      \figlab{sswe_tc5_hars2_day05}
    }
    \subfigure[Day 10]{
      \includegraphics[trim=3.7cm 0.4cm 3.3cm 1cm,clip=true,width=0.3\columnwidth]{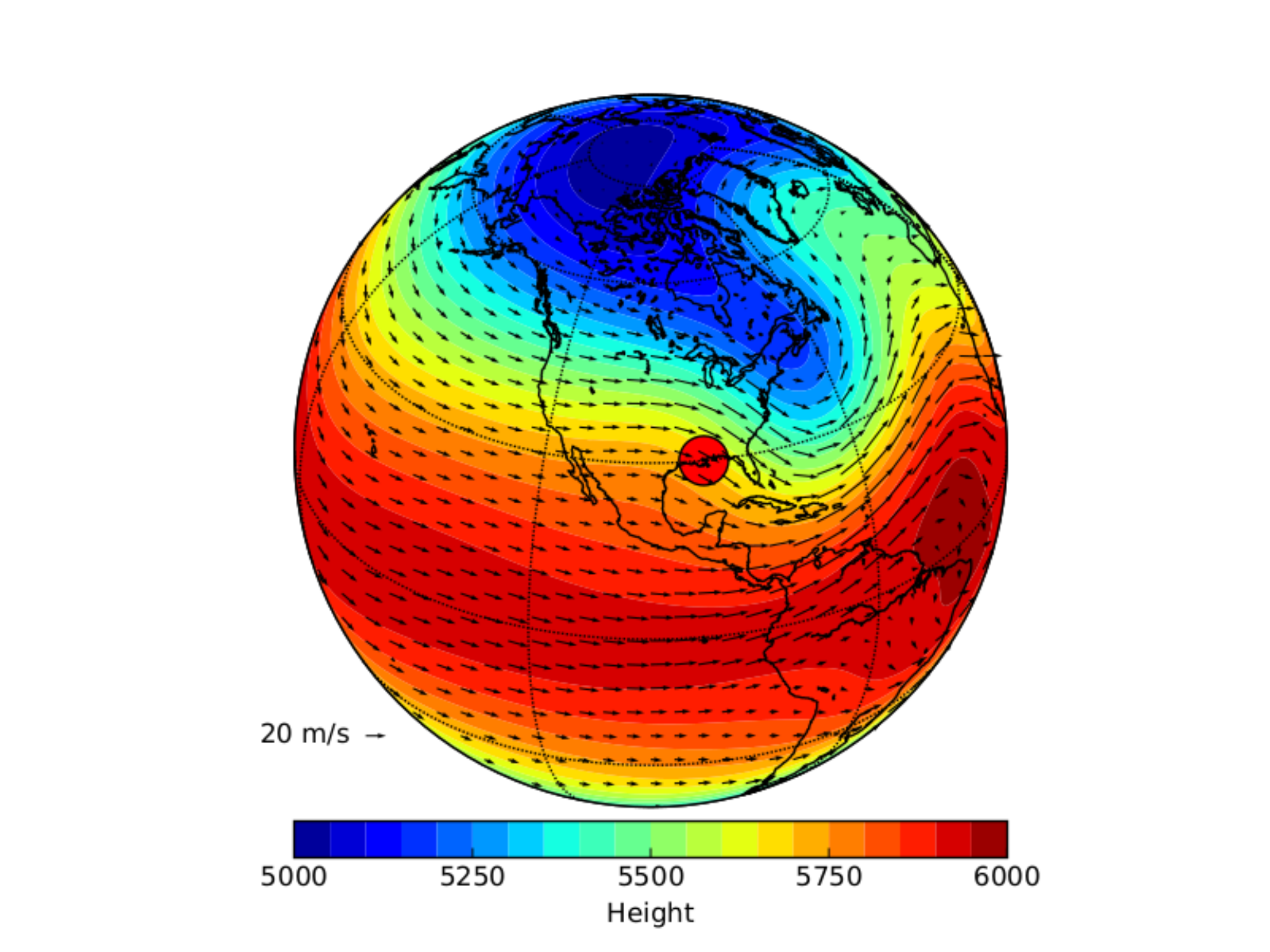}
      \figlab{sswe_tc5_hars2_day10}
    }
    \subfigure[Day 15]{
      \includegraphics[trim=3.7cm 0.4cm 3.3cm 1cm,clip=true,width=0.3\columnwidth]{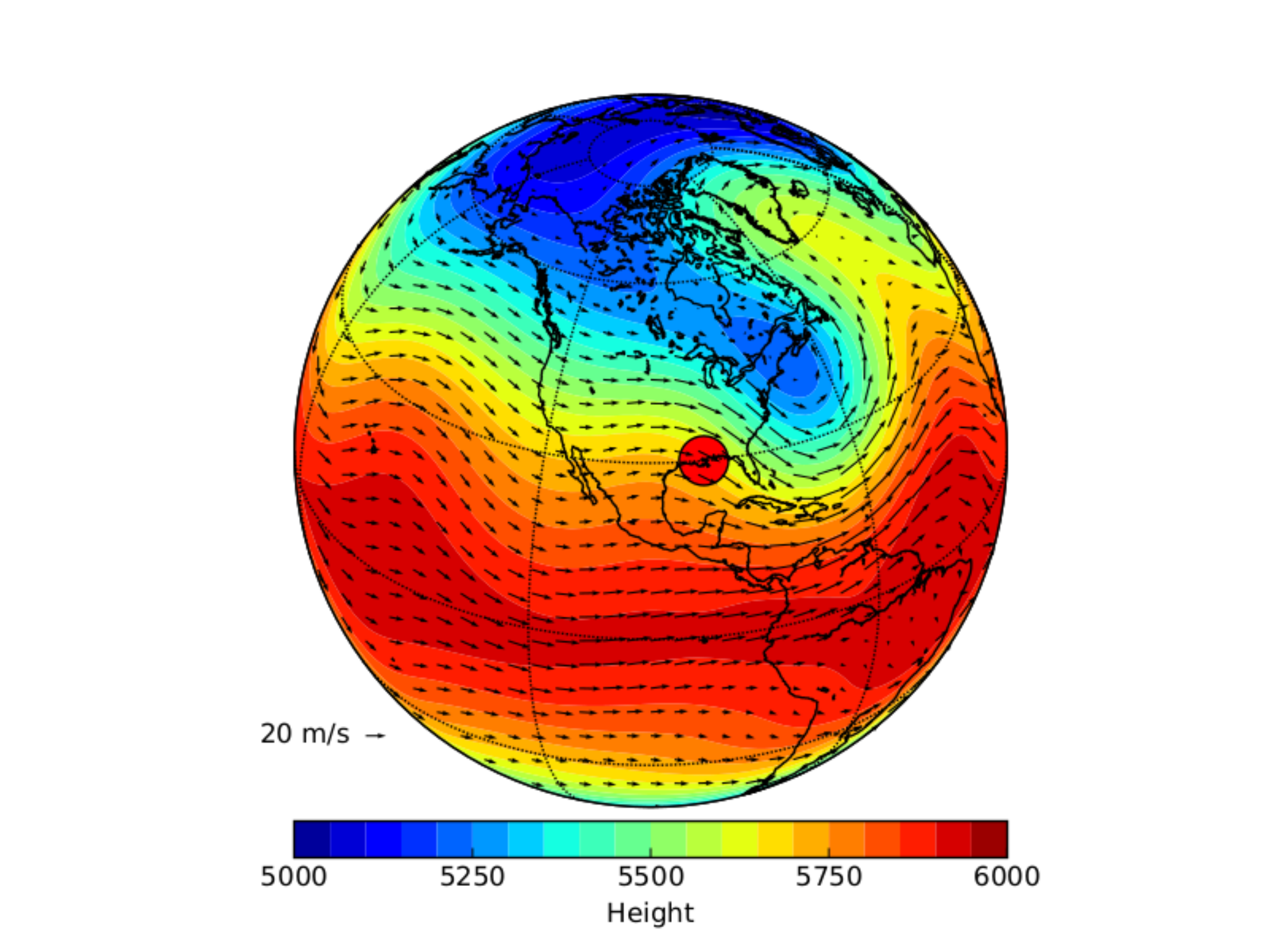}
      \figlab{sswe_tc5_hars2_day15}
    }
    \caption{Flow over an isolated mountain (red circle) computed with the ARS2 HDG-DG scheme:
      shown are the total water height after (a) 5 days, (b) 10 days,
      and (c) 15 days. The numerical experiments are performed 
      on a grid with $N_e=384$ elements, solution order $p=8$, and
      Cr $=1.2$. Contour levels are from $5000$m to
      $6000$m with 21 levels.} 
    \figlab{sswe_tc5_hars2}
  \end{figure}
  
  We compare the height field of ARS2 HDG-DG with that of RK2 DG in Figure \figref{sswe_tc5_h}. 
    We take $\triangle t=43.2$ seconds (Cr=0.15) 
    for RK2 DG. The height field of ARS2 HDG-DG is in good agreement
    with that of RK2 DG: the relative difference in the height field
    is $\mc{O}\LRp{10^{-3}}$.

  \begin{figure}[h!t!b!]
    \centering
      \includegraphics[trim=1.0cm 14.5cm 6.0cm 5.5cm,clip=true,width=0.46\columnwidth]{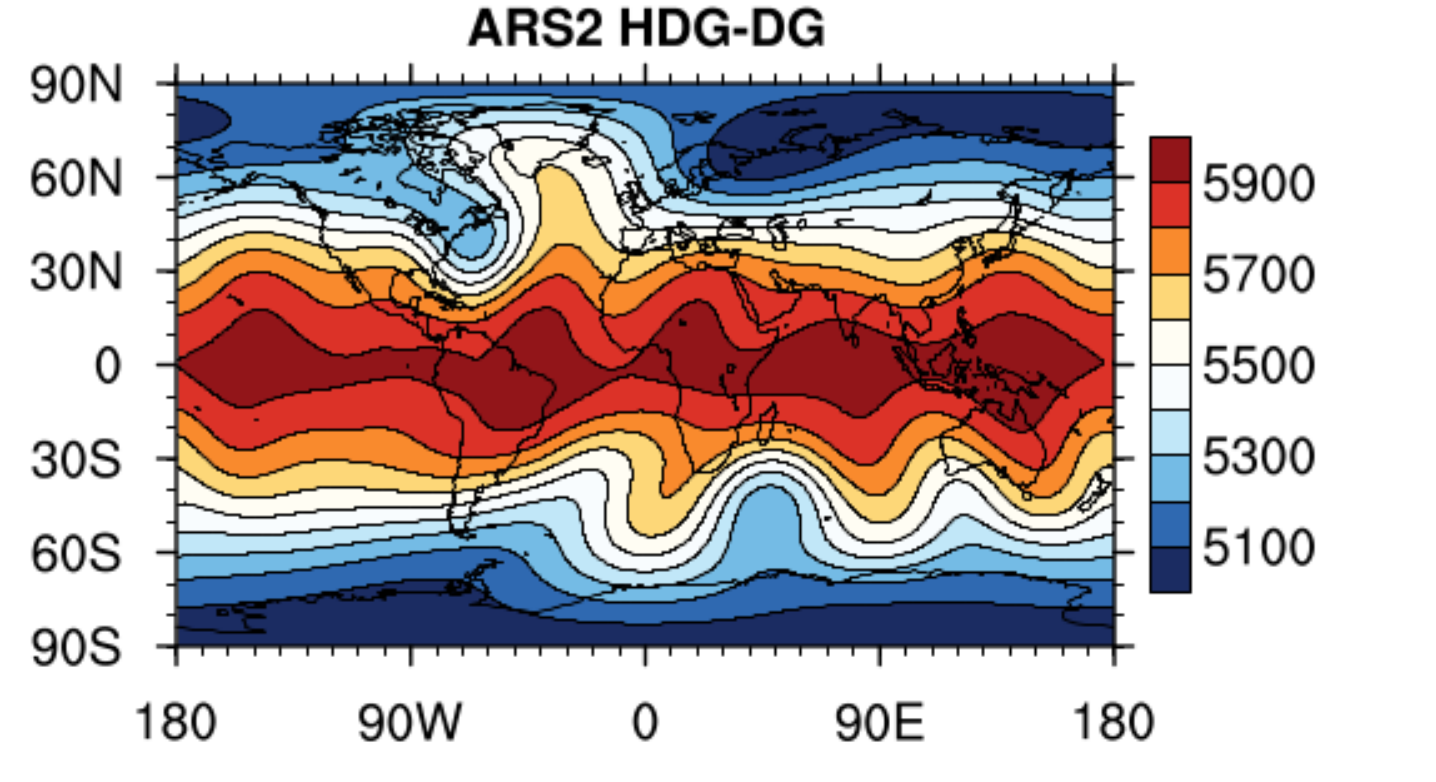}\hspace{-2mm}
      \includegraphics[trim=1.0cm 14.5cm 6.0cm 5.5cm,clip=true,width=0.46\columnwidth]{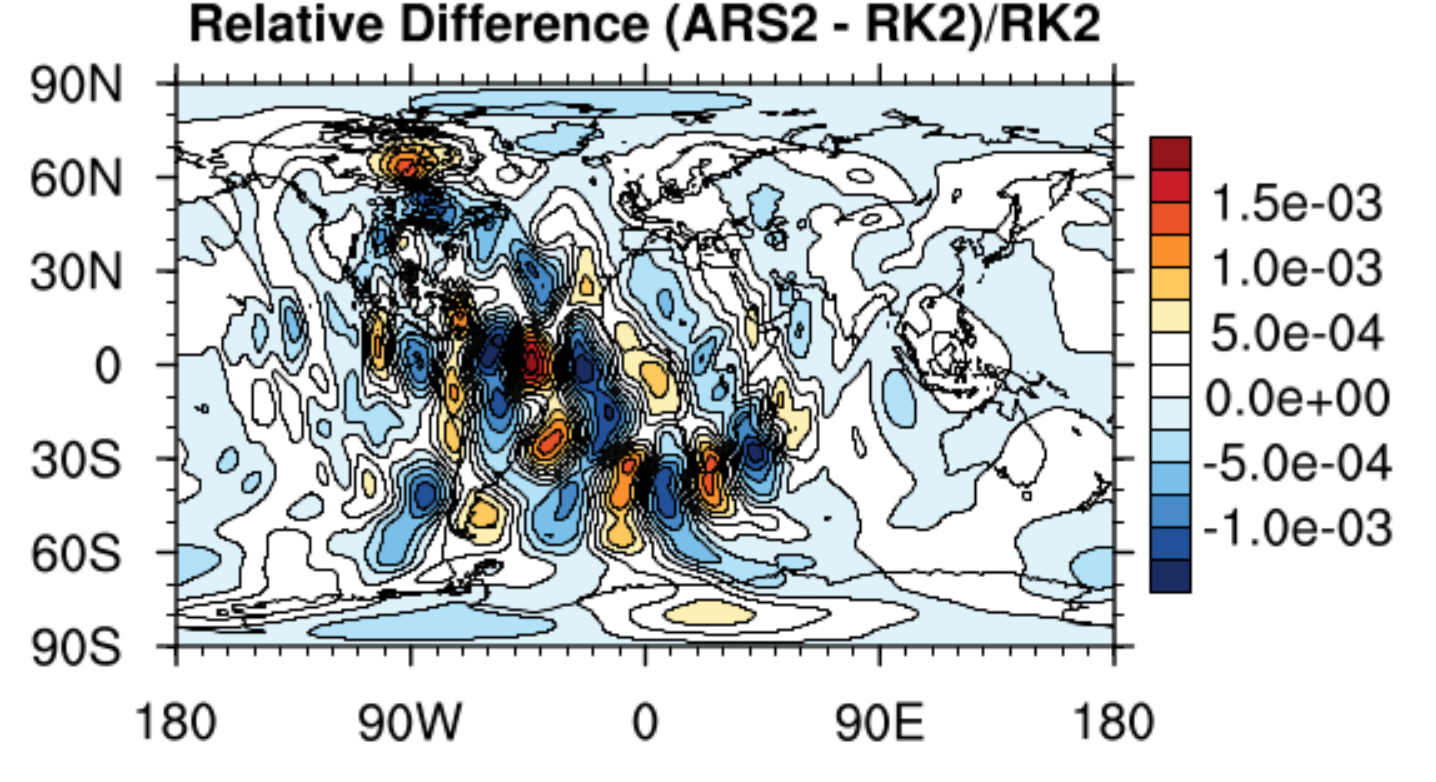}
    \caption{The zonal flow over an isolated mountain:
total height field computed from ARS2 HDG-DG (left) at day 15 with Cr of $1.46$; 
the relative difference (right) with RK2 in the height field.}
      \figlab{sswe_tc5_h}
  \end{figure}

\subsection{Rossby-Haurwitz wave}
We next consider the Rossby-Haurwitz wave test in \cite{williamson1992standard}. 
The Rossby-Haurwitz wave is an exact solution of the nonlinear barotropic vorticity equation \cite{haurwitz1940motion}, but not an exact solution of the shallow water system \cite{nair2005discontinuous}. The wave number is chosen to be 4.

To simulate this test case, we use the ARS2 HDG-DG scheme on a grid with $N_e=864$ elements ($12\times12\times6$), solution order  $p=5$, and  with a time-step size of 345.6 seconds (i.e. Cr$=1.2$). The height fields at days 0, 7 and 14  are shown in Figure \figref{sswe_tc6_hars2}. 
We also compare the results of ARS2 HDG-DG with those of RK2 DG in
Figure \figref{sswe_tc6_h} after 14 days. For RK2 DG, we take the time-step size to be $43.2$ seconds (Cr=$0.14$) for stability. The height field of ARS2 HDG-DG is  in good agreement with that of RK2 DG. In particular, the relative difference in the height field is $\mc{O}\LRp{10^{-3}}$. 

  \begin{figure}[h!t!b!]
    \centering
    \subfigure[Day 0]{
      \includegraphics[trim=3.7cm 0.4cm 3.3cm 1cm,clip=true,width=0.3\columnwidth]{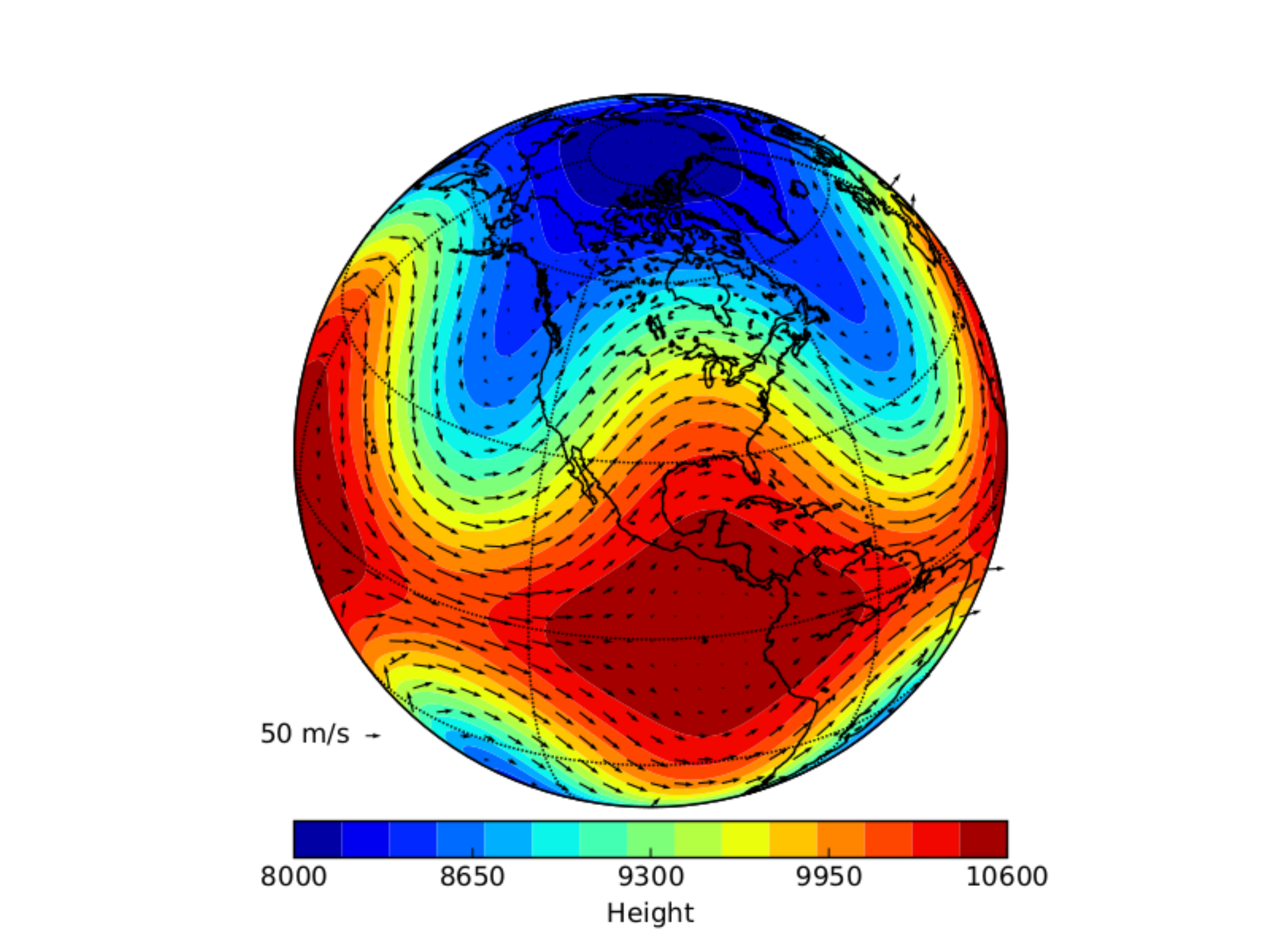}
      \figlab{sswe_tc6_hars2_day0}
    }
    \subfigure[Day 7]{
      \includegraphics[trim=3.7cm 0.4cm 3.3cm 1cm,clip=true,width=0.3\columnwidth]{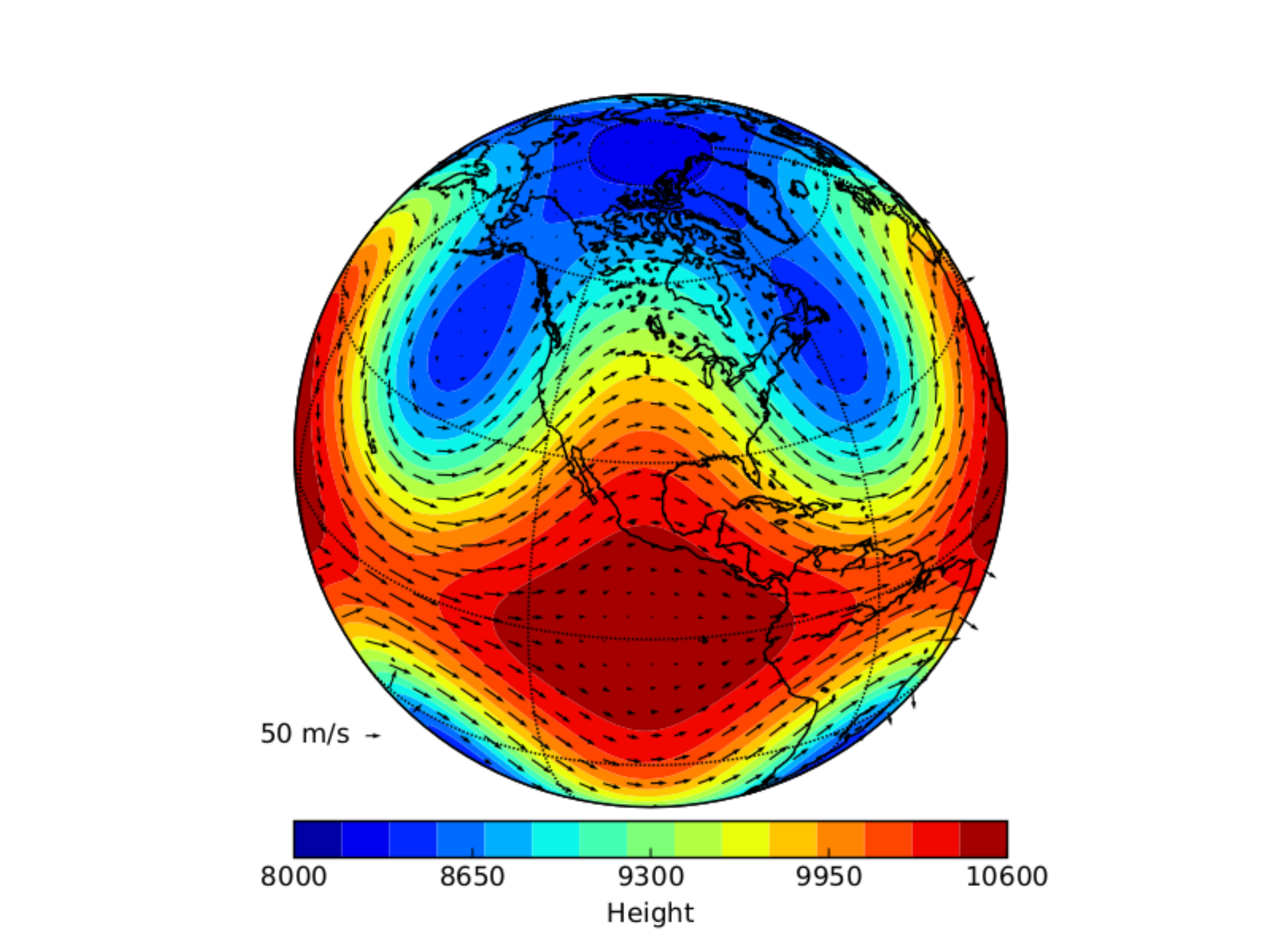}
      \figlab{sswe_tc6_hars2_day7}
    }
    \subfigure[Day 14]{
      \includegraphics[trim=3.7cm 0.4cm 3.3cm 1cm,clip=true,width=0.3\columnwidth]{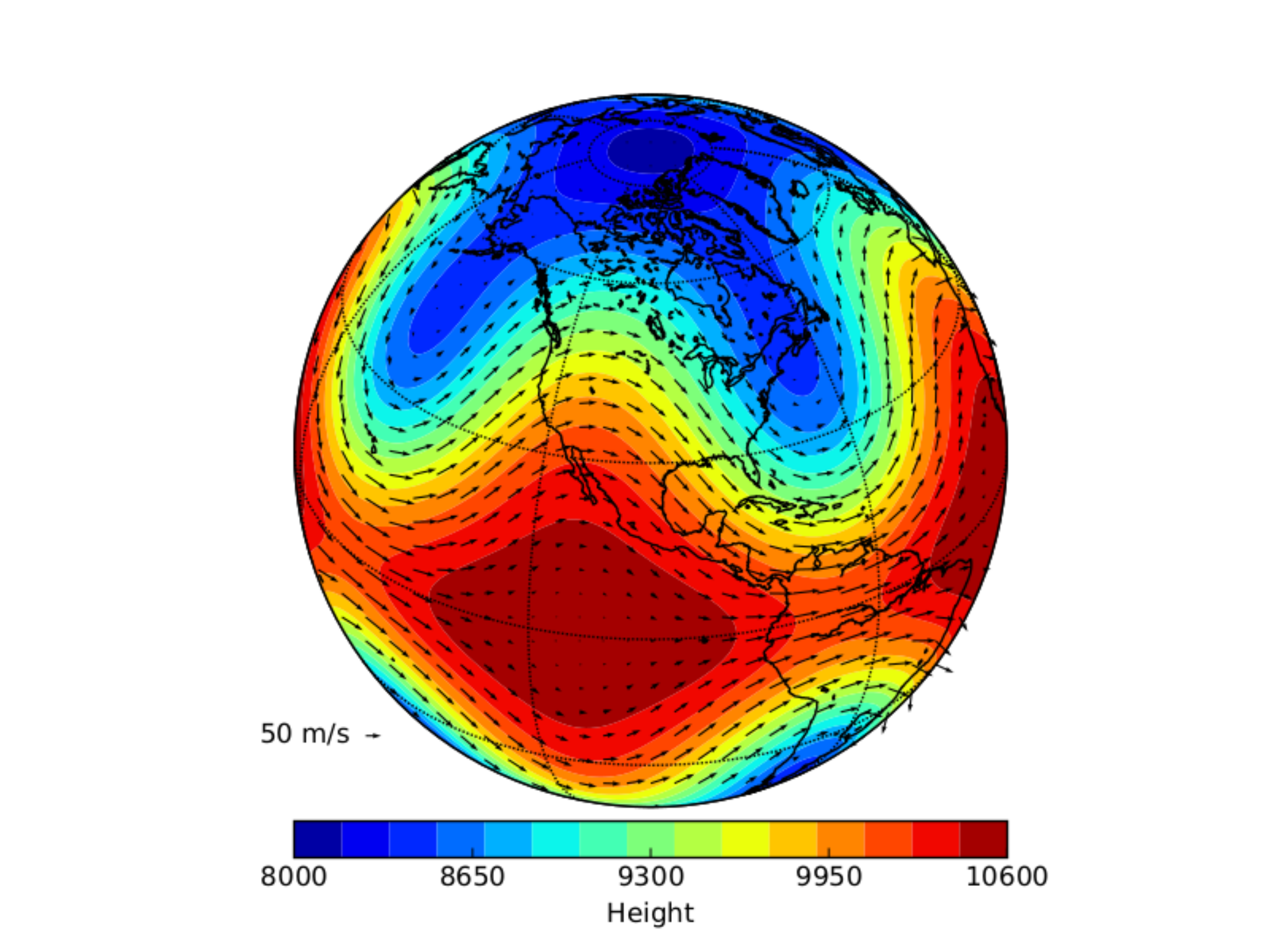}
      \figlab{sswe_tc6_hars2_day14}
    }
    \caption{Rossby-Haurwitz wave: the total height field computed from ARS2 HDG-DG after a) 0 days, (b) 7 days, and (c) 14 days. The numerical experiment is performed on the grid with $N_e=864$ elements, solution order $p=5$, and Cr $=1.2$. Contour levels are from $8000$m to $10600$m with the step size of $173$m.}
    \figlab{sswe_tc6_hars2}
  \end{figure}

  \begin{figure}[h!t!b!]
    \centering
      \includegraphics[trim=1.0cm 14.5cm 6.0cm 5.5cm,clip=true,width=0.46\columnwidth]{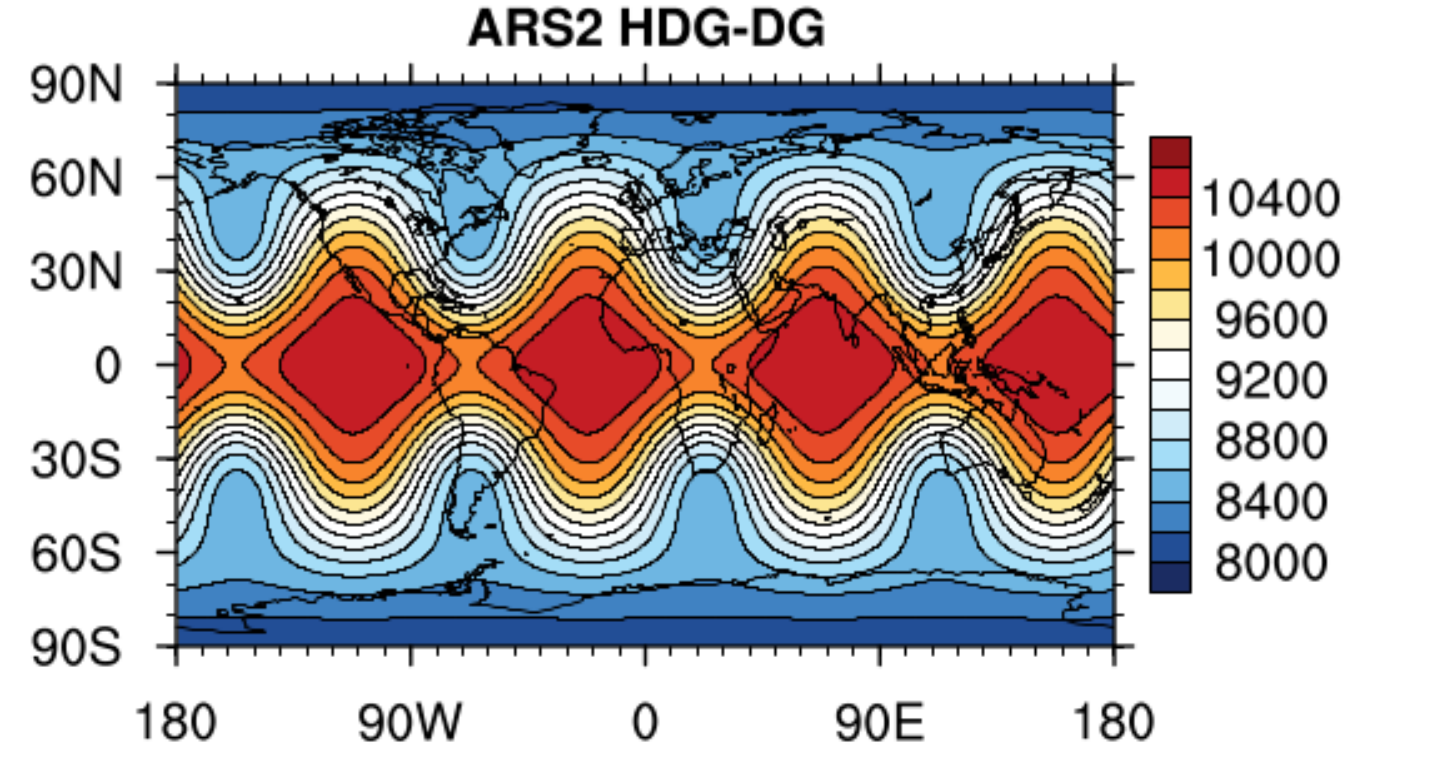}\hspace{-2mm}
      \includegraphics[trim=1.0cm 14.5cm 6.0cm 5.5cm,clip=true,width=0.46\columnwidth]{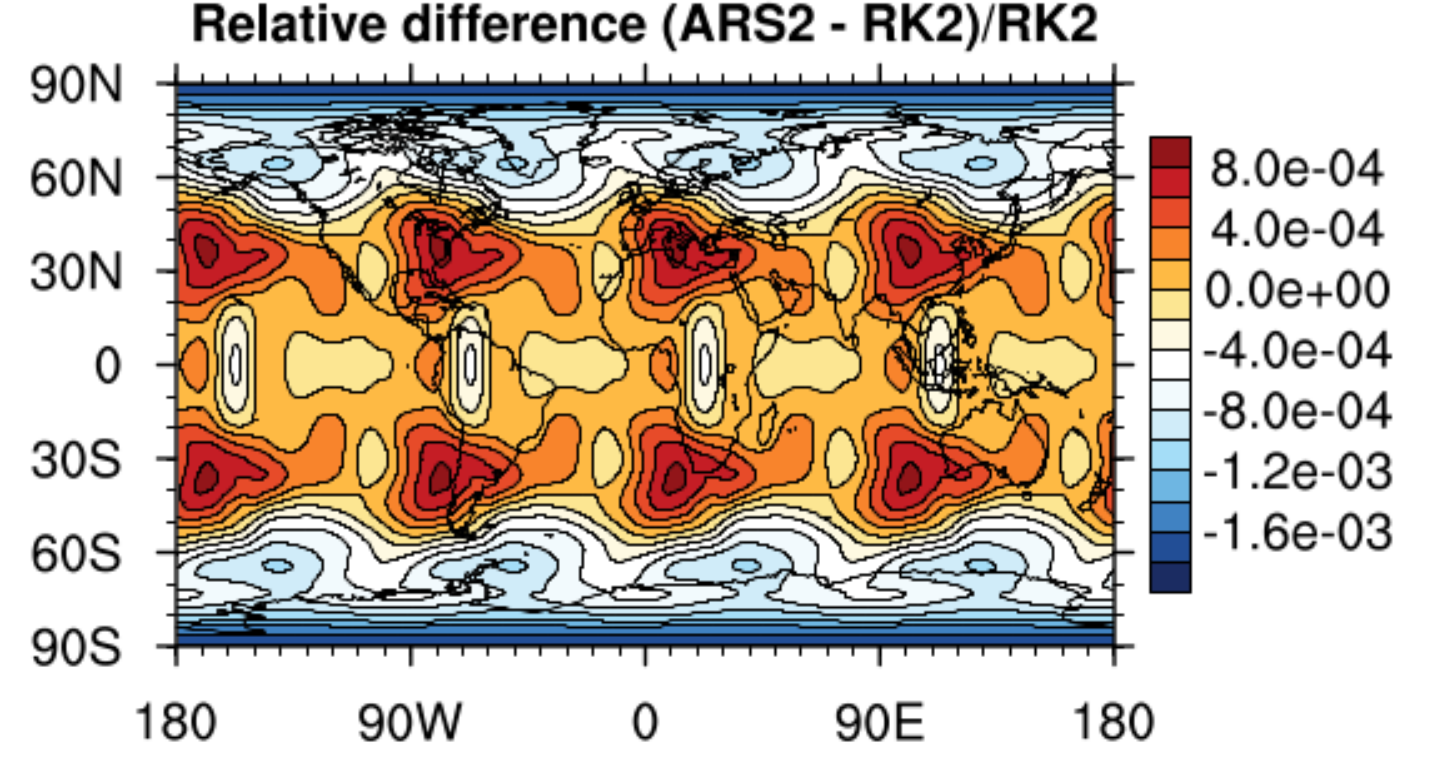}
    \caption{Rossby-Haurwitz wave: the total height field computed from ARS2 HDG-DG (left) after 14 days with Cr of $1.2$;
     the relative difference (right) with RK2 in the height field. 
     }
    \figlab{sswe_tc6_h}
  \end{figure}

\subsection{Barotropic instability test}

In this section, we consider the barotropic instability test in
\cite{galewsky2004initial}.  A zonal jet, a wind field along a
  latitude line and geostrophically balanced height field, is
  initialized in the northern hemisphere. Then, the height field is
  perturbed by adding a smoothly localized bump to the center of the
  jet, which causes barotropic waves to evolve in time.
Figure \figref{sswe_tc8a_hars2} shows the relative vorticity field of the barotropically unstable flow at days 4, 5 and 6. The numerical experiment is conducted on the grid with $N_e=5400$ elements ($30 \times 30 \times 6$), solution order $p=4$, and time-step size of $173$ seconds. The vorticity field computed from ARS2 HDG-DG is comparable to that of \cite{marras2015simulation}, which use high-order continuous and discontinuous Galerkin methods with explicit time-integration.

  \begin{figure}[h!t!b!]
    \centering
    \subfigure[Day 4]{
      \includegraphics[trim=3.43cm 0.4cm 4.1cm 1cm,clip=true,width=0.28\columnwidth]{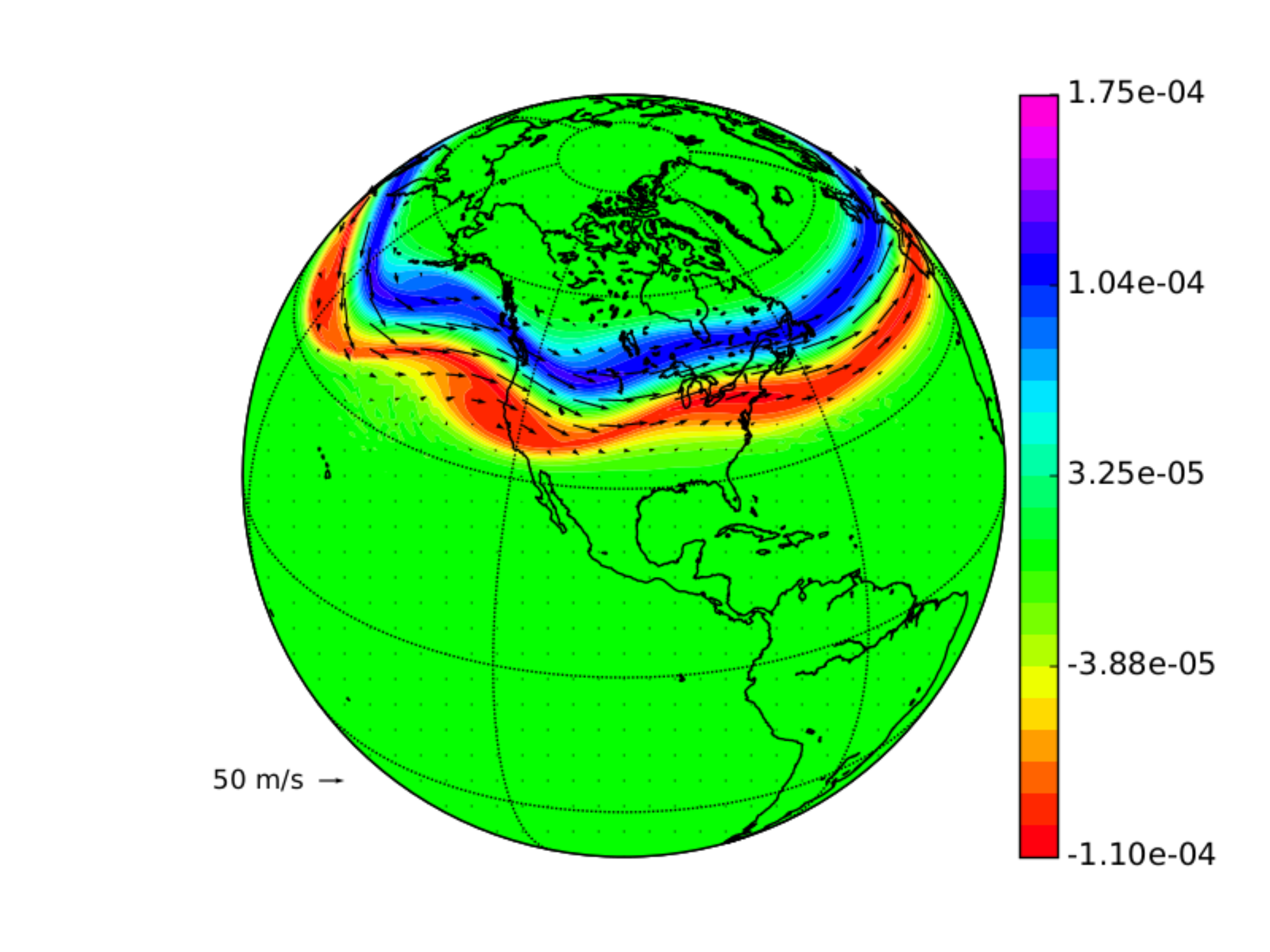}

      \figlab{sswe_tc8a_hars2_day4}
    }
    \subfigure[Day 5]{
      \includegraphics[trim=3.43cm 0.4cm 4.1cm 1cm,clip=true,width=0.28\columnwidth]{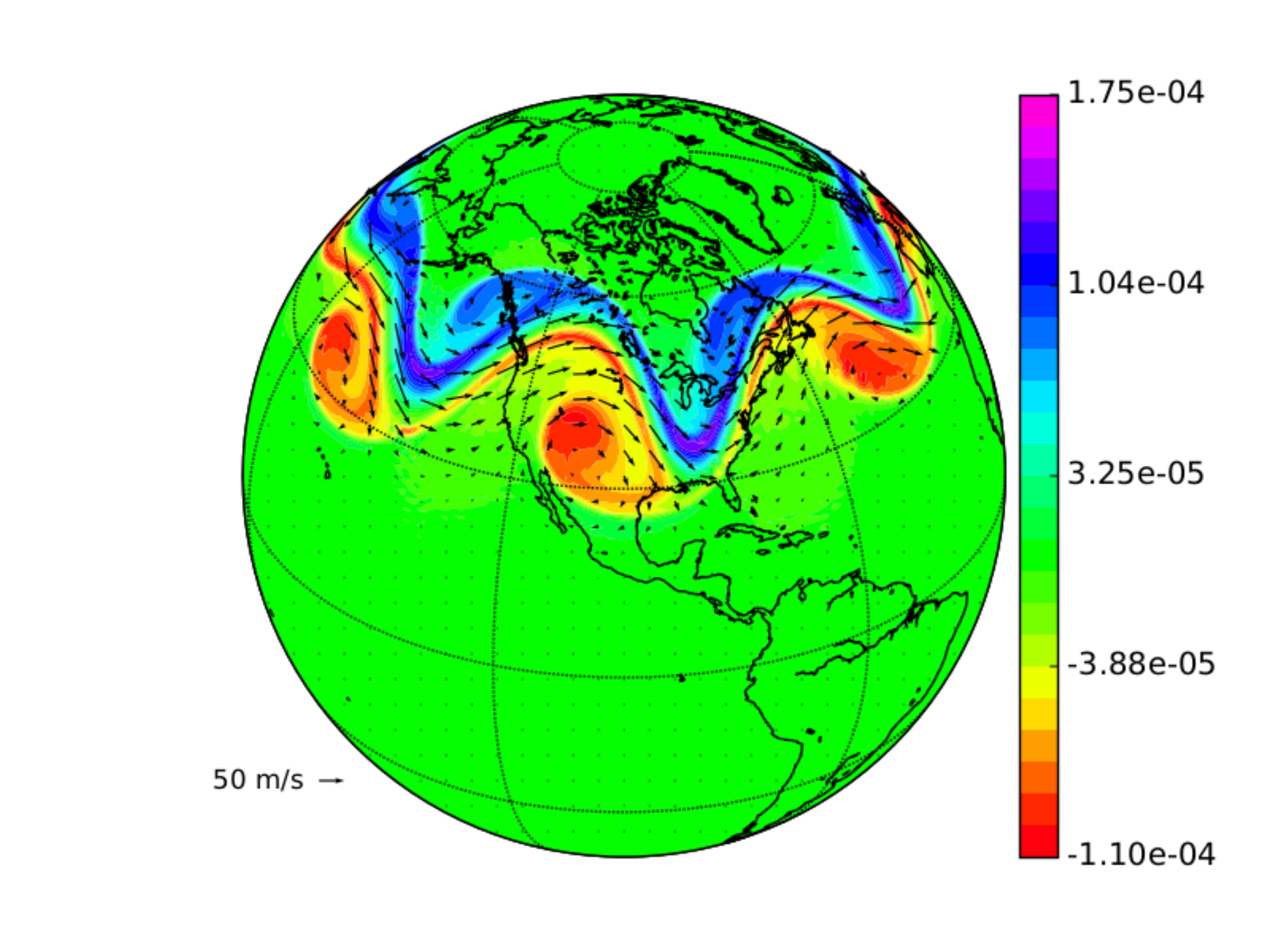}
      \figlab{sswe_tc8a_hars2_day5}
    }
    \subfigure[Day 6]{
      \includegraphics[trim=3.43cm 0.4cm 0.8cm 1cm,clip=true,width=0.35\columnwidth]{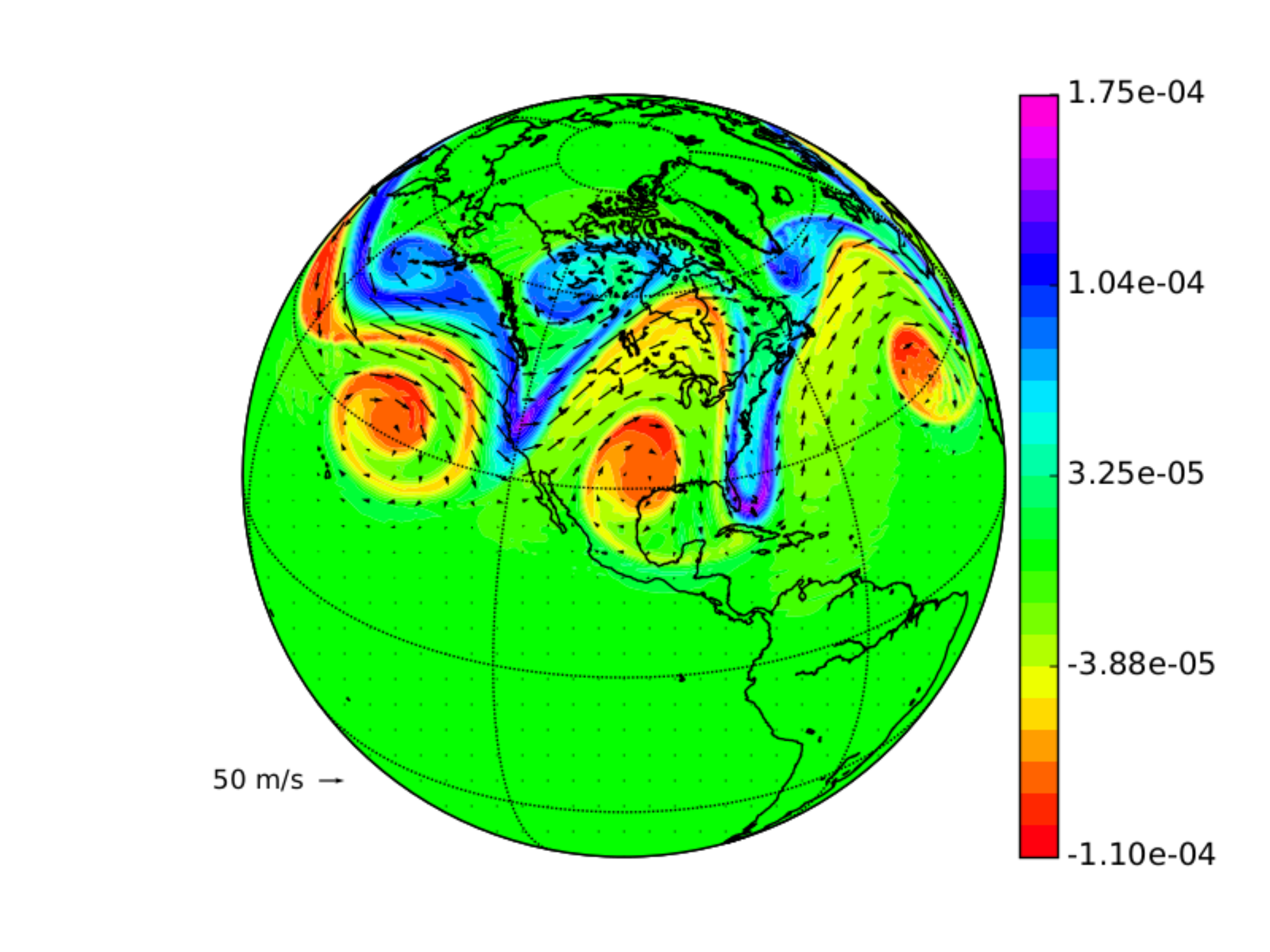}
      \figlab{sswe_tc8a_hars2_day6}
    }
    \caption{Barotropic instability test: relative vorticity field of the ARS2 HDG-DG at (a) 4 days, (b) 5 days, and (c) 6 days. The numerical experiment was performed on the grid of $N_e=5400$ and $p=4$ with a time-step size of 173 seconds (Cr=0.94). The vorticity ranges from $-1.1\times 10^{-4}$ to $1.75\times 10^4$.}
    \figlab{sswe_tc8a_hars2}
  \end{figure}

We also compare the results of ARS2 HDG-DG with those of RK2 DG in
Figure \figref{sswe_tc8a_vort} after 6 days. For RK2 DG, we take the
time-step size to be $21.6$ seconds (i.e Cr =$0.12$) for
stability. The vorticity field of ARS2 HDG-DG is in  good agreement
with that of RK2 DG. Indeed, the difference in the vorticity field is
$O(10^{-6})$.

  \begin{figure}[h!t!b!]
    \centering
    \includegraphics[trim=1.0cm 6.3cm 2.6cm 6cm,clip=true,width=0.77\columnwidth]{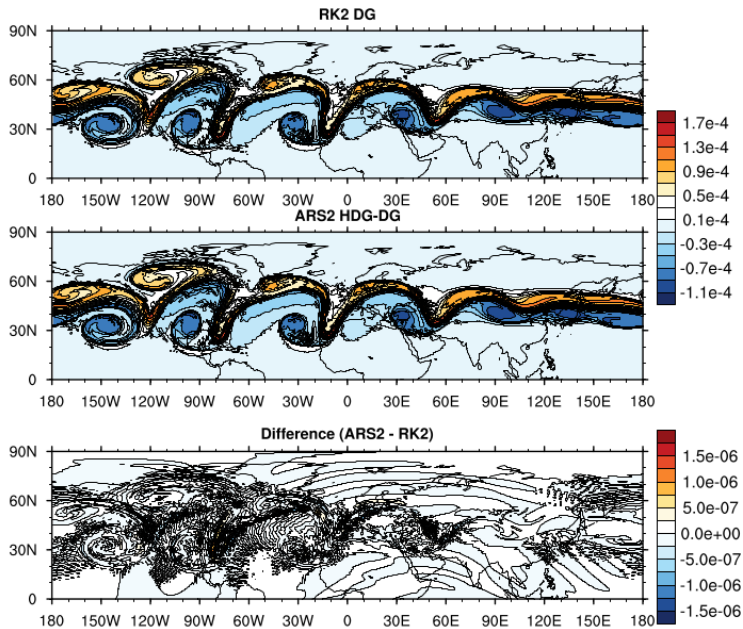}
    \caption{Barotropic instability test: relative vorticity fields of RK2 DG (top) and ARS2 HDG-DG (middle) at 6 days with Cr of $0.94$ and $0.12$, respectively; the difference (bottom) in the vorticity field between ARS2 HDG-DG and RK2 DG. 
    }
    \figlab{sswe_tc8a_vort}
  \end{figure}

\section{Conclusions}
\seclab{conclusion}

   In this paper, we are interested in subcritical shallow water
   systems in which gravity wave velocity is faster than convection
   speed. We start by decomposing the original flux into a linear part
   (obtained from linearizing the flux at the lake at rest condition)
   containing the fast gravity wave and a nonlinear part for which the
   fastest wave is removed. We spatially discretize the former using an HDG
   method, and the latter using a DG approach. This enables us to
   develop an IMEX HDG-DG framework in which we integrate the DG
   discretization explicitly and the HDG discretization
   implicitly. The purpose of our coupled approach is fourfold: (i)
   to step over the fast waves using larger time step sizes (compared to fully explicit methods) without facing instability;
   (ii) to avoid expensive Newton-type iterations (compared to fully implicit methods) for each time step; (iii) to
   take advantage of the DOF reduction in HDG method (relative to DG
   approaches) to further reduce the cost of linear solves; and (iv) to preserve high-order accuracy in both space and time.

   Numerical results have shown that while fully explicit DG
   approaches are stable with small time-step sizes, our IMEX HDG-DG
   method is stable for orders of magnitude larger time-step sizes. We
   have shown that only one forward and one backward substitutions are
   required for each stage per time-step. The numerical results also show that
   our approach achieves the expected high-order accuracy both in space
   and time. Ongoing work is to further improve the efficiency of the
   IMEX HDG-DG approach by developing preconditioned iterative methods
   for the linear solve, and to implement the scheme on parallel
   computing systems. Future developments also include construction of
   IMEX HDG-DG approach for nonhydrostatic equations (stratified
   compressible Euler/Navier-Stokes systems) where it is expected that our approach will yield greater benefits due to the stiffness of the acoustic waves (waves that carry little energy yet dominate the time-step restriction).  Of interest will be the
   rigorous convergence analysis of the IMEX-HDG-DG scheme.
  Due to the similarity between the weak Galerkin methods \cite{wang2013weak,wang2014weak} and HDGs, our ongoing work is to extend the IMEX idea to the weak Galerkin framework.

\bibliography{imexpswe_siam}

\end{document}